\documentclass[12pt]{article}

\usepackage{amsmath,graphicx}
\usepackage{simplewick}
\usepackage{epsf}
\usepackage{graphicx,epsfig}
\usepackage{amsfonts}
\usepackage{amssymb}
\usepackage[nosort]{cite}
\usepackage{setspace}

\def\bk{{\bf k}}
\def\bp{{\bf p}}
\def\bq{{\bf q}}
\def\bx{{\bf x}}
\def\by{{\bf y}}

\def\CA{{\cal A}}

\def\CH{{\cal H}}
\def\CL{{\cal L}}
\def\CO{{\cal O}}

\def\ttau{{\tilde \tau}}

\def\mpl{M_{\rm P}}
\def\half{\frac{1}{2}}

\makeatletter
\renewcommand\section{\@startsection {section}{1}{\z@}%
                                 {-3.5ex \@plus -1ex \@minus -.2ex}%
                                   {2.3ex \@plus.2ex}%
                                   {\normalfont\large\bfseries}}
\renewcommand\subsection{\@startsection{subsection}{2}{\z@}%
                                   {-3.25ex\@plus -1ex \@minus -.2ex}%
                                     {1.5ex \@plus .2ex}%
                                     {\normalfont\bfseries}}
\renewcommand\subsubsection{\@startsection{subsubsection}{3}{\z@}%
                                   {-3.25ex\@plus -1ex \@minus -.2ex}%
                                     {1.5ex \@plus .2ex}%
                                     {\normalfont\itshape}}
\makeatother

\newcommand{\Letter}{
\setlength{\textwidth}{16.5cm}
   \setlength{\textheight}{22.6cm}
    \hoffset=-0.5in
\voffset=-2.1cm }

\Letter

\begin{document}
\newcommand{\be}{\begin{equation}}
\newcommand{\ee}{\end{equation}}
\newcommand{\bea}{\begin{eqnarray}}
\newcommand{\eea}{\end{eqnarray}}
\newcommand{\barr}{\begin{array}}
\newcommand{\earr}{\end{array}}

\thispagestyle{empty}
\begin{flushright}
%\parbox[t]{1.5in}{hep-th/yymmnnn}
\end{flushright}

\vspace*{0.3in}
\begin{spacing}{1.1}

\begin{center}
{\large \bf Primordial Non-Gaussianities from Inflation Models}

\vspace*{0.5in} {Xingang Chen}
\\[.3in]
{\em Center for Theoretical Cosmology, \\
Department of Applied Mathematics and Theoretical Physics, \\
University of Cambridge, Cambridge CB3 0WA, UK
} \\[0.3in]
\end{center}

\begin{center}
{\bf
Abstract}
\end{center}
\noindent
This is a pedagogical review on primordial non-Gaussianities from inflation models. We introduce formalisms and techniques that are used to compute such quantities. We review different mechanisms which can generate observable large non-Gaussianities during inflation, and distinctive signatures they leave on the non-Gaussian profiles. They are potentially powerful probes to the dynamics of inflation. We also provide a non-technical and qualitative summary of the main results and underlying physics.

\vfill

\newpage
\setcounter{page}{1}

\tableofcontents

\newpage

\section{Introduction}
\setcounter{equation}{0}

An ambitious goal of modern cosmology is to understand the origin of our Universe, all the way to its very beginning. To what extend this can be achieved largely depends on what type of observational data we are able to get. Thanks to many modern experiments, we are really making progress in this direction.

One of the representative experiments is the Wilkinson Microwave Anisotropy Probe (WMAP) satellite \cite{Komatsu:2010fb,Gold:2010fm,Larson:2010gs,Weiland:2010ij,Jarosik:2010iu,Bennett:2010jb}. It measures the light coming from the last scattering surface about 13.7 billions years ago. This cosmic microwave background (CMB) is emitted at about 379,000 years after the Big Bang, when electrons and protons combine to form neutral hydrogen atoms and photons start to travel freely through the space. Our Universe was very young at that moment and the large scale fluctuations were still developing at linear level. So the CMB actually carries valuable information much earlier than itself, which can potentially tell us about the origin of the Big Bang.

There are two amazing facts about the CMB temperature map. On the one hand, it is extremely isotropic, despite the fact that the causally connected region at the time when CMB formed spans an angle of only about $0.8$ degree on the sky today. On the other hand, we do observe small fluctuations, with $\Delta T/T \sim 10^{-5}$.

The inflationary scenario \cite{Guth:1980zm,Linde:1981mu,Albrecht:1982wi} naturally solves the above two puzzles. It was proposed nearly thirty years ago to address some of the basic problems of the Big Bang cosmology, namely, why the universe is so homogeneous and isotropic. In this scenario, our universe was once dominated by dark energy and had gone through an accelerated expansion phase, during which a Hubble size patch was stretched by more than 60 efolds or so. Inhomogeneities and large curvature were stretched away by this inflationary epoch, making our current observable universe very homogenous and flat. In the mean while, the fields that were responsible for and participated in this inflationary phase did have quantum fluctuations. These fluctuations also got stretched and imprinted at superhorizon scales. Later they reentered the horizon and seeded the large scale structures today \cite{Mukhanov:1981xt,Hawking:1982cz,Starobinsky:1982ee,Guth:1982ec,Bardeen:1983qw}.

The inflationary scenario has several generic predictions on the properties of the density perturbations that seed the large scale structures:
\begin{itemize}
\item
They are
primordial. Namely, they were laid down at superhorizon scales and entering the horizon after the Big Bang.
\item
They are approximately scale-invariant. This is because, during the 60 efolds, each mode experiences the similar expansion when they are stretched across the horizon.
\item
They are approximately Gaussian. In simplest slow-roll inflation models, the inflaton is freely propagating in the inflationary background at the leading order. This is also found to be true in most of the other models and for different inflationary mechanisms. So the tiny primordial fluctuations can be treated as nearly Gaussian.
\end{itemize}

The CMB temperature anisotropy is the ideal data that we can use to test these predictions. The obvious first step is to analyze their two-point correlation functions, i.e.~the power spectrum. All the above predictions are verified to some extent \cite{Komatsu:2010fb}. The presence of the baryon acoustic oscillations proves that the density perturbations are indeed present at the superhorizon scales and reentering the horizon as the horizon expands after the Big Bang. The spectral index, $n_s=0.963 \pm 0.012$, is very close to one, therefore, the density perturbations are nearly scale invariant. Several generic types of non-Gaussianities are constrained to be less than one thousandth of the leading Gaussian component.

But is this enough?

Experimentally, the amplitude and the scale-dependence of the power spectrum consist of about $1000$ numbers for WMAP. For the Planck satellite, this will be increased up to about $3000$. However, we have about one million pixels in the WMAP temperature map alone, and 60 millions for Planck. So the information that we got so far is highly compressed comparing to what the data could offer in principle. This high compression is only justified if the density perturbations are Gaussian within the ultimate sensitivities of our experiments, so all the properties of the map is determined by the two-point function. Otherwise, we are expecting a lot more information in the non-Gaussian components.

Theoretically, inflation still remains as a paradigm. We do not know what kind of fields are responsible for the inflation. We do not know their Lagrangian. We also would like to distinguish inflation from other alternatives. Being our very first data on quantum gravity, we would like to extract the maximum number of information from the CMB map to understand aspects of the quantum gravity. All these motivate us to go beyond the power spectrum.

To give an analogy, in particle physics, two-point correlation functions of fields describe freely propagating particles in Minkowski spacetime. More interesting objects are their higher order correlations. Measuring these are the goals of particle colliders. Similarly, the power spectrum here describes the freely propagating particles in the inflationary background. To find out more about their interaction details and break the degeneracies among models, we need higher order correlation functions, namely non-Gaussianities. So the role non-Gaussianities play for the very early universe is similar to the role colliders play for particle physics.

With these motivations in mind, in this review, we explore various mechanisms that can generate potentially observable primordial non-Gaussianities, and are consistent with the current results of power spectrum. We will not take the approach of reviewing models one by one. Rather, we divide them into different categories, such that models in each category share the same physical aspect which leaves a unique fingerprint on primordial non-Gaussianities.
On the one hand, if any such non-Gaussianity is observed, we know what we have learned concretely in terms of fundamental physics; on the other hand, explicit forms of non-Gaussianities resulted from this exploration provide important clues on how they should be searched in data.
Even if the primordial density perturbations were perfectly Gaussian, to test it, we would still go through these analyses until various well-motivated non-Gaussian forms are properly constrained.

\subsection{Road map}

The following is the outline of the review. For readers who would like to get a quick and qualitative understanding of the main results instead of technical details, we also provide a shortcut after the outline.

In Sec.~\ref{Sec:inflation}, we review the essential features of the inflation model and density perturbations.

In Sec.~\ref{Sec:in-in_and_correl}, we review the first-principle in-in formalism and related techniques that will be used to calculate the correlation functions in time-dependent background.

In Sec.~\ref{Sec:Nogo}, we compute the scalar three-point function in the simplest slow-roll model. We list the essential assumptions that lead to the conclusion that the non-Gaussianities in this model is too small to be observed.

In Sec.~\ref{Sec:BeyondNogo}, we review aspects of inflation model building, emphasizing various generic problems which suggest that the realistic model may not be the algebraically simplest. We also introduce some terminologies used to describe properties of non-Gaussianities.

Sec.~\ref{Sec:Single}, \ref{Sec:quasi} and \ref{Sec:multifield} contain the main results of this review. We study various mechanisms that can lead to large non-Gaussianities, and their distinctive predictions in terms of the non-Gaussian profile.

In Sec.~\ref{Sec:sum_and_dis}, we give a qualitative summary of the main results in this review. Before conclusion, we discuss several useful relations among different non-Gaussianities.

Here is a road map for readers who wish to have a non-technical explanation and understanding of our main results.
After reading the short review on the inflation model and density perturbations in Sec.~\ref{Sec:inflation}, one may read the first and the last paragraph of Sec.~\ref{Sec:Nogo} to get an idea of the no-go statement, and then directly proceed to read Sec.~\ref{Sec:BeyondNogo}. After that, one may jump to Sec.~\ref{Sec:sum_and_dis} where the main results are summarized in non-technical terms.

\medskip

The subject of the primordial non-Gaussianities is a fast-growing one. There exists many nice reviews and books in this and closely related subjects. The introductions to inflation and density perturbations can be found in many textbooks \cite{Kolb:1988aj,Linde:2005ht,Liddle:2000cg,Dodelson:2003ft,Mukhanov:2005sc,Weinberg:2008zzc} and reviews \cite{Wands:2007bd,Malik:2008im,Kinney:2009vz,Baumann:2009ds,Langlois:2010xc}. Inflationary model buildings in particle physics, supergravity and string theory are reviewed in Ref.~\cite{Lyth:1998xn,Quevedo:2002xw,HenryTye:2006uv,Cline:2006hu,Burgess:2007pz,McAllister:2007bg,Mazumdar:2010sa}. Comprehensive reviews on the developments of theories and observations of primordial non-Gaussianities up to mid 2004 can be found in Ref.~\cite{Komatsu:2002db,Bartolo:2004if}. Recent comprehensive reviews on theoretical and observational developments on the bispectrum detection in CMB and large scale structure has appeared in Ref.~\cite{Liguori:2010hx,Verde:2010wp}.
A recent comprehensive review on non-Gaussianities from the second order post inflationary evolution of CMB, which acts as contaminations of the primordial non-Gaussianities, has appeared in Ref.~\cite{Bartolo:2010qu}.
A recent review on how primordial non-Gaussianities can be generated in alternatives to inflation can be found in Ref.~\cite{Lehners:2010fy}.

\section{Inflation and density perturbations}
\label{Sec:inflation}

In this section, we give a quick review on basic elements of inflation and density perturbations. We consider the simplest slow-roll inflation. The action is
\bea
S = \frac{\mpl}{2} \int d^4x \sqrt{-g} R +
\int d^4x \sqrt{-g} \left[ -\half g^{\mu\nu} \partial_\mu \phi \partial_\nu \phi - V(\phi) \right] ~.
\eea
The first term is the Einstein-Hilbert action. The second term describes a canonical scalar field coupled to gravity through the metric $g_{\mu\nu}$. This is the inflaton, which stays on the potential $V(\phi)$ and creates the vacuum energy that drives the inflation.
We first study the zero-mode background evolution of the spacetime and the inflaton. The background metric is
\bea
ds^2 = g_{\mu\nu}dx^\mu dx^\nu = -dt^2 + a^2(t) d\bx^2 ~,
\eea
where $a(t)$ is the scale factor and $\bx$ is the comoving spatial coordinates.
The equations of motion are
\begin{align}
& H^2 = \frac{1}{3\mpl^2} ( \half \dot\phi_0^2 + V ) ~,
\label{H_eqn}
\\
& \dot H = - \frac{\dot\phi_0^2}{2 \mpl^2} ~,
\label{continuity_eqn}
\\
& \ddot \phi_0 + 3H\dot\phi_0 + V' = 0 ~.
\label{phi_0_eqn}
\end{align}
The first equation determines the Hubble parameter $H$, which is the expansion rate of the universe. The second equation is the continuity condition. The third equation describes the evolution of the inflaton. Only two of them are independent.

The requirement of having at least $\CO(60)$ e-fold of inflation imposes some important conditions. By definition, to have this amount of inflation, the Hubble parameter cannot change much within a Hubble time $H^{-1}$. This gives the first condition
\bea
\epsilon \equiv - \frac{\dot H}{H^2} \ll \CO(1) ~.
\label{sr_cond_1}
\eea
We also require that the parameter $\epsilon$ does not change much within a Hubble time,
\bea
\eta \equiv \frac{\dot \epsilon}{\epsilon H} \ll \CO(1) ~.
\label{sr_cond_2}
\eea
In principle, $\eta$ can be close to $\CO(1)$ but $\epsilon$ kept small. In such a case, $\epsilon$ grows exponentially with efolds and the inflation period tends to be shorter. More importantly, such a case will not generate a scale-invariant spectrum, as we will see shortly, thus cannot be responsible for the CMB. The two conditions (\ref{sr_cond_1}) and (\ref{sr_cond_2}) are called the {\em slow-roll conditions}. Using the background equations of motion, we can see that the slow-roll conditions impose restrictions on the rolling velocity of the inflaton. The first condition (\ref{sr_cond_1}) implies that
\bea
\frac{\dot\phi_0^2}{2 H^2 \mpl^2} =\epsilon \ll \CO(1) ~.
\label{epsilon_def2}
\eea
So the energy driving the inflation on the right hand side of (\ref{H_eqn}) is dominated by the potential. Adding the second condition (\ref{sr_cond_2}) further implies that
\bea
\frac{\ddot\phi_0}{\dot\phi_0 H} = -\epsilon + \frac{\eta}{2}
\ll \CO(1) ~.
\label{eta_def2}
\eea
So the first term $\ddot \phi_0$ in (\ref{phi_0_eqn}) is negligible and the evolution of the zero-mode inflaton is determined by the attractor solution \bea
3H\dot\phi_0 + V' =0 ~.
\label{phi_0_attractor}
\eea
Using (\ref{phi_0_attractor}), the slow-roll conditions can also be written in a form that restricts the shape of the potential,
\begin{align}
\epsilon_V \equiv \frac{\mpl^2}{2} \left( \frac{V'}{V} \right)^2 \ll \CO(1) ~,
\quad
\eta_V \equiv \mpl^2 \frac{V''}{V} \ll \CO(1) ~.
\label{slowV}
\end{align}
They are related to $\epsilon$ and $\eta$ by
\bea
\epsilon = \epsilon_V ~,~~~~ \eta = -2 \eta_V + 4 \epsilon_V ~.
\label{slow_relation}
\eea
So the shape of the potential has to be rather flat relative to its height.
We emphasize that, although in this example several definitions of the slow-roll conditions are all equivalent, the definition (\ref{sr_cond_1}) and (\ref{sr_cond_2}) are more general. In other cases that we will encounter later in this review, these two conditions are still necessary to ensure a prolonged inflation and generate a scale-invariant spectrum, but the others no longer have to be satisfied. For example, the shape of potential can be steeper, or the inflationary energy can be dominated by the kinetic energy.

Now let us study the perturbations. To keep things simple but main points illustrated, in this section, we will ignore the perturbations in the gravity sector and only perturb the inflaton,
\bea
\phi(\bx,t) = \phi_0(t) + \delta \phi(\bx,t) ~.
\eea
We also ignore terms suppressed by the slow-roll parameters, which we often denote collectively as $\CO(\epsilon)$. For example, the mass of the inflaton is $V'' \sim \CO(\epsilon) H^2$ and will be ignored.
The quadratic Lagrangian for the perturbation theory is simply
\bea
L = \int d^3x \left[ \frac{a^3}{2} \dot{\delta\phi}^2 - \frac{a}{2} (\partial_i \delta\phi)^2 \right] ~,
\eea
and the equation of motion follows,
\bea
\ddot {\delta\phi}(\bk,t) + 3 H \dot{\delta\phi}(\bk,t) + \frac{k^2}{a^2} \delta\phi(\bk,t) =0 ~.
\label{deltaphi_eqn}
\eea
where we have written it in the comoving momentum space,
\bea
\delta\phi(\bk,t) = \int d^3x \delta\phi(\bx,t) e^{i\bk \cdot \bx} ~.
\eea
The solution to the differential equation (\ref{deltaphi_eqn}), $u(\bk,t)$, is called the mode function. It is not difficult to check that
\bea
a^3 u(\bk,t) \dot u^*(\bk,t) - {\rm c.c.} = t \textrm{-} {\rm independent~const.} ~.
\label{uustar}
\eea
To quantize the perturbations according to the canonical commutation relations between $\delta\phi$ and its conjugate momentum $\delta\pi\equiv \partial L/\partial \delta\dot\phi$,
\begin{align}
& [ \delta\phi(\bx,t), \delta\pi(\by,t) ] = i\delta (\bx-\by) ~,
\nonumber \\
& [ \delta\phi(\bx,t), \delta\phi(\by,t) ] = 0 ~,
\quad
[ \delta\pi(\bx,t), \delta\pi(\by,t) ] = 0 ~,
\label{commutation_phipi}
\end{align}
we decompose
\begin{align}
& \delta\phi = u(\bk,t) a_{\bk} + u^*(-\bk,t) a^\dagger_{-\bk} ~,
\label{delta_phi_decomp}
\\
& \delta\pi = a^3 \dot u(\bk,t) a_{\bk} + a^3 \dot u^*(-\bk,t) a^\dagger_{-\bk} ~,
\label{delta_pi_decomp}
\end{align}
with the commutation relations
\begin{align}
& [ a_\bp, a^\dagger_{-\bq} ] = (2\pi)^3 \delta^3(\bp+\bq) ~,
\nonumber \\
& [ a_\bp, a_{-\bq} ] = 0 ~,
\quad
[ a^\dagger_\bp, a^\dagger_{-\bq} ] = 0 ~.
\label{commutation_aadagger}
\end{align}
One can check that the commutation relations (\ref{commutation_phipi}) and (\ref{commutation_aadagger}) are equivalent because of (\ref{uustar}), given that the constant on the right hand side of (\ref{uustar}) is specified to be $i$. This gives the normalization condition for the mode function.

We now write down the mode function explicitly by solving (\ref{deltaphi_eqn}),
\bea
u(\bk,\tau) = C_+ \frac{H}{\sqrt{2k^3}} ( 1+ik\tau) e^{-ik\tau}
+ C_- \frac{H}{\sqrt{2k^3}} ( 1-ik\tau) e^{ik\tau}
~,
\label{u_superposition}
\eea
where we have used the conformal time $\tau$ defined as $dt\equiv ad\tau$. The infinite past corresponds to $\tau \to - \infty$ and the infinite future $\tau \to 0$. We also used the relation $\tau = -1/Ha +\CO(\epsilon)$. This mode function is a superposition of two linearly independent solutions with the normalization condition
\bea
|C_+|^2-|C_-|^2 = 1
\label{C_normal}
\eea
followed from (\ref{uustar}). Consider the limit in which the mode is well within the horizon, i.e.~its wavelength $a/k$ much shorter than the Hubble length $1/H$, and consider a time period much shorter than a Hubble time. In these limits, the mode effectively feels the Minkowski spacetime, and the first component in (\ref{u_superposition}) with the positive frequency asymptotes to the vacuum mode of the Minkowski spacetime as we can see from (\ref{C_normal}). We choose this component as our vacuum choice, and it is usually called the Bunch-Davies state. The annihilation operator $a_\bp$ annihilates the corresponding Bunch-Davies vacuum, $a_\bp |0\rangle =0$.

The mode function
\bea
u(\bk,\tau) = \frac{H}{\sqrt{2k^3}} ( 1+ik\tau) e^{-ik\tau}
\label{u_BD}
\eea
has the following important properties.
It is oscillatory within the horizon $k|\tau| \gg 1$. As it gets stretched out of the horizon $k|\tau| \ll 1$, the amplitude becomes a constant and frozen. Physically this means that, if we look at different comoving patches of the universe that have the superhorizon size, and ignore the shorter wavelength fluctuations, they all evolve classically but with different $\delta\phi$. This difference makes them arrive at $\phi_f$, the location of the end of inflation, at different times. This space-dependent time difference $\delta t \approx \delta \phi/\dot \phi_0$ leads to the space-dependent inflationary e-fold difference
\bea
\zeta \approx H\delta t \approx H \frac{\delta \phi}{\dot \phi_0} ~.
\label{zeta_phi_relation}
\eea
Again we ignore terms that are suppressed by the slow-roll parameters.
This e-fold difference is the conserved quantity after the mode exits the horizon, and remains so until the mode re-enters the horizon sometime after the Big Bang. It is the physical quantity that we can measure, for example, by measuring the temperature anisotropy in the CMB, $\zeta\approx -5 \Delta T/T$ \cite{Sachs:1967er}. The information about the primordial inflation is then encoded in the statistical properties of this variable. So we would like to calculate the correlation functions of this quantity. Using (\ref{zeta_phi_relation}), (\ref{delta_phi_decomp}), (\ref{u_BD}) and (\ref{epsilon_def2}), we get the following two-point function,
\bea
\langle \zeta^2 \rangle \equiv \langle 0| \zeta(\bk_1,0) \zeta(\bk_2,0) |0\rangle = \frac{P_\zeta}{2 k_1^3} (2\pi)^5 \delta(\bk_1 + \bk_2) ~,
\label{2pt_def}
\eea
where $P_\zeta$ is defined as the power spectrum and in this case it is
\bea
P_\zeta = \frac{H^2}{8\pi^2 \mpl^2 \epsilon} ~.
\eea
The spectrum index is defined to be
\bea
n_s-1 \equiv \frac{d\ln P_\zeta}{d\ln k} = -2\epsilon -\eta ~,
\eea
where the relation $d\ln k = H dt$ is used. If $n_s=1$, the spectrum is scale invariant. The current data from CMB tells us $n_s = 0.963\pm 0.012$ \cite{Komatsu:2010fb}. So as we have mentioned, this requires a small $\eta$, which is also a value that tends to give more e-folds of inflation.

If this were the end of story, all the primordial density perturbations would be determined by this two-point function and they are Gaussian. The rest of the review will be devoted to making the above procedure rigorous and to the calculations of higher order non-Gaussian correlation functions in this and various other models.

\section{In-in formalism and correlation functions}
\label{Sec:in-in_and_correl}
\setcounter{equation}{0}

In this section, we review the in-in formalism and the related techniques that are used to calculate the correlation functions in time-dependent background. The main procedure is summarized in the last subsection.

\subsection{In-in formalism}
\label{Sec:in-in}

We start with the in-in formalism \cite{inin,Schwinger:1960qe,Bakshi:1962dv,Keldysh:1964ud,Calzetta:1986ey,Weinberg:2005vy}, following Weinberg's presentation \cite{Weinberg:2005vy}.

We are interested in the correlation functions of superhorizon primordial perturbations generated during inflation. So our goal is to calculate the expectation value of an operator $Q$, which is a product in terms of field perturbations $\delta\phi_a$ and $\delta \pi_a$, at the end of inflation. The subscript $a$ labels different fields. In inflation models, these fields are, for example, the fluctuations of the scalars and metric and their conjugate momenta. In the Heisenberg picture,
\bea
\langle Q \rangle \equiv \langle \Omega| Q(t) |\Omega \rangle ~,
\label{expect_value}
\eea
where $t$ is the end of inflation, $|\Omega\rangle$ is the vacuum state for this interacting theory at the far past $t_0$.

We start by looking at how the time-dependence in $Q(t)$ is generated.

The Hamiltonian of the system
\bea
H\left[ \phi(t),\pi(t) \right]
\equiv \int d^3x {\cal H} \left[ \phi_a(\bx,t),\pi_a(\bx,t) \right]
\label{Hamiltonian}
\eea
is a functional of the fields $\phi_a(\bx,t)$ and their conjugate momenta $\pi_a(\bx,t)$ at a fixed time $t$. On the left hand side of (\ref{Hamiltonian}) we have suppressed the variable $\bx$ and index $a$ which are integrated or summed over. The $\phi_a(\bx,t)$ and $\pi_a(\bx,t)$ satisfy the canonical commutation relations
\begin{align}
& \left[ \phi_a(\bx,t),\pi_b(\by,t)\right] = i\delta_{ab} \delta^3(\bx-\by) ~,
\nonumber \\
& \left[ \phi_a(\bx,t),\phi_b(\by,t)\right] =
\left[ \pi_a(\bx,t),\pi_b(\by,t) \right] =0 ~,
\label{comm_relation_0}
\end{align}
and their evolution is generated by $H$ following the equations of motion,
\begin{align}
& \dot \phi_a(\bx,t) = i \left[ H[\phi(t),\pi(t)],\phi_a(\bx,t) \right] ~,
& \dot \pi_a(\bx,t) = i \left[ H[\phi(t),\pi(t)],\pi_a(\bx,t) \right] ~.
\label{evolution_eom}
\end{align}

We consider a time-dependent background, $\bar\phi_a(\bx,t)$ and $\bar\pi_a(\bx,t)$ which are c-numbers and commute with everything, and the perturbations, $\delta\phi_a(\bx,t)$ and $\delta\pi_a(\bx,t)$,
\begin{align}
\phi_a(\bx,t) \equiv \bar\phi_a(\bx,t) + \delta\phi_a(\bx,t) ~,
\quad
\pi_a(\bx,t) \equiv \bar\pi_a(\bx,t) + \delta\pi_a(\bx,t) ~.
\end{align}
The background evolution is determined by the classical equations of motion,
\bea
\dot{\bar\phi}_a(\bx,t) = \frac{\partial \CH}{\partial \bar\pi_a} ~,
\quad
\dot{\bar\pi}_a(\bx,t) = - \frac{\partial \CH}{\partial \bar\phi_a} ~.
\label{bkgd_eom}
\eea
The commutation relations (\ref{comm_relation_0}) become those for the perturbations,
\begin{align}
& \left[ \delta\phi_a(\bx,t),\delta\pi_b(\by,t)\right] = i\delta_{ab} \delta^3(\bx-\by) ~,
\nonumber \\
& \left[ \delta\phi_a(\bx,t),\delta\phi_b(\by,t)\right] =
\left[ \delta\pi_a(\bx,t),\delta\pi_b(\by,t) \right] =0 ~.
\label{comm_relation}
\end{align}
We expand the Hamiltonian as
\begin{align}
H\left[ \phi(t),\pi(t) \right] =&~ H \left[ \bar\phi(t),\bar\pi(t) \right]
+ \sum_a \int d^3x \frac{\partial \CH}{\partial \bar\phi_a(\bx,t)} \delta\phi_a(\bx,t)
+ \sum_a \int d^3x \frac{\partial \CH}{\partial \bar\pi_a(\bx,t)} \delta\pi_a(\bx,t)
\nonumber \\
& + \tilde H \left[ \delta\phi(t),\delta\pi(t);t \right] ~,
\label{Hexpand}
\end{align}
where we use $\tilde H$ to denote terms of quadratic and higher orders in perturbations.

Using (\ref{bkgd_eom}), (\ref{comm_relation}) and (\ref{Hexpand}), the equations of motion (\ref{evolution_eom}) become
\begin{align}
\delta\dot\phi_a(\bx,t) = i \left[ \tilde H \left[ \delta\phi(t),\delta\pi(t);t \right],\delta\phi_a(\bx,t) \right] ~,
\quad
\delta\dot\pi_a(\bx,t) = i \left[ \tilde H \left[ \delta\phi(t),\delta\pi(t);t \right],\delta\pi_a(\bx,t) \right] ~.
\label{delta_eom}
\end{align}
So the evolution of the perturbations, $\delta\phi_a$ and $\delta\pi_a$, is generated by $\tilde H$. It is straightforward to verify that the solutions for (\ref{delta_eom}) are
\begin{align}
\delta\phi_a(\bx,t) = U^{-1}(t,t_0) \delta\phi_a(\bx,t_0) U(t,t_0) ~,
\quad
\delta\pi_a(\bx,t) = U^{-1}(t,t_0) \delta\pi_a(\bx,t_0) U(t,t_0) ~,
\end{align}
where $U$ satisfies
\bea
\frac{d}{dt}U(t,t_0) = - i \tilde H \left[ \delta\phi(t_0),\delta\pi(t_0);t \right]  U(t,t_0)
\label{dUdt}
\eea
with the condition at an initial time $t_0$ being
\bea
U(t_0,t_0)=1 ~.
\eea

To have a systematic scheme to do the perturbation theory, we split $\tilde H$ into two parts,
\bea
\tilde H \left[ \delta\phi(t),\delta\pi(t);t\right]
= H_0 \left[ \delta\phi(t),\delta\pi(t);t\right]
+ H_I \left[ \delta\phi(t),\delta\pi(t);t\right] ~.
\label{Hsplit}
\eea
The $H_0$ is the quadratic kinematic part, which in the perturbation theory will describe the leading evolution of fields. Fields whose evolution are generated by $H_0$ are called the {\em interaction picture fields}. We add a superscript "$I$" to label such fields. They satisfy
\begin{align}
\delta\dot\phi^I_a(\bx,t) = i \left[ H_0 \left[ \delta\phi^I(t),\delta\pi^I(t);t \right],\delta\phi^I_a(\bx,t) \right] ~,
\quad
\delta\dot\pi^I_a(\bx,t) = i \left[ H_0 \left[ \delta\phi^I(t),\delta\pi^I(t);t \right],\delta\pi^I_a(\bx,t) \right] ~.
\label{delta_eom_I}
\end{align}
The solutions are
\begin{align}
\delta\phi_a^I(\bx,t) = U_0^{-1}(t,t_0) \delta\phi_a(\bx,t_0) U_0(t,t_0) ~,
\quad
\delta\pi_a^I(\bx,t) = U_0^{-1}(t,t_0) \delta\pi_a(\bx,t_0) U_0(t,t_0) ~,
\end{align}
where $U_0$ satisfies
\bea
\frac{d}{dt}U_0(t,t_0) = - i H_0 \left[ \delta\phi(t_0),\delta\pi(t_0);t \right]  U_0(t,t_0)
\label{dU0dt}
\eea
with
\bea
U_0(t_0,t_0)=1 ~.
\eea

So the idea is to encode the leading kinematic evolution in terms of the interaction picture fields, and calculate the effects of the interaction through the series expansion in terms of powers of $H_I$. To do this, we rewrite (\ref{expect_value}) as
\bea
&&\langle \Omega| Q\left[\delta\phi_a(\bx,t),\delta\pi_a(\bx,t)\right] |\Omega\rangle
\nonumber \\
&=&\langle \Omega| U^{-1}(t,t_0) ~Q\left[\delta\phi_a(\bx,t_0),\delta\pi_a(\bx,t_0)\right]~
U(t,t_0) |\Omega\rangle
\nonumber \\
&=& \langle \Omega| F^{-1}(t,t_0) U_0^{-1}(t,t_0)
~Q\left[\delta\phi_a(\bx,t_0),\delta\pi_a(\bx,t_0)\right]~
U_0(t,t_0) F(t,t_0) |\Omega\rangle
\nonumber \\
&=& \langle \Omega| F^{-1}(t,t_0) ~Q\left[\delta\phi^I_a(\bx,t),\delta\pi^I_a(\bx,t)\right]~
F(t,t_0) |\Omega\rangle ~,
\eea
where
\bea
F(t,t_0) \equiv U_0^{-1}(t,t_0) U(t,t_0) ~.
\eea
Using (\ref{dUdt}), (\ref{dU0dt}) and (\ref{Hsplit}), we get
\begin{align}
\frac{d}{dt}F(t,t_0) =& -i U_0^{-1}(t,t_0) H_I\left[ \delta\phi(t_0),\delta\pi(t_0);t\right] U_0(t,t_0) ~ F(t,t_0)
\nonumber \\
=& -i H_I\left[ \delta\phi^I(t),\delta\pi^I(t);t\right] ~ F(t,t_0)
\nonumber \\
\equiv & -i H_I(t) F(t,t_0) ~,
\label{dFdt}
\end{align}
with
\bea
F(t_0,t_0)=1 ~.
\label{F_initial}
\eea
The solution to (\ref{dFdt}) and (\ref{F_initial}) can be written in the following way,
\bea
F(t,t_0) = T \exp \left( -i \int_{t_0}^t H_I(t) dt \right) ~,
\label{F_def}
\eea
where the operator $T$ means that, in each term in the Taylor series expansion of the exponential, the time variables have to be time-ordered. The operator $\bar T$ will be used to mean the reversed time-ordering.

In summary, the expectation value (\ref{expect_value}) is
\begin{align}
\langle Q \rangle =& \langle \Omega| F^{-1}(t,t_0) Q^I(t) F(t,t_0) |\Omega\rangle ~,
\nonumber \\
=& \langle \Omega| \left[ \bar T \exp \left( i \int_{t_0}^t H_I(t) dt \right) \right]
~Q^I(t)~
\left[ T \exp \left( -i \int_{t_0}^t H_I(t) dt \right)\right] |\Omega\rangle ~.
\label{in-in_final}
\end{align}
Notice that in
\begin{align}
H_I(t) &\equiv H_I\left[ \delta\phi^I(t), \delta\pi^I(t);t\right] ~,
\\
Q^I(t) &\equiv Q\left[\delta\phi^I_a(\bx,t),\delta\pi^I_a(\bx,t)\right] ~,
\end{align}
all the field perturbations are in the interaction picture.

\medskip
The perturbation theory is also often done in terms of the Lagrangian formalism. In the following, we show that they are equivalent.
In the above, we perform perturbations on the Hamiltonian, and define $\delta\pi_a$ by perturbing $\pi_a \equiv \partial L/\partial \dot\phi_a$, (here we use $\partial$ to denote the functional derivative,)
\bea
\delta\pi_a = \frac{\partial L}{\partial \dot\phi_a}(\phi_a,\dot\phi_a)
- \frac{\partial L}{\partial \dot{\bar\phi}_a}(\bar\phi_a, \dot{\bar\phi}_a) ~.
\eea
The Hamiltonian $\tilde H$ is defined by (\ref{Hexpand}). So using the definition
\bea
H \equiv \int d^3x \frac{\partial L}{\partial\dot\phi_a} \dot\phi_a - L ~,
\eea
together with the classical equations of motions (\ref{bkgd_eom}) and $\dot{\bar\pi}_a = \partial L/\partial \bar\phi_a$, the definition (\ref{Hexpand}) for $\tilde H$ becomes
\begin{align}
\tilde H = \int d^3x \frac{\partial L}{\partial \dot\phi_a} (\phi_a,\dot\phi_a) \delta\dot\phi_a
+ \int d^3x \frac{\partial L}{\partial \bar\phi_a}(\bar\phi_a, \dot{\bar\phi}_a) \delta\phi_a
-L(\phi_a,\dot\phi_a) + L(\bar\phi_a,\dot{\bar\phi}_a) ~.
\end{align}
If we perturb the Lagrangian directly, we keep
the part of the Lagrangian that is quadratic and higher in perturbations $\delta\phi_a$ and $\delta\dot\phi_a$,
\begin{align}
\tilde L (\delta\phi_a, \delta\dot\phi_a, t)
\equiv L (\phi_a,\dot\phi_a)
%\nonumber \\
- L (\bar\phi_a, \dot{\bar\phi}_a)
- \int d^3x \frac{\partial L}{\partial \bar\phi_a}(\bar\phi_a, \dot{\bar\phi}_a) \delta\phi_a
- \int d^3x \frac{\partial L}{\partial \dot{\bar\phi}_a}(\bar\phi_a, \dot{\bar\phi}_a) \delta\dot\phi_a ~.
\label{tL_def}
\end{align}
The $\delta\pi_a$ is defined directly as
\begin{align}
\delta\pi_a \equiv \frac{\partial \tilde L}{\partial (\delta\dot\phi_a)}
= \frac{\partial L}{\partial \dot\phi_a}(\phi_a,\dot\phi_a)
- \frac{\partial L}{\partial \dot{\bar\phi}_a}(\bar\phi_a, \dot{\bar\phi}_a) ~,
\end{align}
where in the second step Eq.~(\ref{tL_def}) has been used. So these two definitions of $\delta \pi_a$ are equivalent.
The Hamiltonian $\tilde H$ is defined through $\tilde L$,
\begin{align}
\tilde H \equiv \int d^3x \frac{\partial \tilde L}{\partial \delta \dot\phi_a} \delta\dot\phi_a - \tilde L ~.
\end{align}
Again, using (\ref{tL_def}), we can see that the two definitions of $\tilde H$ are equivalent.

\subsection{Mode functions and vacuum}
\label{Sec:mode_functions}

The Hamiltonian $H_0$ in the above formalism is typically chosen to be the quadratic kinematic terms for field perturbations $\delta\phi_a$ without mixing,
\bea
H_0 = \int d^3x \sum_a \left[ \frac{1}{2A} \delta\pi_a^2 + \frac{B}{2} (\partial_i\delta\phi_a)^2 + \frac{C}{2} \delta\phi_a^2 \right] ~.
\eea
So they describe free fields propagating in the time-dependent background.
The $A$, $B$ and $C$ are some time-dependent background fields, and they are all positive. The solutions to the equations of motion (\ref{delta_eom_I}) in momentum space, $u_a(\bk,t)$, are called the {\em mode functions}, where $\bk$ denotes the comoving momentum. They satisfy the Wronskian condition
\bea
A u_a(\bk,t) \dot u_a^*(\bk,t) - {\rm c.c.} = i ~,
\quad ({\rm no~sum~over~}a) ~.
\label{Wronskian_cond}
\eea
Note that we have specified the time-independent constant on the right hand side of (\ref{Wronskian_cond}) to be $i$ for the same reason that we see in Sec.~\ref{Sec:inflation}. Namely, we decompose $\delta\phi^I_a$ as
\bea
\delta\phi^I_a (\bk,t) = u_a(\bk,t) a_a(\bk) + u_a^*(-\bk,t) a_a^\dagger(-\bk) ~,
\label{decomp_deltaphi}
\eea
where the annihilation and creation operators satisfy the following relations,
\bea
&& [a_a(\bk), a_b^\dagger(-\bp)] = (2\pi)^3 \delta_{ab} \delta^3(\bk+\bp)
~,
\nonumber \\
&&
[a_a(\bk), a_b(-\bp)] = 0 ~, \quad
[a_a^\dagger(\bk), a_b^\dagger(-\bp)] = 0 ~.
\label{comm_relation_a}
\eea
These commutation relations are equivalent to (\ref{comm_relation}) because of (\ref{Wronskian_cond}), but the constant needs to be $i$. This gives the normalization condition for the mode functions.

Being the solutions of the second order differential equation, generally the mode function is a linear superposition of two independent solutions.
So we need to specify the initial condition. For inflation models, as long as the field theory applies, one can always find an early time at which the physical momentum of the mode is much larger than the Hubble parameter and study a time interval much less than a Hubble time. Under these conditions, the equations of motion approach to those in the Minkowski limit, in which the mode function is a linear superposition of two independent plane waves, one with positive frequency and another negative. The ground state in the Minkowski spacetime is the positive one. The mode function which approaches this positive frequency state in the Minkowski limit is called the Bunch-Davies state. In physical coordinates, this limit is proportional to
$e^{-ik_{\rm ph}t}$, (for $k_{\rm ph} \gg m$),
where $k_{\rm ph}$ is the physical momentum.
In terms of the conformal time $\tau \equiv \int dt/a(t)$ and the comoving momentum coordinate $k \equiv k_{\rm ph}/ a(t)$ which we often use, this limit is proportional to $e^{-ik\tau}$. We have seen an example in Sec.~\ref{Sec:inflation} and will see more similar examples later with different $A$, $B$ and $C$.
The corresponding vacuum $|0\rangle$ is the Bunch-Davies vacuum and annihilated by $a_a(\bk)$ defined in (\ref{decomp_deltaphi}), $a_a(\bk)|0\rangle=0$.

We also would like to write the vacuum of the interacting theory (\ref{in-in_final}) in terms of the vacuum of the free theory $|0\rangle$ defined above.
Unlike the scattering theory where the vacuum of the free theory is generally different from the vacuum of the interaction theory, the process that we are studying here do not generate any non-trivial vacuum fluctuations through interactions. This is a direct consequence of the identity
\bea
F^{-1} F =1 ~.
\eea
So we can replace
$|\Omega \rangle$ in (\ref{in-in_final}) with the Bunch-Davies vacuum $|0\rangle$ that we have specified above.

The integrand $H_I(t)$ in (\ref{F_def}) is highly oscillatory in the infinite past due to the behavior of the mode function $\propto e^{-ik\tau}$. Their contribution to the integral is averaged out. For the Bunch-Davies vacuum, this regulation can be achieved by introducing a small tilt to the integration contour $\tau_0 \to -\infty(1+i\epsilon)$ or performing a Wick rotation $\tau \to i\tau$. The imaginary component turns the oscillatory behavior into exponentially decay, making the integral well defined.

\subsection{Contractions}
\label{Sec:contractions}

When evaluating (\ref{in-in_final}), one encounters (anti-)time-ordered integrals, of which the integrands are products of fields, such as $\delta\phi^I_a$ and $\delta\pi^I_a$, or $\delta\phi^I_a$ and $\delta\dot\phi^I_a$, sandwiched between the vacua. In contrast to the Minkowski space, in the inflationary background, we do not have a simple analogous Feynman propagator which takes care of the time-ordering. Therefore we will just evaluate the integrands, but leave the complication of the time-ordering to the final integration.

To evaluate the integrand, one can shift around the orders of fields in that product, following the rules of the commutation relations. A {\em contraction} is defined to be a non-zero commutator between the following components of two fields,
$\left[ \delta\phi_a^+, \delta\phi_b^- \right]$, where $\delta\phi_a^+$ and $\delta\phi_b^-$ denote the first and second term on the right hand side of (\ref{decomp_deltaphi}), respectively.
After normal ordering, namely moving annihilation operators to the right-most and creation operators to the left-most so that they give zeros hitting the vacuum, it is not difficult to convince oneself that the only terms left are those with all fields contracted.
Feynman diagrams can be used to keep track of what kind of contractions are necessary.

\begin{figure}[t]
\begin{center}
%\epsfxsize=10cm
%\epsfbox{Fep.eps}
\epsfig{file=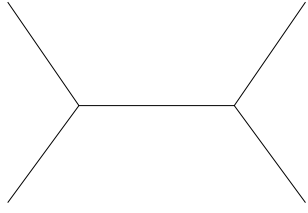, width=4cm}
\end{center}
\caption{An example of Feynman diagram.}
\label{Fig:Fdiagram_ex}
\end{figure}

In the following we demonstrate this using a simple example. We consider a field $\delta\phi^I$ and quantize it as usual,
\bea
\delta\phi^I (\bk,t) \equiv \delta\phi^+ + \delta\phi^- = u(\bk,t) a_\bk +
u^*(-\bk,t) a^\dagger_{-\bk} ~.
\eea
So a contraction between the two
terms, $\delta\phi(\bk,t')$ on the left and $\delta\phi(\bp,t'')$ on
the right, is defined to be
\bea
[\delta\phi^+(\bk,t'), \delta\phi^-(\bp,t'')] = u(\bk,t')
u^*(-\bp,t'') (2\pi)^3 \delta^3(\bk+\bp) ~.
\label{contraction_ex}
\eea
For example, we want to compute a contribution to the four-point function $\langle \delta\phi^4 \rangle$ from a tree-diagram containing two three-point interactions of the following form,
\bea
H^I \propto \int \prod_{i=1}^3 d\bp_i \delta\dot\phi^I(\bp_1,t) \delta\dot\phi^I(\bp_2,t) \delta\dot\phi^I(\bp_3,t) ~.
\label{3point_ex}
\eea
These two $H^I$'s come from expanding $F^{-1}$ or $F$ in (\ref{in-in_final}).
The corresponding Feynman diagram is Fig.~\ref{Fig:Fdiagram_ex}.

In Fig.~\ref{Fig:Fdiagram_ex}, the two cubic vertices each represent the three-point interaction (\ref{3point_ex}). Each line represents a contraction. The four outgoing legs connect to the four $\delta\phi(\bp_i,t)$ $(i=1,2,3,4)$ in $\langle \delta\phi^4 \rangle$.  The following is a term from the perturbative series expansion of (\ref{in-in_final}). We demonstrate in the following one set of contractions represented by the diagram in Fig.~\ref{Fig:Fdiagram_ex},

{\footnotesize
\bea
&&
\contraction{\delta\dot\phi^I(} {\bp_1 }{ ,t') \delta\dot\phi^I(\bp_2,t')
\delta\dot\phi^I(\bp_3,t') \delta\phi^I( } { \bk_1 }
\contraction[2ex]{\delta\dot\phi^I(\bp_1,t') \delta\dot\phi^I(} {\bp_2} {,t')
\delta\dot\phi^I(\bp_3,t') \delta\phi^I(\bk_1,t) \delta\phi^I(} {\bk_2}
\contraction[1ex]{\delta\dot\phi^I(\bp_1,t') \delta\dot\phi^I(\bp_2,t')
\delta\dot\phi^I(\bp_3,t') \delta\phi^I(\bk_1,t) \delta\phi^I(\bk_2,t) \delta\phi^I(} {\bk_3}
{,t) \delta\phi^I(\bk_4,t) \delta\dot\phi^I(} {\bq_1}
\contraction[2ex]{\delta\dot\phi^I(\bp_1,t') \delta\dot\phi^I(\bp_2,t')
\delta\dot\phi^I(\bp_3,t') \delta\phi^I(\bk_1,t) \delta\phi^I(\bk_2,t) \delta\phi^I(\bk_3,t)
\delta\phi^I(} {\bk_4} {,t) \delta\dot\phi^I(\bq_1,t'') \delta\dot\phi^I(} {\bq_2}
\bcontraction[1ex]{\delta\dot\phi^I(\bp_1,t') \delta\dot\phi^I(\bp_2,t')
\delta\dot\phi^I(} {\bp_3} {,t') \delta\phi^I(\bk_1,t) \delta\phi^I(\bk_2,t) \delta\phi^I(\bk_3,t)
\delta\phi^I(\bk_4,t) \delta\dot\phi^I(\bq_1,t'') \delta\dot\phi^I(\bq_2,t'')
\delta\dot\phi^I(} {\bq_3}
\delta\dot\phi^I(\bp_1,t') \delta\dot\phi^I(\bp_2,t')
\delta\dot\phi^I(\bp_3,t') \delta\phi^I(\bk_1,t) \delta\phi^I(\bk_2,t) \delta\phi^I(\bk_3,t)
\delta\phi^I(\bk_4,t) \delta\dot\phi^I(\bq_1,t'') \delta\dot\phi^I(\bq_2,t'')
\delta\dot\phi^I(\bq_3,t'')
\nonumber\\
&=&
[\delta\dot\phi^+(\bp_1,t'), \delta\phi^-(\bk_1,t)]
[\delta\dot\phi^+(\bp_2,t'), \delta\phi^-(\bk_2,t)]
[\delta\phi^+(\bk_3,t), \delta\dot\phi^-(\bq_1,t'')]
[\delta\phi^+(\bk_4,t), \delta\dot\phi^-(\bq_2,t'')]
\nonumber \\
&&[\delta\dot\phi^+(\bp_3,t'), \delta\dot\phi^-(\bq_3,t'')] ~.
\nonumber
\eea
}
Note that all terms are contracted. The result can be further evaluated using (\ref{contraction_ex}).
After integration over momenta indicated in (\ref{3point_ex}), the final momentum conservation will always manifest itself as $(2\pi)^3 \sum_i(\bk_i)$.
There are other sets of contractions represented by the same diagram for the same term. Namely, there are three ways of picking two of the three $\bp_i$'s ($\bq_i$'s),
so we have a symmetry factor 9; also, there are 24 permutations of the four
$\bk_i$'s. We need to sum over all these possibilities. We also need to sum over all possible terms containing two $H^I$'s in the perturbative series, which are not listed here, with their corresponding time-ordered integral structure.

\subsection{Three forms}
\label{Sec:three_forms}

Now we deal with the time-ordered integrals in the series expansion. There are two ways to expand (\ref{in-in_final}).

In the first form, we simply expand the exponential in (\ref{F_def}). For example, for an even $n$, the $n$th order term is
\begin{align}
&i^n (-1)^{n/2}\int_{t_0}^{t}d\bar t_1 \int_{t_0}^{\bar t_1}d\bar{t}_2 \cdots
\int_{t_0}^{\bar t_{n/2-1}}d\bar{t}_{n/2}
\int_{t_0}^{t}dt_1 \int_{t_0}^{t_1}dt_2 \cdots
\int_{t_0}^{t_{n/2-1}}dt_{n/2} \nonumber\\
\times&
\langle H_I(\bar t_{n/2})\cdots
 H_I(\bar t_{1}) Q_I(t) H_I(t_1)\cdots H_I(t_{n/2}) \rangle
 \nonumber\\
+&2{\rm Re}\sum_{m=1}^{n/2} i^n (-1)^{m+n/2} \int_{t_0}^{t}d\bar t_1
 \int_{t_0}^{\bar t_1}d\bar{t}_2\cdots
\int_{t_0}^{\bar t_{n/2-1-m}}d\bar{t}_{n/2-m}
\int_{t_0}^{t}dt_1 \int_{t_0}^{t_1}dt_2 \cdots
\int_{t_0}^{t_{n/2-1+m}}dt_{n/2+m} \nonumber\\ \times&
\langle H_I(\bar t_{n/2-m})\cdots
 H_I(\bar t_{1}) Q_I(t) H_I(t_1)\cdots H_I(t_{n/2+m}) \rangle ~.
\label{inin_factorizedeven}
\end{align}
Each term in the above summation contains two factors of multiple integrals, one from $F^{-1}$ and another from $F$. Each multiple integral is time-ordered or anti-time-ordered, but there is no time-ordering between the two. We call this representation the {\em factorized form}.

In the second form, we rearrange the factorized form so that all the time variables are time-ordered, and all the integrands are under a unique integral. The $n$th order term in this form is \cite{Weinberg:2005vy}
\begin{align}
  i^n \int_{t_0}^t dt_1 \int_{t_0}^{t_1} dt_2 \cdots \int_{t_0}^{t_{n-1}}dt_n \left\langle
\left[ H_I(t_n),\left[H_I(t_{n-1}), \cdots ,
\left[ H_I(t_1),Q_I(t) \right]
\cdots \right] \right]
\right\rangle~.
\label{inin_commutator}
\end{align}
We call this representation the {\em commutator form}.

Each representation has its computational advantages and disadvantages.

The factorized form is most convenient to achieve the UV ($t_i\to t_0$) convergence. As mentioned, after we tilt or rotate the integration contour into the positive imaginary plane for the left integral, and negative imaginary plane for the right integral, all the oscillatory behavior in the UV becomes well-behaved exponential decay. However this form is not always convenient to deal with the IR ($t_i \to t$) behavior. For cases where the correlation functions have some non-trivial evolution after modes exit the horizon, as typically happens for inflation models with multiple fields, the convergence in the IR is slow. Cancellation of spurious leading contributions from different terms in the sum (\ref{inin_factorizedeven}) can be very implicit in this representation, and could easily lead to wrong leading order results in analytical estimation or numerical evaluation.

The commutator form is most convenient to get the correct leading order behavior in the IR. The mutual cancellation between different terms are made explicit in terms of the nested commutators, before the multiple integral is performed. However, such a regrouping of integrands makes the UV convergence very implicit. Recall that the contour deformation is made to damp the oscillatory behavior in the infinite past. In the commutator form, for any individual term in the integrand, we can still generically choose a unique convergence direction in terms of contour deformation. Although the directions are different for different terms, they can be separately chosen for each of them. But now the problem is, if these different terms have to be grouped as in the nested commutator so that the IR cancellation is explicit, the two directions get mixed. Hence the explicit IR cancellation is incompatible with the explicit UV convergence in this case.

To take advantage of both forms, we introduce a cutoff $t_c$ and write the IR part of the in-in formalism in terms of the commutator form and the UV part in terms of the factorized form \cite{Chen:2009zp},
\begin{align}
\sum_{i=1}^n \int_{t_c}^t dt_1 \cdots \int_{t_c}^{t_{i-1}} dt_i
~{\rm \{commutator~form\}} ~
\int_{-\infty}^{t_c} dt_{i+1} \cdots
\int_{-\infty}^{t_{n-1}} dt_n
~{\rm \{factorized~form\} } ~.
\label{inin_mixedform}
\end{align}
This representation is called the {\em mixed form}. This form is particularly efficient in numerical computations when combined with the Wick-rotations in the UV.

We will not always encounter all these subtleties in every model, but there does exist such interesting examples, as we will see in Sec.~\ref{Sec:int}.

\subsection{Summary}
\label{Sec:in-in_sum}

To end this section, we summarize the procedure that we need to go through to calculate the correlation functions in the in-in formalism.

Starting with the Lagrangian $L[\phi(t),\dot\phi(t)]$, we perturb it around the homogenous solutions $\bar\phi_a$ and $\dot{\bar\phi}_a$ of the classical equations of motion,
\bea
\phi_a(\bx,t) = \bar\phi_a(t) + \delta\phi_a(\bx,t) ~,
\quad
\dot\phi_a(\bx,t) = \dot{\bar\phi}_a(t) + \delta\dot\phi_a(\bx,t) ~.
\eea
Keep the part of the Lagrangian that is quadratic and higher in perturbations and denote it as $\tilde L$.
Define the conjugate momentum densities as $\delta\pi_a = \partial \tilde L/\partial (\delta\dot\phi_a)$. We can also equivalently expand the Hamiltonian $H[\phi(t),\pi(t)]$ by perturbing $\phi_a(\bx,t)$ and $\pi_a(\bx,t)$.

Work out the Hamiltonian in terms of $\delta \phi_a$ and $\delta
\pi_a$, and separate them into the quadratic kinematic part $H_0$, which describes the free fields in the time-dependent background, and the interaction
part $H_I$. Relabel $\delta\phi_a$'s and $\delta\pi_a$'s
in the Hamiltonian density as the interaction picture fields, $\delta\phi^I_a$'s and
$\delta\pi^I_a$'s. These latter variables satisfy the equations of
motion followed from the $H_0$. We quantize $\delta\phi^I_a$ and $\delta\pi^I_a$ in terms of the annihilation and creation operators as in (\ref{decomp_deltaphi}) and (\ref{comm_relation_a}). The mode functions $u_a(\bk,t)$ are solutions of the equations of motion from $H_0$, normalized according to the Wronskian conditions (\ref{Wronskian_cond}) and specified by an initial condition such as the Bunch-Davies vacuum.
The correlation function for $Q(t)$ is given by
\bea
\langle Q(t) \rangle \equiv
\langle 0| \left[ \bar T \exp \left( i \int_{t_0}^t H_I(t) dt \right) \right]
~Q^I(t)~
\left[ T \exp \left( -i \int_{t_0}^t H_I(t) dt \right)\right] |0 \rangle ~,
\label{inin_summary}
\eea
where $Q(t)$ is a product in terms of $\delta\phi_a^I(\bx,t)$ and $\delta\pi^I(\bx,t)$.
If we want to work with $\delta \phi_a^I$ and $\delta\dot\phi_a^I$ instead of $\delta\phi_a^I$ and $\delta\pi_a^I$, we
replace $\delta \pi^I_a$ with $\dot{\delta\phi^I_a}$ using the relation $\dot {\delta\phi^I_a} = \partial
H_0/\partial (\delta \pi^I_a)$.

Choose appropriate forms in Sec.~\ref{Sec:three_forms} and series-expand the integrand in powers of $H_I$ to the desired orders. Perform contractions defined in Sec.~\ref{Sec:contractions} for each term in this expansion. Each term gives a non-zero contribution only when all fields are contracted. Draw Feynman diagrams that represent the correlation functions, and use them as a guidance to do contractions. Finally sum over all possible contractions and perform the time-ordered integrations.

\section{A no-go theorem}
\setcounter{equation}{0}
\label{Sec:Nogo}

Simplest inflation models generate negligible amount of non-Gaussianities that are well below our current experimental abilities \cite{Maldacena:2002vr,Acquaviva:2002ud}. By simplest, we mean
\begin{itemize}
\item
single scalar field inflation
\item
with canonical kinetic term
\item
always slow-rolls
\item
in Bunch-Davies vacuum
\item
in Einstein gravity.
\end{itemize}
This list is extracted based on Maldacena's computation of three-point functions in an explicit slow-roll model \cite{Maldacena:2002vr}. We now review this proof. The notations here follow Ref.~\cite{Seery:2005wm,Chen:2006nt} and will be consistently used later in this review.

The Lagrangian for the {\em single scalar field inflation} with {\em canonical kinetic term} is the following,
\bea
S= \int d^4x \sqrt{-g} \left[ \frac{\mpl}{2} R + X - V(\phi) \right] ~,
\label{action_slow-roll}
\eea
where $\phi$ is the inflaton field, $X=-\frac{1}{2} g^{\mu\nu} \partial_\mu \phi \partial_\nu \phi$ is the canonical kinetic term and $V$ is the slow-roll potential. The first term is the {\em Einstein gravity} and $\mpl=(8\pi G)^{-1/2}$ is the reduced Planck mass. For convenience we will set the reduced Planck mass $\mpl=1$. The signature of the metric is $(-1,1,1,1)$.

The inflaton starts near the top of the potential and slowly rolls down. As we have reviewed in Sec.~\ref{Sec:inflation}, to ensure that the inflation lasts for at least $\CO(60)$ efolds, the potential is required to be flat so that the slow-roll parameters (\ref{slowV})
are both much less than one most of the time. The energy of the universe is dominated by the potential energy, and the inflaton follows the slow-roll attractor solution (\ref{phi_0_attractor}).
Also as discussed in Sec.~\ref{Sec:inflation}, we will use the following more general slow-roll parameters,
\begin{align}
\epsilon = -\frac{\dot{H}}{H^2} ~,
~~~
\eta = \frac{\dot{\epsilon}}{\epsilon H} ~.
\label{slow}
\end{align}

To study the perturbation theory, it is convenient to use the ADM formalism, in which the metric takes the form
\begin{equation}
d s^2=-N^2dt^2+h_{ij}(dx^i+N^i dt)(dx^j+N^j dt) ~.
\label{ADMmetric}
\end{equation}
The action becomes
\bea
S= \half \int dt dx^3 \sqrt{h} N (R^{(3)} + 2X-2V)
+ \half \int dt dx^3 \sqrt{h} N^{-1} (E_{ij}E^{ij} -E^2) ~,
\label{ADMaction}
\eea
where the index of $N^i$ can be lowered by the 3d metric $h_{ij}$ and
$R^{(3)}$ is the 3d Ricci scalar constructed from $h_{ij}$. The
definition of $E_{ij}$ and $E$ are
\bea
E_{ij} &=& \half (\dot h_{ij} - \nabla_i N_j - \nabla_j N_i) ~,
\cr
E &=& E_{ij} h^{ij} ~.
\eea
In the ADM formalism, the variables $N$ and $N^i$ are Lagrangian multipliers whose equations of motion are easy to solve.
In single field inflation, we have only one physical scalar perturbation \cite{Mukhanov:1990me}.
We choose the uniform inflaton gauge (also called the comoving gauge) in which the scalar perturbation $\zeta$ appears in the three dimensional metric $h_{ij}$ in the following form,
\begin{equation}
h_{ij}=a^2 e^{2\zeta}\delta_{ij} ~,
\label{h_uni_inf}
\end{equation}
and the inflaton fluctuation $\delta\phi$ vanishes.
The $a(t)$ is the homogeneous scale
factor of the universe, so $\zeta$ is a space-dependent rescaling factor.
In this review we do not consider the tensor perturbations.

We plug (\ref{ADMmetric}) and (\ref{h_uni_inf}) into the action (\ref{ADMaction}) and solve the constraint equations for the Lagrangian multipliers $N$ and $N^i$. We then plug them back to the action and expand up to the cubic order in $\zeta$ in order to calculate the three-point functions. To do this, in the ADM formalism, it is enough to solve $N$ and $N^i$ to the first order in $\zeta$.
This is because their third order perturbations will multiply the zeroth order constraint equation which vanishes, and their second order perturbations will multiply the first order constraint equation which again vanishes.
After some lengthy algebra, we obtain the following expansions,
\begin{align}
S_2 = \int dt d^3x~
\left[ a^3 \epsilon \dot\zeta^2- a \epsilon
(\partial \zeta)^2 \right] ~,
\label{action_quad}
\end{align}
\begin{align}
S_3 = & \int dtd^3x \bigg[
a^3\epsilon^2 \zeta\dot{\zeta}^2+a\epsilon^2 \zeta(\partial\zeta)^2-
2a \epsilon\dot{\zeta}(\partial
\zeta)(\partial \chi)
\nonumber
\\
&+
\frac{a^3\epsilon}{2}\dot\eta \zeta^2\dot{\zeta}
+\frac{\epsilon}{2a}(\partial\zeta)(\partial
\chi) \partial^2 \chi +\frac{\epsilon}{4a}(\partial^2\zeta)(\partial
\chi)^2
\nonumber \\
& + f(\zeta)\frac{\delta L}{\delta \zeta} \bigg|_1 \bigg] ~,
\label{action_cubic}
\end{align}
where
\begin{eqnarray}
\chi &=& a^2 \epsilon \partial^{-2} \dot \zeta ~, \\
\frac{\delta
L}{\delta\zeta}\bigg|_1 &=& 2 a
\left( \frac{d\partial^2\chi}{dt}+H\partial^2\chi
-\epsilon\partial^2\zeta \right) ~, \\
f(\zeta)&=& \frac{\eta}{4}\zeta^2+ {\rm terms~with~derivatives~on~\zeta
}
~.
\label{redefinition}
\end{eqnarray}
Here $\partial^{-2}$ is the inverse Laplacian and $\delta
L/\delta\zeta|_1$ is the variation of the quadratic action with
respect to the perturbation $\zeta$. We now can follow Sec.~\ref{Sec:in-in_and_correl} and proceed to calculate the correlation functions. For simplicity, we will always neglect the superscript "$I$" on various interaction picture fields.

We restrict to the case where the slow-roll parameters are {\em always small and featureless}.
We first look at the quadratic action. In this case, we can analytically
solve the equation of motion followed from (\ref{action_quad}) in terms of the Fourier mode of $\zeta$,
\bea
u_k = \int d^3x \zeta(t,\bx) e^{-i \bk\cdot\bx} ~,
\eea
and get the mode function
\bea
u_k= u({\bf k},\tau) = \frac{i H}{\sqrt{4\epsilon
k^3}}(1+i k\tau)e^{-i k\tau} ~,
\label{uk_sr}
\eea
where $\tau \equiv \int dt/a \approx -(aH)^{-1}$ is the conformal time. The normalization is determined by the Wronskian condition (\ref{Wronskian_cond}). We have chosen the {\em Bunch-Davies vacuum} by imposing the condition that the mode function approaches the vacuum state of the Minkowski spacetime in the short wavelength limit $k/a\gg 1/H$,
\bea
u_k \to -\frac{H\tau}{\sqrt{4\epsilon k}} e^{-ik\tau} ~.
\eea
The dynamical behavior of $\zeta$ that has been explained around Eq.~(\ref{u_BD}) and (\ref{zeta_phi_relation}) is made precise here. In particular, $\zeta$ is exactly massless without dropping any $\CO(\epsilon)$ suppressed terms. In addition, from (\ref{h_uni_inf}), we can see that, for superhorizon modes, the only effect of $\zeta$ is to provide a homogeneous spatial rescaling. And $\zeta$ is the only scalar perturbation. So the fact that $\zeta$ is frozen after horizon exit will not be changed by higher order terms.

If we choose the spatially flat gauge, we make $\zeta$ disappear and the scalar in this perturbation theory becomes the perturbation of $\phi$.
The relation between $\zeta$ and $\delta \phi$ in (\ref{zeta_phi_relation}) (with $\CO(\epsilon)$ corrections) is thus a gauge transformation through a space-dependent time-shift.

We quantize the field as
\bea
\zeta(\bk,\tau) = u_k a_\bk + u^*_k a^\dagger_{-\bk} ~,
\eea
with the canonical commutation relation $[a_\bk, a^\dagger_{\bk'}] = (2\pi)^3 \delta^3(\bk-\bk')$.
We can easily compute the two-point function at the tree level,
\bea
\langle \zeta(\bk_1)\zeta(\bk_2) \rangle = \frac{P_\zeta}{2k_1^3} (2\pi)^5 \delta^3(\bk_1+\bk_2) ~,
\eea
where
\bea
P_\zeta = \frac{H^2}{8\pi^2 \epsilon} ~.
\eea
Since $\zeta$ remains constant after it exits the horizon, the $H$ and $\epsilon$ are both evaluated near the horizon exit.

We next look at the cubic action. For single field models, $H_{I,3} = -L_3$. Keeping in mind that $\chi$ is proportional to $\epsilon$, one can see that the first line of (\ref{action_cubic}) is proportional to $\epsilon^2$.
For the featureless potential, $\dot\eta = \CO(\epsilon^2)$, where $\epsilon$ collectively denotes either $\epsilon$ or $\eta$. So the second line of (\ref{action_cubic}) is proportional to $\epsilon^3$, and negligible. The third line can be absorbed by a field redefinition $\zeta \to \zeta_n + f(\zeta_n)$. The only term in $f(\zeta_n)$ that will contribute to the correlation function is written out explicitly in (\ref{redefinition}). All the others involve derivatives of $\zeta$ so vanish outside the horizon. Thus this redefinition eventually introduces an extra term
\begin{eqnarray}
\langle\zeta(\bk_1)\zeta(\bk_2)\zeta(\bk_3)\rangle
&=&\langle\zeta_n(\bk_1)\zeta_n(\bk_2)\zeta_n(\bk_3)\rangle
\nonumber \\
&+& \frac{\eta}{4}
(\langle \zeta_n^2(\bk_1) \zeta_n(\bk_2) \zeta_n(\bk_3) \rangle
+ {\rm 2~perm.}) + \CO(\eta^2 (P^\zeta_k)^3) ~.
\label{3ptredef}
\end{eqnarray}
According to (\ref{inin_summary}), we expand the exponential to the first order in $H_{I,3}$ to get the leading result,
\bea
\langle \zeta_n^3  \rangle = -i \langle 0| \int_{t_0}^t dt [ \zeta_n(\bk_1)\zeta_n(\bk_2)\zeta_n(\bk_3), H_{I,3} ] |0\rangle ~.
\eea

To estimate the order of magnitude of the bispectrum, we only need to keep track of the factors of $H$ and $\epsilon$. For example, from the first term in (\ref{action_cubic}), we have $\int dt H_3(t) \supset -\int dx^3 d\tau a^2 \epsilon^2 \zeta \zeta'^2$, where we used the conformal time $\tau$ and the prime denotes the derivative to $\tau$. Using $a\propto H^{-1}$, $\zeta\propto\zeta' \propto H/\sqrt{\epsilon}$, we see that this three-point vertex contributes $\propto H\sqrt{\epsilon}$. Together with the three external legs $\zeta^3$ and the definition $P_\zeta \propto H^2/\epsilon$, we get
\bea
\langle \zeta^3 \rangle = -i \int dt \langle [\zeta^3,H_{I,3}(t)] \rangle \propto \frac{H^4}{\epsilon}
\propto \CO(\epsilon) P_\zeta^2 ~,
\label{3pt_sr_estimate}
\eea
Similar results can be obtained for the other two terms in the first line of (\ref{action_cubic}).
As we will define more carefully later, the size of the three-point function is conventionally characterized by the number $f_{NL}$, which is defined as $\langle \zeta^3 \rangle \sim f_{NL} P_\zeta^2$.
So the contribution from the first line of (\ref{action_cubic}) is $f_{NL} = \CO(\epsilon)$.
The extra term due to the redefinition (\ref{3ptredef}) contributes $f_{NL} =\CO(\eta)$.
This completes the order-of-magnitude estimate. To get the full non-Gaussian profile, we need to compute the integrals explicitly and get
\bea
\langle \zeta(\textbf{k}_1)\zeta(\textbf{k}_2)\zeta(\textbf{k}_3)\rangle
&=&
(2\pi)^7\delta^3(\textbf{k}_1+\textbf{k}_2+\textbf{k}_3)
P_\zeta^2
\frac{1}{\prod_i k_i^2}~ S ~,
\eea
where
\begin{align}
S &= \frac{\epsilon}{8} \left[ -\left( \frac{k_1^2}{k_2k_3} + {\rm 2~perm.} \right)
+ \left( \frac{k_1}{k_2} + {\rm 5~perm.}\right)
+ \frac{8}{K} \left( \frac{k_1k_2}{k_3} + {\rm 2~perm.} \right) \right]
\nonumber \\
&+ \frac{\eta}{8} \left( \frac{k_1^2}{k_2k_3} + {\rm 2~perm.} \right) ~,
\label{slow_roll_S}
\end{align}
where $K=k_1+k_2+k_3$ and the permutations stand for those among $k_1$, $k_2$ and $k_3$.

The slow-roll parameters are of order $\CO(0.01)$, so $f_{NL} \sim \CO(0.01)$ for these models.
Even if we start with Gaussian primordial perturbations, non-linear effects in CMB evolution will generate $f_{NL} \sim \CO(1)$ \cite{Bartolo:2010qu}, and a similar number for large scale structures due to the non-linear gravitational evolution or the galaxy bias \cite{Liguori:2010hx}.
It seems unlikely that we can disentangle all these contaminations and detect such small primordial non-Gaussianities in the near future.

\section{Beyond the no-go}
\label{Sec:BeyondNogo}
\setcounter{equation}{0}

\subsection{Inflation model building}
\label{Sec:Model_building}

The following are two examples of slow-roll potentials in the simplest inflation models that we studied in Sec.~\ref{Sec:Nogo},
\bea
V_{\rm small} &=& V_0 - \half m^2 \phi^2 ~,
\label{V_small}
\\
V_{\rm large} &=& \half m^2 \phi^2 ~.
\label{V_large}
\eea
The first type (\ref{V_small}) belongs to the small field inflation models. The slow-roll conditions (\ref{slowV}) require the potential to be flat enough relative to its height, i.e.~the mass of the inflaton should satisfy $m\ll H$. The second type (\ref{V_large}) belongs to the large field inflation models. The potential also needs to be flat relative to its height, but here one achieves this by making the field range $\phi$ very large, typically $\phi \gg \mpl$. The other conditions that we listed in Sec.~\ref{Sec:Nogo} should also be satisfied by these models. These are the classic examples, which exhibit algebraic simplicities and illustrate many essential features of inflation.

However, when it comes to the more realistic model building in a UV complete setup, such as in supergravity and string theory, situations get much more complicated. For example, it is natural that we encounter multiple light and heavy fields, and the potentials for them form a complex landscape. These multiple fields live in an internal space, whose structure can be very sophisticated. In string theory, some of them manifest themselves as extra dimensions and can have intricate geometry and warping. All these elements have to coexist with the inflationary background that introduces profound back-reactions.

Even with varieties of model building ingredients, it has been proven to be very subtle to construct an explicit and self-consistent inflation model. Indeed various problems have been noticed over the years in the course of the inflation model building. For example,
\begin{itemize}
\item {\bf The $\eta$-problem for slow-roll inflation} \cite{Copeland:1994vg}.
As we have seen, in order to have slow-roll inflation \cite{Linde:1981mu,Albrecht:1982wi}, the mass of the inflaton field has to be light enough, $m\ll H$, to maintain a flat potential. However, in the inflationary background, the natural mass of a light particle is of order $H$. This can be seen in many ways, and in some ideal situations they are equivalent to each other. For example, one way to see this is to consider the coupling between the Ricci scalar and the inflaton, $\sim R\phi^2$. In the inflationary background $R\sim H^2$. Unless we have good reasons to set the coefficient of this term to be much less than one, it will give inflaton a mass of order $H$, spoiling the inflation. Another way to see this is to note that the effective potential in supergravity takes the form
$V=V_0\exp(K/\mpl^2)\times {\rm other~terms}$. Here schematically $K\sim \phi^2 + \cdots$ is the Kahler potential and its dependence on $\phi$ is normalized as such to give the canonical kinetic term for $\phi$. So the first term in the expansion of $V$ is of order $V_0\phi^2/\mpl^2 \sim H^2 \phi^2$ and model independent. Therefore, either symmetry needs to be imposed or other tuning contributions introduced to solve this $\eta$-problem.

\item{\bf The $h$-problem for DBI inflation} \cite{Chen:2008hz}.
DBI inflation \cite{Silverstein:2003hf} is invented to generate inflation by a different mechanism. It makes use of the warped space in the internal field space \cite{Randall:1999ee,Giddings:2001yu}. These warped space impose speed-limits for the scalar field, so even if the potential is steep, the inflaton is not allowed to roll down the potential very quickly. A canonical example of warped space is
\bea
ds^2 = h(r)^2(-dt^2 + a(t)^2 d\bx^2)+ h(r)^{-2}dr^2 ~,
\label{metric_warp}
\eea
where $r$ is the extra-dimension (or internal space), $h(r)=r/R$ is the warp factor and $R$ is the length scale of the warped space.
The position of a $3+1$ dimensional brane in the $r$-coordinate is the inflaton.
So the inflaton velocity is limited by the speed-limit in the $r$-direction, $h^2$.
In order to provide a speed-limit that is small enough for inflation, the warp factor has to be small enough, $h\ll HR$. However one of the Einstein equations with the metric (\ref{metric_warp}) takes the following form,
\bea
(dh/dr)^2 - H^2 h^{-2} = \frac{1}{R^2}+{\rm other~source~terms} ~,
\label{h_eq}
\eea
where the second term on the left hand side is due to the back-reaction of the inflationary spacetime. It is easy to see that the naive $h=r/R$ should be modified for $h<HR$, precisely where the inflation is supposed to happen. Without contributions from other source terms, such a deformed geometry closes up too quickly and leads to an unacceptable probe-brane back-reaction if we demand the inflaton still follow the speed-limit. Therefore, either symmetry, or tuning using other source terms from the right hand side of (\ref{h_eq}), is necessary to solve this $h$-problem.
The $\eta$-problem and $h$-problem are closely related in an AdS/CFT setup.

\item{\bf The field range bound} \cite{Chen:2006hs,Baumann:2006cd}.
Large field inflation models require the field range to be much larger than $\mpl$. In supergravity and string theory, starting from a ten-dimensional theory with 10-dim Planck mass $M_{10}$, the 4-dim Planck mass $\mpl$ is obtained by integrating out the six extra-dimensions,
\bea
&& M_{(10)}^8 \int d^6yd^4x \sqrt{-G_{(10)}} R_{(10)}
\nonumber \\
&\supset&
M_{(10)}^8 V_{(6)} \int d^{4}x \sqrt{-g_{(4)}} R_{(4)}
\equiv \mpl^2 \int d^{4}x \sqrt{-g_{(4)}} R_{(4)} ~,
\eea
where we use $L$ and $V_{(6)} \sim L^6$ to denote the size and volume of the extra-dimensions, respectively. The field range $\Delta\phi$ often appears as the distance in the extra dimensions, $\Delta\phi \sim \Delta L \cdot M_{(10)}^2$, with the factor $M_{(10)}^2$ being the proportional coefficient. Clearly, $\Delta L \lesssim L$. If the field range manifests itself within a warped throat with a length scale $R$, we still require $R<L$, and so $\Delta L\lesssim L$. Together with $\mpl=M_{(10)}^4 L^3$, we get
\bea
\Delta\phi \lesssim \mpl/(M_{(10)}L)^2 ~.
\label{field_bound}
\eea
We further note that the microscopic length scale $L$ has to be much larger than the 10-dim Planck length $M_{(10)}^{-1}$ for the field theory to make sense. So $M_{(10)} L \gg 1$, and the field range $\Delta\phi$ in these models is generically sub-Planckian. For example, for a warped throat with charge $N$, $(M_{(10)} L)^2 \gtrsim (M_{(10)} R)^2 \sim N^{1/2}$, we have
\bea
\Delta\phi \lesssim \mpl/\sqrt{N} ~.
\label{field_bound2}
\eea
We have ignored a detailed numerical coefficient appearing on the right hand side of (\ref{field_bound2}), which is model dependent. For example, considering the volume $V_{(6)}$ to be the sum of the throat and a generic bulk volume, it is $\CO(0.01)$ \cite{Chen:2006hs}; considering an extreme case where the throat does not attach to a bulk, it is $\CO(1)$ \cite{Baumann:2006cd}.
Notice that, due to the dependence of $\mpl$ on the volume $V_{(6)}$, increasing the volume only makes the bound tighter.

\item{\bf The variation of potential} \cite{Lyth:1996im}.
Even in cases where there is no fundamental restriction on the excursion of fields, one encounters problems constructing the large field inflationary potential. Large field potentials that arise from a fundamental theory take the following general from,
\bea
V(\phi) = \sum_{n=0}^\infty \lambda_n m_{\rm fund}^{4-n} \phi^n ~,
\label{V_general}
\eea
where $m_{\rm fund}$ represents typical scales in the theory. For field theory descriptions to hold, such scales are much less than $\mpl$. For example, $m_{\rm fund}$ can be the higher dimensional Planck mass, string mass, or their warped scales. The $\lambda_n$'s are dimensionless couplings of order $\CO(1)$. Unless some symmetries are present to forbid an infinite number of terms in (\ref{V_general}), or a high degree of fine-tuning is assumed, the shape of potential (\ref{V_general}) varies over a scale of order $m_{\rm fund} \ll \mpl$. This variation is too dramatic for the potential to be a successful large field slow-roll potential.

\end{itemize}

None of the arguments in the above list is meant to show that the specific type of inflation is impossible. In fact, these have been the driving forces for the ingenuity and creativity in the field of inflation model building. This list is used to demonstrate some typical examples of complexities in reality. Often times, solving one problem will be companied by other structures that make the model step beyond the simplest one. So we may want to keep an open mind that the algebraic simplicity may not mean the simplicity in Nature.

Following is a partial list of possibilities that allow us to go beyond the no-go theorem in Sec.~\ref{Sec:Nogo}.
\begin{itemize}
\item
Instead of single field inflation, we can consider quasi-single field or multifield inflation models (Sec.~\ref{Sec:quasi} \& \ref{Sec:multifield}).

\item
Instead of canonical kinetic terms, there are models where the higher derivative kinetic terms dominate the dynamics (Sec.~\ref{Sec:Eq}).

\item
Instead of following the attractor solution such as the slow-roll precisely, features can be present in the potentials or internal space, that temporarily break the attractor solution, or cause small but persistent perturbations on the background evolution (Sec.~\ref{Sec:sin} \& \ref{Sec:res}).

\item
Instead of staying in the Bunch-Davies vacuum, other excitations can exist due to, for example, boundary conditions or low scales of new physics (Sec.~\ref{Sec:folded}).

\item
Although strong constraints, from experimental results and theoretical consistencies, exist on non-Einstein gravities, early universe may provide an opportunity for their appearance. We use this category to include a variety of possibilities, such as modified gravities, non-commutativity, non-locality and models beyond field theories.

\end{itemize}

There are also strong motivations from data analyses for us to search and study different large non-Gaussianities. The signal-to-noise ratio in the CMB data is not large enough for us to detect primordial non-Gaussianities model-independently. A well established method is to start with a theoretical non-Gaussian ansatz, and construct optimal estimators that compare theory and data by taking into accounts all momenta configurations. This then gives constraints on the parameters characterizing the theoretical ansatz. Therefore, the following two important possibilities exist. First, the primordial non-Gaussianities exist in data could be missed if we did not start with a right theoretical ansatz. Second, even if a non-Gaussian signal were detected with one ansatz, it does not mean that we have found the right one. So different well-motivated non-Gaussian templates are needed for clues on how corresponding data analyses should be formed.
From a different perspective, even if the primordial density perturbations were Gaussian, we would still do the similar amount of work and reach the conclusion after various well-motivated non-Gaussian forms are properly constrained.

\subsection{Shape and running of bispectra}

In this review, we will be mainly interested in the three-point correlation functions of the scalar primordial perturbation $\zeta$. They are also called the {\em bispectra}. In this subsection, we introduce some simple terminologies that we often encounter in studies of bispectra.

The three-point function is a function of three momenta, $\bk_1$, $\bk_2$ and $\bk_3$, which form a triangle due to the translational invariance. Assuming also the rotational invariance, we are left with three variables, which are their amplitudes, $k_1$, $k_2$ and $k_3$, satisfying the triangle inequalities. The information is encoded in a function $S(k_1,k_2,k_3)$ that we define as below,
\bea
\langle \zeta^3 \rangle \equiv S(k_1,k_2,k_3) \frac{1}{(k_1k_2k_3)^2} \tilde P_\zeta^2 (2\pi)^7 \delta^3 ( \sum_{i=1}^3 \bk_i ) ~,
\label{3pt_def}
\eea
where $\tilde P_\zeta$ is the fiducial power spectrum, and we fix it to be a constant $\tilde P_\zeta\equiv P_\zeta(k_{\rm wmap}) = 6.1\times 10^{-9}$, where $k_{\rm wmap}=0.027 {\rm Mpc}^{-1}$. We have chosen the above definition so that it can be uniformly applied to different types of bispectra that we will encounter in this review. In literature, different notations have been used. The differences are simple and harmless. For example, different functions such as $\CA=k_1k_2k_3 S$ or $F=S/(k_1k_2k_3)^2$ are sometimes defined. We choose $S$ since it is dimensionless and, for scale-invariant bispectra, it is invariant under a rescaling of all momenta. This quantity is the combination that is used to plot the profiles of bispectra in literature any way, despite of different conventions. Also, the precise power spectrum $P_\zeta$ instead of $\tilde P_\zeta$ is often used in the definition (\ref{3pt_def}). Here we absorb the momentum dependence of $P_\zeta$ in $S$. This is because the three-point function is an independent statistic relative to the two-point. In cases where both the power spectrum and bispectrum have strong scale dependence, it is not convenient if they are defined in an entangled way.

Under different circumstances, different properties of $S$ are emphasized. The conventions involved may not always be precisely consistent with each other, since they are chosen to best describe the case at hand. Following are some typical examples.

The dependence of $S$ on $k_1$, $k_2$ and $k_3$ is usually split into two kinds.

\begin{figure}[t]
\begin{center}
%\epsfxsize=10cm
%\epsfbox{Fep.eps}
\epsfig{file=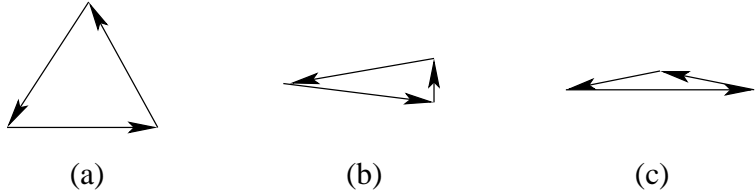, width=10cm}
\end{center}
%\medskip
\caption{Momentum configurations: (a) equilateral, (b) squeezed, (c) folded.}
\label{Fig:triangles}
\end{figure}

One is called the {\em shape} of the bispectrum. This refers to the dependence of $S$ on the momenta ratio $k_2/k_1$ and $k_3/k_1$, while fixing the overall momentum scale $K=k_1+k_2+k_3$. Several special momentum configurations are shown in Fig.~\ref{Fig:triangles}.

Another is called the {\em running} of the bispectrum. This refers to the dependence of $S$ on the overall momentum scale $K=k_1+k_2+k_3$, while fixing the ratio $k_2/k_1$ and $k_3/k_1$.

For bispectra that are approximately scale-variant, the shape is a more important property \cite{Babich:2004gb,Chen:2006nt}.
We will encounter such cases in Sec.~\ref{Sec:Eq}, \ref{Sec:int} \& \ref{Sec:local}.
The amplitude, also called the {\em size}, of the bispectra is often denoted as $f_{NL}$ by matching
\bea
S(k_1,k_2,k_3) \xrightarrow[\rm limit]{k_1=k_2=k_3}
\frac{9}{10} f_{NL} ~.
\label{fNLdef}
\eea
In this case, $f_{NL}$ is approximately a constant but can also have a mild running, i.e.~a weak dependence on the overall momentum $K$ \cite{Chen:2005fe,Byrnes:2009pe}.
An index $n_{NG}-1 \equiv d\ln f_{NL}/d \ln k$ is introduced to describe this scale dependence.
The power spectrum also has a mild running, $P_\zeta = (k/k_0)^{n_s-1} \tilde P_\zeta$. In this review, when we give explicit forms of $S$ in the approximately scale-invariant cases, for simplicity we mostly ignore these mild scale dependence and concentrate on shapes.
Shapes of bispectra have been given names according to the overall dependence of $S$ on momenta. For example, for the {\em equilateral bispectrum}, $S$ peaks at the equilateral triangle limit and vanishes as $\sim k_3/k_1$ in the squeezed triangle limit ($k_3\ll k_1=k_2$). The {\em local bispectrum} peaks at the squeezed triangle limit in the form $\sim (k_3/k_1)^{-1}$, such as the two shape components in (\ref{slow_roll_S}).
To visualize the shapes, we often draw 3D plots $S(1,x_2,x_3)$, where $x_2$ and $x_3$ vary from $0$ to $1$ and satisfy the triangle inequality $x_2+x_3 \ge 1$.

There are also cases where the running becomes the most important property, while the shape is relatively less important \cite{Chen:2006xjb,Chen:2008wn}. In such cases, the bispectra are mostly functions of $K$. So $f_{NL}$ defined in (\ref{fNLdef}) has strong scale dependence. Instead, one can choose a constant $f_{NL}$ to describe the overall running amplitude. We will encounter such cases in Sec.~\ref{Sec:sin} \& \ref{Sec:res}. In these cases, the shape plot $S(1,x_2,x_3)$ may look nontrivial but this is because it does not fix $K$.

The above dissection will become less clean for cases where both properties become important.

One thing is clear. The $f_{NL}$, that is always used to quantify the level of non-Gaussianities, is only sensible with an extra label that specifies, at least qualitatively, the profile of the momentum dependence, such as shapes and runnings.

It is useful to quantify the correlations between different non-Gaussian profiles, because as we mentioned in data analyses an ansatz can pick up signals that are not completely orthogonal to it. In real data analyses this is performed in the CMB $l$-space. To have a simple but qualitative analogue in the $k$-space, we define the inner product of the two shapes as
\bea
S \cdot S' \equiv \int_{V_k} S(k_1,k_2,k_3)S'(k_1,k_2,k_3) w(k_1,k_2,k_3) dk_1 dk_2 dk_3 ~,
\label{inner_product}
\eea
and normalize it to get the shape correlator \cite{Babich:2004gb,Fergusson:2008ra}
\bea
C(S,S') \equiv \frac{S \cdot S'}{(S\cdot S)^{1/2} (S'\cdot S')^{1/2}} ~.
\label{correlator}
\eea
Following Ref.~\cite{Fergusson:2008ra}, we choose the weight function to be
\bea
w(k_1,k_2,k_3) = \frac{1}{k_1+k_2+k_3} ~,
\eea
so that the $k$-scaling is close to the $l$-scaling in the data analyses estimator. Later in this review, when we use this correlator to estimate the correlations between shapes, we take the ratio between the smallest and largest $k$ to be $2/800$, close to that in WMAP.
A more precise correlator should be computed in the $l$-space in the same way that the estimator is constructed. We refer to Ref.~\cite{Liguori:2010hx} for more details.

In typical data analyses \cite{Komatsu:2001rj,Komatsu:2003iq,Creminelli:2006gc,Smith:2006ud,Yadav:2007ny}, the estimator involves a triple integral of the bispectrum over the three momenta $k_i$. To have practical computational costs, it is necessary that this integral can be factorized into a multiplication of three integrals, each involves only an individual $k_i$. This requires the bispectrum to be of the form $f_1(k_1) f_2(k_2) f_3(k_3)$, or a sum of such forms. Such a form is called the {\em factorizable form} or {\em separable form}. The factor $K^{-n}$ may be tolerable since it can be written as $(1/\Gamma(n))\int_0^\infty t^{n-1} e^{-Kt}$. If the analytical result is too complicated, to make contact with experiments we will try to construct simple factorizable ansatz or template to capture the main features of the original one.
New methods that are applicable to non-factorizable bispectrum forms and are more model-independent are under active development \cite{Fergusson:2009nv}.

\section{Single field inflation}
\label{Sec:Single}
\setcounter{equation}{0}

In this section, we relax several restrictions of the no-go theorem on single field inflation models and study how large non-Gaussianities can arise. We present the formalisms and compute the three-point functions. We emphasize how different physical processes during inflation are imprinted as distinctive signatures in non-Gaussianities. Obviously, any mechanism that works for single field inflation can be generalized to multi-field inflation models.

\subsection{Equilateral shape: higher derivative kinetic terms}
\label{Sec:Eq}

In this subsection, we study large non-Gaussianities generated by non-canonical kinetic terms in general single field inflation models, following Ref.~\cite{Chen:2006nt}.

Consider the following action for the general single field inflation \cite{Garriga:1999vw},
\bea
S= \int d^4x \sqrt{-g} \left[ \frac{\mpl}{2} R + P(X,\phi) \right] ~.
\label{action_single}
\eea
Comparing to (\ref{action_slow-roll}), we have replace the canonical form $X-V$ with an arbitrary function of $X\equiv -\half g^{\mu\nu} \partial_\mu\phi \partial_\nu \phi$ and $\phi$.
This is the most general Lorentz-invariant Lagrangian as a function of $\phi$ and its first derivative.
It is useful to define several quantities that characterize the differential properties of $P$ with respect to $X$ \cite{Garriga:1999vw,Seery:2005wm},
\bea
c_s^2 &=& \frac{P_{,X}}{P_{,X}+2XP_{,XX}} ~, \\
\Sigma &=& X P_{,X}+2X^2P_{,XX}  = \frac{H^2\epsilon}{c_s^2} ~,\\
\lambda &=& X^2P_{,XX}+\frac{2}{3}X^3P_{,XXX} ~,
\eea
where $c_s$ is called the sound speed and the subscript "$X$" denotes the derivative with respect to $X$.
The third derivative is enough since we will only study the three-point function here.

It is a non-trivial question which forms of $P$ will give rise to inflation. The model-independent approach we take here is to list the conditions that an inflation model has to satisfy, no matter which mechanism is responsible for it. Namely, we generalize the slow-roll parameters in (\ref{slow}) to the following slow-variation parameters
\bea
\epsilon = -\frac{\dot H}{H^2} ~, \quad
\eta = \frac{\dot\epsilon}{\epsilon H} ~, \quad
s=\frac{\dot c_s}{c_s H} ~,
\label{slow_variation}
\eea
and require them to be small most of the time during the inflation.
The smallness of these parameters guarantees the Hubble constant $H$, the parameter $\epsilon$ and the sound speed $c_s$ to vary slowly in terms of the Hubble time. Similar to the arguments given in the case of slow-roll inflation in Sec.~\ref{Sec:inflation}, these are necessary to ensure a prolonged inflation as well as an approximately scale-invariant power spectrum that we observed in the CMB.

Following the same procedure that is outlined in Sec.~\ref{Sec:Nogo}, we get the quadratic and cubic action for the scalar perturbation $\zeta$ \cite{Maldacena:2002vr,Seery:2005wm,Chen:2006nt}. The quadratic part is
\bea
S_2 = \int dt d^3x~
\left[ a^3  \frac{\epsilon}{c_s^2}\dot\zeta^2- a \epsilon
(\partial \zeta)^2 \right] ~.
\label{action_quad_single}
\eea
If the slow-variation parameters are always small and featureless, we can analytically solve the equation of motion followed from (\ref{action_quad_single}) and get the following mode function,
\begin{eqnarray}
u_k (\tau) = \frac{i H}{\sqrt{4\epsilon
c_s k^3}}(1+i k c_s\tau)e^{-i k c_s\tau} ~.
\label{uk_single}
\end{eqnarray}
Notice the appearance of $c_s$ comparing to (\ref{uk_sr}). The two-point function is
\bea
\langle \zeta(\bk_1)\zeta(\bk_2) \rangle = \frac{P_\zeta}{2k_1^3} (2\pi)^5 \delta^3(\bk_1+\bk_2)
\eea
with the power spectrum
\bea
P_\zeta = \frac{H^2}{8\pi^2 \epsilon c_s} ~,
\eea
where the variables are evaluated at the horizon crossing of the corresponding $k$-mode.

To calculate the bispectrum, we look at the cubic action. In the following, we list three terms that are most interesting for this subsection,
\begin{align}
S_3 = \int dt d^3x & \bigg\{
- \frac{a^3\epsilon}{Hc_s^2} \left[\left(1-\frac{1}{c_s^2}\right)+\frac{2\lambda}{\Sigma}\right] \dot\zeta^3
+ \frac{3a^3\epsilon}{c_s^2} \left(1-\frac{1}{c_s^2}\right) \zeta\dot\zeta^2
-a\epsilon \left(1-\frac{1}{c_s^2}\right) \zeta \left(\partial\zeta\right)^2
\nonumber \\
& + \cdots
\bigg\} ~.
\label{action_cubic_single}
\end{align}
The full terms can be found in Eq.~(4.26)-(4.28) in Ref.~\cite{Chen:2006nt}.

The order of magnitude contribution from these three terms can be estimated similarly as we did in (\ref{3pt_sr_estimate}), but now we not only keep factors of $H$ and $\epsilon$, but also factors of $c_s$. Take the first term as an example, we write it in terms of the conformal time,
\bea
\int d\tau H_3(\tau) \supset  \int d\tau d^3x \frac{a\epsilon}{Hc_s^2} \left[ \left(1-\frac{1}{c_s^2}\right)+\frac{2\lambda}{\Sigma}\right] \zeta^{\prime 3} ~.
\label{H_3_general_example}
\eea
Comparing (\ref{uk_single}) with (\ref{uk_sr}), we see that there is an extra factor of $c_s$ companying $\tau$. So we estimate $d\tau \propto c_s^{-1}$ and $a \approx -(H\tau)^{-1} \propto c_s H^{-1}$. Also, $\zeta \propto H/\sqrt{\epsilon c_s}$, but $\zeta'\propto c_s \zeta$. Overall, the vertex (\ref{H_3_general_example}) contributes
\bea
\propto
\frac{H}{\sqrt{\epsilon c_s} }
\left[ \left(1-\frac{1}{c_s^2} \right)+\frac{2\lambda}{\Sigma} \right] ~.
\eea
Multiplying the three external legs $\zeta^3$, and using the definition
\bea
\langle \zeta^3 \rangle \sim f_{NL} P_\zeta^2
\eea
and $P_\zeta \propto (H/\sqrt{\epsilon c_s})^2$,
we get
\bea
f_{NL} \sim \CO(\frac{1}{c_s^2}) + \CO(\frac{\lambda}{\Sigma}) ~.
\eea
The other two terms are similar. A detailed calculation reveals
\begin{eqnarray}
\langle \zeta(\textbf{k}_1)\zeta(\textbf{k}_2)\zeta(\textbf{k}_3)\rangle
&=&
(2\pi)^7\delta^3(\textbf{k}_1+\textbf{k}_2+\textbf{k}_3)
(P_\zeta)^2
\frac{1}{\prod_i k_i^2} \cr &\times&
(S_\lambda +S_c + S_o +S_\epsilon +S_\eta +S_s) ~,
\label{3pointFinal}
\end{eqnarray}
where we have decomposed the shape of the three-point function
into six parts. The first two come from the leading order terms that we listed in (\ref{action_cubic_single}),
\bea
S_\lambda &=& \left(\frac{1}{c_s^2}-1
- \frac{2\lambda}{\Sigma} \right)
\frac{3k_1k_2k_3}{2K^3} ~,
\label{S_lam} \\
S_c &=&
\left(\frac{1}{c_s^2}-1\right)
\left(-\frac{1}{K}\sum_{i>j}k_i^2k_j^2+\frac{1}{2K^2}
\sum_{i\neq j}k_i^2k_j^3+\frac{1}{8}\sum_{i}k_i^3 \right) \frac{1}{k_1k_2k_3} ~.
\label{S_c}
\eea
In terms of $f_{NL}$ their sizes are
\begin{align}
f_{NL}^\lambda &= \frac{5}{81}
\left(\frac{1}{c_s^2}-1-\frac{2\lambda}{\Sigma} \right) ~,
\label{fNLlambda} \\
f_{NL}^c &= -\frac{35}{108} \left(\frac{1}{c_s^2}-1 \right) ~.
\label{fNLc}
\end{align}
The next four terms come from the subleading terms that we did not list explicitly in (\ref{action_cubic_single}) as well as the subleading contributions from the first two terms. Their orders of magnitude are
\begin{align}
f_{NL}^o &= \CO \left( \frac{\epsilon}{c_s^2} ~,~
\frac{\epsilon\lambda}{\Sigma} \right) ~,
\nonumber \\
f_{NL}^{\epsilon,\eta,s} &= \CO (\epsilon,\eta,s) ~.
\end{align}
The detailed profiles can be found in Ref.~\cite{Chen:2006nt}.

\begin{figure}
\begin{center}
%\epsfxsize=10cm
%\epsfbox{Fep.eps}
\epsfig{file=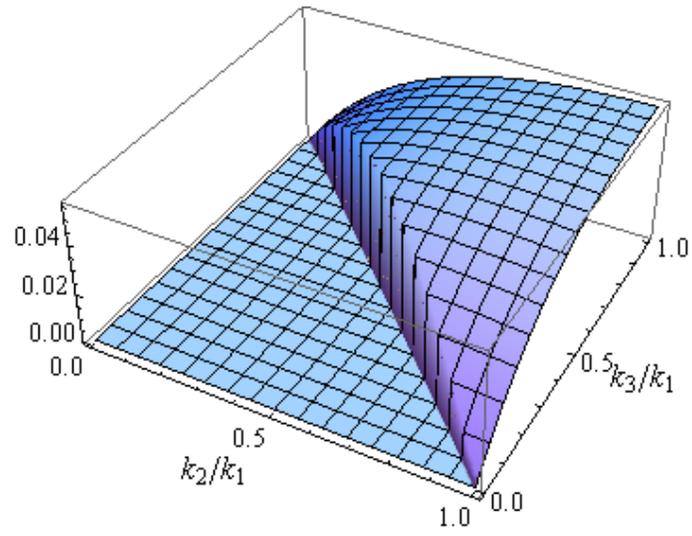, width=9cm}
\end{center}
\caption{Shape of $S_\lambda$ in (\ref{S_lam}).}
\label{Fig:shape_eq1}
\end{figure}

\begin{figure}
\begin{center}
%\epsfxsize=10cm
%\epsfbox{Fep.eps}
\epsfig{file=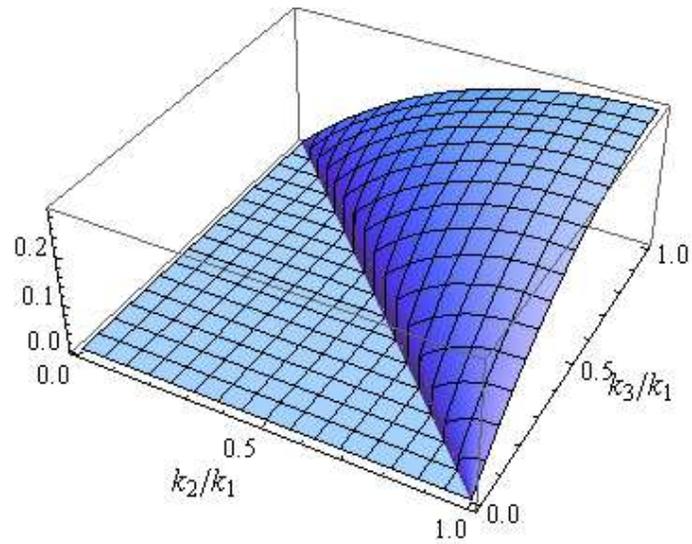, width=9cm}
\end{center}
\caption{Shape of $S_c$ in (\ref{S_c}).}
\label{Fig:shape_eq2}
\end{figure}

The full results we obtained can be used in different regimes.

\begin{itemize}

\item
If we look at the limit, $c_s \ll 1$ or $\lambda/\Sigma \gg 1$,
the leading order results give two shape components, $S_\lambda$ and $S_c$. This result can also be obtained using a simple method of considering only the fluctuations in scalar field while neglecting those in gravity \cite{Alishahiha:2004eh,Gruzinov:2004jx}. Intuitively, this is because the higher derivative terms are responsible for the generation of large non-Gaussianities, and the gravity contribution is expected to be small as we saw in Sec.~\ref{Sec:Nogo}.
Therefore one expands $P(X,\phi)$ using $\phi(\bx,t)=\phi_0(t)+\delta\phi(\bx,t)$. The derivatives of $P$ with respect to $\phi$ are ignored because the inflation and scale invariance imposes an approximate shift symmetry on $P$ in terms of inflaton $\phi$. We then get two terms in the cubic Lagrangian density
\bea
\CL_3 = a^3 \left( \half P_{,XX} \dot\phi_0 + \frac{1}{6}P_{,XXX} \dot\phi_0^3 \right) \dot{\delta\phi}^3 - \frac{a}{2} P_{,XX}\dot\phi_0 \dot{\delta\phi} (\nabla\delta\phi)^2 ~.
\eea
This gives two leading bispectra the same as (\ref{S_lam}) and (\ref{S_c}).
The approach that we present here gives a rigorous justification to such an method. The subleading order component $S_o$ may be observable as well.
At this limit where the higher derivative terms of the inflaton field are dominant, the Lagrangian of the above effective field theory are generalized \cite{Cheung:2007st} to include, for example, the ghost inflation \cite{ArkaniHamed:2003uz} whose Lagrangian cannot be written in a form of $P(X,\phi)$. Another two slightly different equilateral shapes arise. However it is worth to mention that, generally in single field models and Einstein gravity, going beyond $P(X,\phi)$ requires adding either terms that explicitly break the Lorentz symmetry, or terms with higher time derivatives on $\phi$ which cannot be eliminated by partial integration, such as $(\Box\phi)^2$. Different treatment of such terms and discussions on their effects can be found in
\cite{Weinberg:2008hq,deUrries:1998bi,Woodard:2006nt}.

\item
If we take the opposite, slow-roll limit, $c_s \to 1$ and $\lambda/\Sigma \to 0$, we recover the two shape components $S_\epsilon$ and $S_\eta$ that we got in Sec.~\ref{Sec:Nogo}, with unobservable size $f_{NL} \sim \CO(\epsilon)$.

\item
We can also look at the intermediate parameter space. In slow-roll inflation models, one can also add higher derivative terms \cite{Creminelli:2003iq,Seery:2005wm}. But in order not to spoil the slow-roll mechanism, the effect of these terms can only be subdominant. This  corresponds to $c_s \approx 1$ and $\lambda/\Sigma < \CO(1)$. Using the full results, we can see that the size of the non-Gaussianity is $f_{NL} < \CO(1)$.
Therefore it is important to emphasize that, for the class of models we consider here, {\em non-slow-roll} inflationary mechanisms, such as the example that will be given below, are necessary to generate observable large non-Gaussianities.

\item
The other terms that we did not list in (\ref{action_cubic_single}) (see Ref.~\cite{Chen:2006nt}) and their canonical limit (\ref{action_cubic}) are also useful. These terms are exact for arbitrary values of $\epsilon$, $\eta$ and $s$, so the usage of the action is beyond the category of models that we focus on in this subsection. As we will see in Sec.~\ref{Sec:sin} and Sec.~\ref{Sec:res}, it can be applied to the cases of sharp or periodic features where these parameters do not always remain small.

\end{itemize}

In the rest of this subsection, we focus on the first case.

\begin{figure}
\begin{center}
%\epsfxsize=10cm
%\epsfbox{Fep.eps}
\epsfig{file=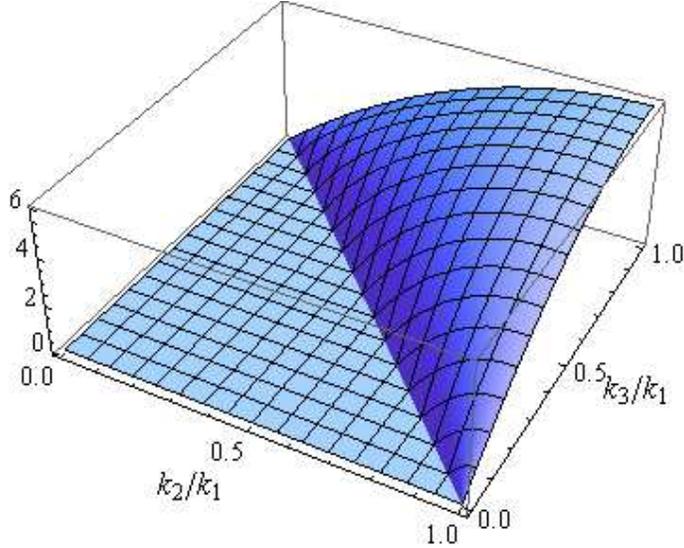, width=9cm}
\end{center}
\caption{An ansatz (\ref{ansatz_eq}) for the equilateral shape.}
\label{Fig:shape_ansatz_eq}
\end{figure}

In Fig.~\ref{Fig:shape_eq1} and \ref{Fig:shape_eq2}, we draw the shapes of $S_\lambda$ and $S_c$. The two shapes are similar. They both peak at the equilateral limit, and behave as $S \sim k_3/k_1$ in the squeezed limit $k_3 \ll k_1=k_2$. We call these shapes the {\em equilateral shapes}. There are some small differences between $S_\lambda$ and $S_c$, e.g.~around the folded triangle limit $k_2+k_3=k_1$.
A factorizable shape ansatz for the equilateral shape that is often used in data analyses is the following \cite{Creminelli:2005hu}:
\bea
S_{\rm ansatz}^{\rm eq} = -6\left(\frac{k_1^2}{k_2k_3} + {\rm 2~perm.}\right)
+ 6 \left( \frac{k_1}{k_2} + {\rm 5~perm.} \right) -12 ~,
\label{ansatz_eq}
\eea
and is shown in Fig.~\ref{Fig:shape_ansatz_eq}. As we can see, it represents the most important features of Fig.~\ref{Fig:shape_eq1} \& \ref{Fig:shape_eq2}.

The shape of $S_o$ is more complicated, but we expect they have the similar shapes as the equilateral one because their squeezed limits behave the same \cite{Chen:2006nt}. The three other shapes $S_\epsilon$, $S_\eta$ and $S_s$ are all close to the local shapes as their squeezed limit scale as $k_1/k_3$ for $k_3 \ll k_1=k_2$.

The scale dependence in $P_\zeta$, $c_s$ and $\lambda/\Sigma$ will introduce mild running for the three-point function. We usually regard only the contributions from $c_s$ and $\lambda/\Sigma$ as the running of the non-Gaussianity.

The underlying physics of the equilateral shape can be readily understood in terms of their generation mechanism. In single field inflation, the long wavelength mode that exits the horizon are frozen and can have little interaction with modes within the horizon. The large interaction only occurs among modes that are crossing the horizon at about the same time. These modes then have similar wavelengths. This is why the shape of the non-Gaussianity peaks at the equilateral limit in momentum space.

This physical origin also suggests the caveat that, as long as there are large interactions involving modes with similar wavelengths, an equilateral-like shape may arise. For example, such cases can happen in multifield models where there are particle creation \cite{Chen:2007gd,Green:2009ds}. (See however \cite{Moss:2007cv}).

\medskip

$\bullet$ {\bf An example: Dirac-Born-Infeld (DBI) inflation.}
An explicit example of the above general results is the DBI inflation \cite{Silverstein:2003hf,Alishahiha:2004eh,Chen:2004gc,Chen:2005ad,Shandera:2006ax,Kecskemeti:2006cg,Shiu:2006kj,Thomas:2007sj,Bean:2007hc,Bean:2007eh}. These inflation models describe a 3+1 dimensional brane moving in warped extra dimensions. The location of the brane is a scalar field in 4d effective field theory, and it is the inflaton. The warped extra dimensions provide a non-trivial internal field space for the inflaton. In terms of the 4d effective field theory, the action is
\bea
- \int d^4x
\sqrt{-g}~ \left[f(\phi)^{-1}
\sqrt{1+f(\phi)g^{\mu\nu} \partial_\mu \phi \partial_\nu \phi}
-f(\phi)^{-1}+V(\phi)\right] ~.
\eea
The non-trivial part is the kinetic term involving the square-root. It can be understood as a generalization of the following two familiar situations. It is a higher dimensional generalization of the action of a relativistic point particle
\bea
\int dt \sqrt{1-f \dot \bx^2} ~,
\eea
where the speed of light $f^{-1/2}$ varies with $\bx$. It is also a relativistic generalization of the usual canonical kinetic term in the non-relativistic limit $|f(\phi)g^{\mu\nu} \partial_\mu \phi \partial_\nu \phi|\ll 1$,
\bea
-\int d^4x \sqrt{-g} \left[ \half g^{\mu\nu}\partial_\mu\phi \partial_\nu\phi + V(\phi) \right] ~.
\eea
Because the speed-limit of the inflaton $f^{-1/2}$ can vary in the internal space, if it can be made small enough near the top of potential where the inflaton is about to roll down, the warped space restricts the rolling velocity even if the potential is too steep for slow-roll inflation to happen. So the inflaton rolls ultra-relativistically, but with very small velocity, and this generates the DBI inflation.

The physical consequence is now easy to obtain using the general results in this subsection. In our notation the Lagrangian is
\bea
P=-f^{-1} \sqrt{1-2fX} + f^{-1} - V ~.
\eea
The sound speed is
\bea
c_s=\sqrt{1-2fX} ~,
\eea
which is the inverse of the Lorentz boost factor $\gamma$, so $c_s\ll 1$.
The component (\ref{fNLlambda}) vanishes identically, and we have a large bispectrum of shape $\CA_c$ with size (\ref{fNLc}).

DBI inflation is still driven by the potential energy. The general single field inflation models also include the k-inflation \cite{ArmendarizPicon:1999rj}, where the inflation is driven by the inflaton kinetic energy. Model construction of single field k-inflation can be found in Ref.~\cite{ArmendarizPicon:1999rj,Li:2008qc,Tolley:2009fg}. The bispectra for such models are computed in Ref.~\cite{Chen:2006nt,Li:2008qc}.

Multifield generalization have been studied in Ref.~\cite{Langlois:2008wt,Langlois:2008qf,Arroja:2008yy,Langlois:2009ej,Mizuno:2009cv,Cai:2009hw,Gao:2009qy}, where this type of kinetic terms are generalized to multiple fields. The three-point functions involving these different fields have the same or similar shapes.

The current CMB constraint on the equilateral ansatz (\ref{ansatz_eq}) is $-214 < f_{NL}^{\rm eq} < 266$ \cite{Komatsu:2010fb}.

\subsection{Sinusoidal running: sharp feature}
\label{Sec:sin}

Although various slow-variation parameters in (\ref{slow_variation}) have to be small most of the time during inflation, they can become temporarily large. Such cases can happen if there are sharp features in inflaton potentials or internal field space, so the behavior of inflatons temporarily deviates from the attractor solution, and then relaxes back within several Hubble time, or stay longer but with small deviation amplitudes. Motivations for such models include the following. It may be possible explanations for features in power spectrum \cite{Starobinsky:1992ts,Adams:2001vc,Hamann:2007pa,Joy:2007na}, and if so the associated non-Gaussianity is a cross-check. And there are brane inflation models that are very sensitive to sharp features present in the potential or in the internal space \cite{Bean:2008na}.

As an example, we study a sharp feature in the slow-roll potential. The fact that a sharp feature in potential can enhance non-Gaussianities has long been anticipated and qualitative estimates have been made by different methods \cite{Kofman:1991qx,Wang:1999vf,Komatsu:2003fd}.
The precise method of analyzing the size, running and shape of such non-Gaussianities \cite{Chen:2006xjb,Chen:2008wn} is made possible with the developments of the formalisms that we reviewed in Sec.~\ref{Sec:in-in}, \ref{Sec:Nogo} and \ref{Sec:Eq}. This will be the subject of this subsection.

We start by studying the behavior of the slow-roll parameters. We use a small step in potential as an example and will ignore numerical coefficients. We use $c\sim \Delta V/V$ to denote the relative height of the step, and $d$ the width of the step. In the attractor solution, the inflaton velocity is given by $\dot\phi \sim V'/H \sim \sqrt{\epsilon V}$. As it falls down the step, the potential energy $cV$ gets converted to the kinetic energy, so we have
\bea
\dot\phi \lesssim \sqrt{V(c+\epsilon)} ~.
\label{dotphi_sharp}
\eea
The amplitude of density perturbations is given by $P_\zeta \sim H^4/\dot\phi^2$, so such a sharp feature causes glitches in the power spectrum. It will leave a dip with relative size $\Delta P_\zeta/P_\zeta \sim\sqrt{1+c/\epsilon}-1$ since $\dot\phi$ increases first, followed by oscillations caused by a non-attractor component of the mode function before it settles down again in the attractor solution.
To fit the CMB data, $\dot\phi$ cannot change much. As we can see, the sensitivity of the power spectrum to the step size $c$ is proportional to $\epsilon$, and we need $c/\epsilon \lesssim 1$. Reducing the width $d$ of sharp feature increases the amplitude of the glitches, but this is only for a large $d$ over which the inflaton spends more than one e-fold to cross. Further reducing $d$ will not change the amplitude of the glitches since (\ref{dotphi_sharp}) is saturated; but the sharpness will determine how deep within the horizon the modes are affected.

So $\epsilon$ does not change much, $\Delta\epsilon \sim \Delta(\dot\phi^2)/H^2 \sim c$. But it changes within a very short period, $\Delta t \sim \Delta\phi/\dot\phi\sim d/\sqrt{V(c+\epsilon)}$. So $\eta$ can be very large,
\bea
\Delta\eta \sim \frac{\Delta\epsilon}{H\epsilon\Delta t}
\sim \frac{c\sqrt{c+\epsilon}}{d\epsilon} ~.
\eea
It is also clear that the feature is associated with a characteristic physical scale and generates a scale-dependent power spectrum and higher order correlation functions.

With these qualitative behavior in mind, we now study the three-point function. An important fact of the formalisms in Sec.~\ref{Sec:Nogo} and \ref{Sec:Single} is that the expansion is exact in terms of the slow-variation parameters. So it is valid even if these parameters are not always small, as long as the expansion in $\zeta\sim \CO(10^{-5})$ is perturbative.

In all terms in the cubic expansion (\ref{action_cubic}), $\zeta$ appears at most with one time derivative; the field redefinition gives a term that is proportional to $\eta$ at the end of the inflation; and the other terms are all suppressed by powers of $\epsilon$, which remains small even in the presence of a sharp feature.
So the most important term  is
\bea
\int dtd^3x \half a^3 \epsilon\dot\eta \zeta^2 \dot\zeta ~,
\eea
in which the coupling is proportional to $\dot\eta$.
The correlation function $\langle\zeta(\bk_1)\zeta(\bk_2)\zeta(\bk_3)\rangle$
is dominated by
\begin{equation}
i \left( \prod_i u_{i}(\tau_{end}) \right)
\int_{-\infty}^{\tau_{end}} d\tau a^2 \epsilon  \eta'
\left( u_{1}^*(\tau) u_{2}^*(\tau) \frac{d}{d\tau} u_{3}^*(\tau)
+ {\rm sym} \right) (2\pi)^3 \delta^3(\sum_i \bk_i) + {\rm c.c.}
~.
\label{cubicterm_sharp}
\end{equation}
Precise evaluation of this expression has to be done numerically. But it is not difficult to see the generic properties of bispectra associated with a sharp feature.

\begin{figure}[t]
\begin{center}
%\epsfxsize=10cm
%\epsfbox{Fep.eps}
\epsfig{file=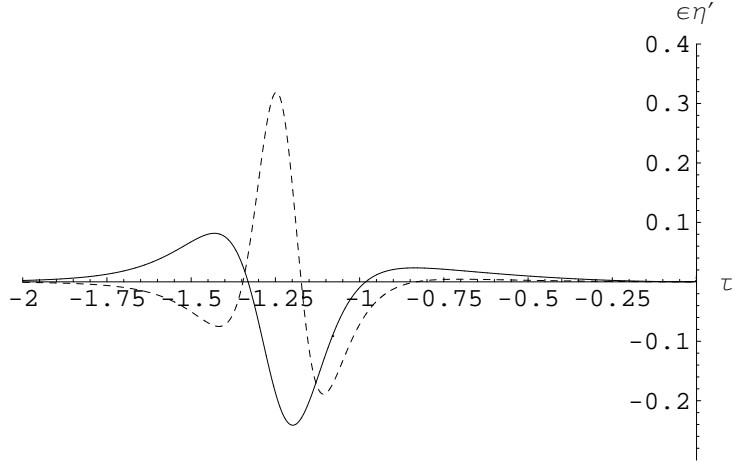, width=10cm}
\end{center}
\medskip
\caption{Behavior of the slow-roll parameters for a step (solid line, $c=0.0018$, $d=0.022\mpl$) and bump (dashed line, $c=0.0005$, $d=0.01\mpl$) sharp feature on $\half m^2\phi^2$ potential. Note that in the absence of the sharp feature, $\epsilon\eta'$ is of order $\CO(10^{-4})$.}
\label{Fig:sharp_eta}
\end{figure}

For long wavelength modes that already crossed the horizon at the time of the sharp feature, $k_i \tau \ll 1$, the mode function is already frozen and the integration (\ref{cubicterm_sharp}) gives vanishing contribution. For short wavelength modes that are still well within the horizon, the modes are not affected if their momenta are larger than the inverse of the time scale characterizing the sharpness of changes in slow-roll parameters. The modes most affected are those which are near the horizon crossing.
These modes are all oscillatory, $\sim e^{-ik_i\tau}$.
As we have studied, $\eta'$ is temporarily boosted, so it can be roughly approximated as several hat-functions that satisfy $\int d\tau \eta' =0$. Examples of such behavior are shown in Fig.~\ref{Fig:sharp_eta}.
If we simply approximate the hat-functions by several delta-functions, $\eta' \propto \delta(\tau-\tau_*)$, the integration (\ref{cubicterm_sharp}) will give something like
\bea
S \sim f_{NL}^{\rm feat} \sin \left(\frac{K}{k_*} + \phi_0 \right) ~,
\label{ansatz_sharp}
\eea
where $k_* \equiv 1/\tau_*$ is the scale corresponding to the location of feature, $\phi_0$ is a phase and
\bea
f_{NL}^{\rm feat} \sim \Delta\eta \sim \CO\left(\frac{c\sqrt{c+\epsilon}}{d\epsilon} \right) ~.
\eea
Comparing with the effect on the power spectrum, one can keep the size of glitches in the power spectrum small while make $f_{NL}$ large, for example, by fixing $c/\epsilon$ and decreasing $d$.

This ansatz describes the most important running behavior of this bispectrum. Notice that the oscillatory frequency in the $k$-space is of order $1/k_*$, which is the scale of the feature. A rescale in $k_*$ can be compensated by a rescale in all $k_i$.
Also notice that the oscillatory frequency, $3/k_*$, along the $k_1=k_2=k_3\equiv k$ direction is $3/2$ of that in the power spectrum\footnote{For power spectrum, the sharp feature introduces a small non-Bunch-Davies component for the mode function. The oscillatory frequency in the power spectrum is determined by the phase of the coefficient for this component. This is obtained through matching conditions across the feature, and the phase is $\sim 2k/k_*$. See Sec.5.3 of Ref.~\cite{Bean:2008na} for an example.

From this result we can see that, observationally, while sharp features located at large scales (such as $\ell\sim 30$) introduce glitches that need to be distinguished from statistical fluctuations, those located at much shorter scales (such as $\ell \sim 1000$) introduce oscillatory modulation that coherently shifts all points over several acoustic peaks in the same direction, which is completely different from statistical fluctuations of data points.}, $2/k_*$.

\begin{figure}[t]
\begin{center}
%\epsfxsize=10cm
%\epsfbox{Fep.eps}
\epsfig{file=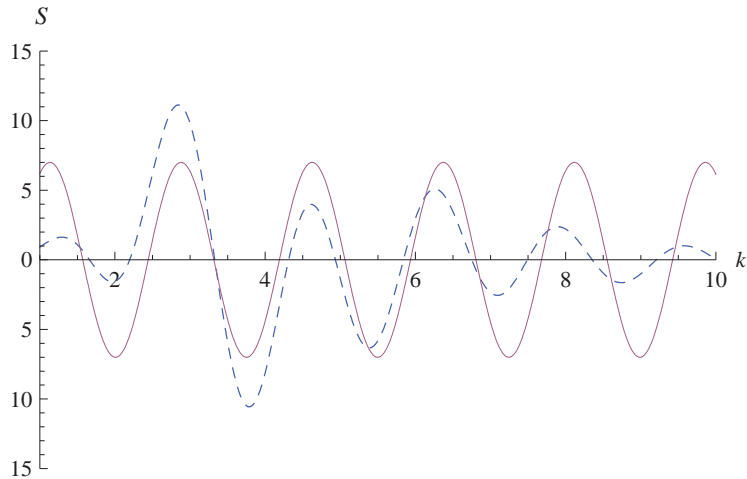, width=10cm}
\end{center}
\medskip
\caption{Numerical result (dashed line) for the bispectrum running for a sharp step ($c=0.0018, d=0.022$) along the $k_1=k_2=k_3\equiv k$ direction, compared with the simple ansatz (\ref{ansatz_sharp}) (solid line).}
\label{Fig:sharp_running}
\end{figure}

In practice, (\ref{ansatz_sharp}) is a crude ansatz that needs to be refined.
First of all, we have only considered the modes that have not exited the horizon. For those did, as we mentioned, their correlation function is as small as usual. The ansatz needs to be cut off for the long wavelength modes $K/k_* \ll 1$. A more detailed analysis \cite{Chen:2008wn} reveals, using the hat functions as an approximation of the slow-roll parameter behavior, that the bispectrum falls off as $K^2$ for these long wavelength modes.
Secondly, the fact that in (\ref{ansatz_sharp}) all short wavelength modes are equally affected is due to the sharp-change approximation. Smoother functions will only affect a finite range of modes within the horizon. So the amplitude of the ansatz should decay and how fast depends on the sharpness of feature.
To take into account both effects, empirically, we can multiply (\ref{ansatz_sharp}) with an envelop function
\bea
\propto (K/k_*)^n e^{-K/mk_*} ~,
\label{sin_envelop}
\eea
where $n$ and $m$ are parameters chosen to fit the numerical results. For example, $n=2$, $m=5$ for Fig.~\ref{Fig:sharp_running}.
Lastly, in the very squeezed limit, $k_3 \ll k_1 |K/k_*|$, $S$ can no longer be approximated as a function of $K$ only and starts to have a non-trivial shape \cite{Chen:2008wn}. Here we concentrate on the signature running behavior.

A numerical result with
\bea
V(\phi)=\half m^2 \phi^2 \left[1+c\tanh\left( \frac{\phi-\phi_s}{d} \right) \right]
\eea
is shown in Fig.~\ref{Fig:sharp_running}.
A subtlety encountered in the numerical integration is how to handle the oscillatory behavior at $\tau\to -\infty$. One can do a tilt into the imaginary plane, $-\tau \to -\infty(1+i\epsilon)$, as prescribed in the analytical procedure in Sec.~\ref{Sec:mode_functions}; or more efficiently, perform integration by part to increase the convergence of the integrand at the $\tau \to -\infty$ end. One may also try the method of Wick rotation, but this will first require solving the background equations of motion in the Wick-rotated space, since here we do not have the analytical expression for the mode function.

Sharp features can also appear elsewhere instead of potentials, for example, in the internal warped space for DBI inflation \cite{Bean:2008na}. The qualitative running behavior in bispectrum is similar, and overall large non-Gaussianities become a superposition of the approximate scale-invariant equilateral shape and the sinusoidal running.

Non-attractor initial conditions can be included as a case of sharp features, except that we only observe the relaxation part.

\subsection{Resonant running: periodic features}
\label{Sec:res}

In this subsection, we consider a different type of features. These features may or may not be sharp, but the most important property is their periodicity. Such features will induce an oscillatory component to the background evolution, in particular, to the couplings in the interaction terms. We denote this oscillatory frequency as $\omega$. We know that each mode oscillates when it starts the life well within the horizon. This frequency keeps on decreasing as the mode gets stretched by the inflation, until it reaches $H$ when the mode becomes frozen.
So the mode scans through all frequencies that is larger than $H$, up to some very high cutoff scale such as $\mpl$.
Therefore, as long as
\bea
\omega >H ,
\eea
the oscillatory frequency of the modes in the integral will hit $\omega$ at some point during the inflation. This cause a resonance between the couplings and modes, hence a constructive contribution to the correlation function \cite{Chen:2008wn}. Without the resonance, as we encountered previously, the highly oscillatory modes simply average out within the horizon. In contrast to the previous mechanisms, here the non-Gaussianities are generated when modes are sub-horizon.

We now study the properties of such a non-Gaussianity, following Ref.~\cite{Chen:2008wn}.

To estimate the integral, we use the unperturbed mode function. Similar to the sharp feature case, we get
\bea
\langle \zeta(\bk_1)\zeta(\bk_2)\zeta(\bk_3) \rangle
&\approx&
i\frac{H^4}{64 \epsilon^3 \prod_i k_i^3}
(2\pi)^3 \delta^3 (\sum_i \bk_i) \nonumber \\ &&
\times
\int_{-\infty}^0
\frac{d\tau}{\tau}
\epsilon \eta'
\left(1-i(k_1+k_2)\tau - k_1k_2\tau^2 \right) k_3^2 e^{i K\tau}\nonumber \\ &&
+ {\rm ~two~perm.} + {\rm c.c.} ~.
\label{res_3pt_terms}
\eea
In this case, we are interested in the region $|K\tau|\gg 1$ in order to have resonance. So the last term dominates as long as the momentum triangle is not too squeezed so one of the $k_i$'s becomes $<1/\tau$ at the resonance point. The oscillatory coupling is dominantly contributed by $\eta'$. The integral is proportional to
\bea
\int d\tau \tau \sin(\omega t) \exp(i K\tau) ~.
\label{res_integral}
\eea
This integral can be done analytically using the relation
$t\approx -H^{-1} \ln (-H\tau)$. But its most important properties can be understood as follows in terms of the physical picture that we described.

First, let us look at its oscillatory running in $K$-space.
The phase of the background repeats itself after $\Delta N_e =2\pi H/\omega$ e-fold, during which the wave-number $K$ changes by $-K \Delta N_e$. So the running of the non-Gaussianity in $K$-space is also oscillatory with the period given by
\bea
\Delta K = K\Delta N_e = 2 \pi K H/\omega ~.
\label{res_freq}
\eea
Note that this period is changing with $K$ in a specific way that we will see more clearly in a moment.

Next, let us look at the size of the non-Gaussianity.
Each $K$-mode briefly resonates with the oscillatory coupling when its frequency sweeps through the resonance frequency $\omega$. Once its frequency differs from $\omega$ by $\Delta \omega$, the integration in the 3pt starts to cancel if is performed over $\Delta t_1\sim \pi/\Delta\omega$. In the meanwhile it takes $\Delta t_2 \sim \Delta\omega/(\omega H)$ to stretch the mode and change its frequency from $\omega$ to $\omega-\Delta\omega$.
Equating $\Delta t_1$ and $\Delta t_2$ gives the time period over which the resonance occurs for this mode,
\bea
\Delta t \sim \sqrt{\frac{\pi}{\omega H}} ~.
\eea
This corresponds to the number of oscillation periods
\bea
\frac{\omega \Delta t}{2\pi} \sim \sqrt{\frac{ \omega}{4\pi H}}
\label{resonant_number}
\eea
that we need to integral over to estimate the resonance contribution.
Note that one period in the integral (\ref{res_integral}) for $K=\omega$ contributes $\pi\tau_*/K$, where $\tau_*$ is evaluated at the resonant point. Multiplying the total number of the resonant periods (\ref{resonant_number}), using the definition (\ref{3pt_def}) and $\tilde P_\zeta \approx H^2/(8\pi^2\epsilon)$, we see that the amplitude of $S(k_1,k_2,k_3)$ is
\bea
f_{NL}^{\rm res} \sim \frac{\sqrt{\pi}}{16} \eta_A'\tau_* \sqrt{\frac{\omega}{H}} \sim \frac{\sqrt{\pi}}{8\sqrt{2}} \frac{\omega^{1/2} \dot\eta_A}{H^{3/2}} ~.
\label{fNL_res}
\eea
Slow-roll parameters acquire small oscillatory components, and here $\eta_A$ denotes the amplitude of such an oscillation. Other pre-factors of $k_i$ are cancelled according the definition of $S$ and the $S$ turns out to be a function of $K$ only. In the last step of (\ref{fNL_res}), we have listed the accurate numerical number, which differs from the estimate by a factor of $\sqrt{2}$.

Summarizing both the running behavior and the amplitude, we get the following ansatz for the bispectrum
\bea
S_{\rm ansatz}^{\rm res} = f_{NL}^{\rm res} \sin \left( C\ln (K/k_*) \right) ~,
\label{ansatz_res}
\eea
where
\bea
C=2\pi K/\Delta K = \omega/H
\eea
and $k_*$ gives a phase.
The argument $C \ln K$ in (\ref{ansatz_res}) appears because of (\ref{res_freq}). This gives a scale dependent oscillatory frequency in the $K$-space. In fact, this kind of dependence makes
the density perturbations in the resonance model semi-scale-invariant. We call it {\em periodic-scale-invariant} -- they are invariant under a discrete subgroup of rescaling. Namely, the ansatz (\ref{ansatz_res}) is invariant if we rescale all $k_i$ by $n\Delta K/K = 2\pi nH/\omega$ e-fold, where $n$ is an integer. Other rescaling causes a phase shift. This property is a direct consequence of the symmetry of the Lagrangian. It is periodic, so invariant under a discrete shift of the inflaton field. This periodic-scale-invariance should also be respected by the full bispectrum results, as well as other correlation functions.

\begin{figure}[t]
\begin{center}
%\epsfxsize=10cm
%\epsfbox{Fep.eps}
\epsfig{file=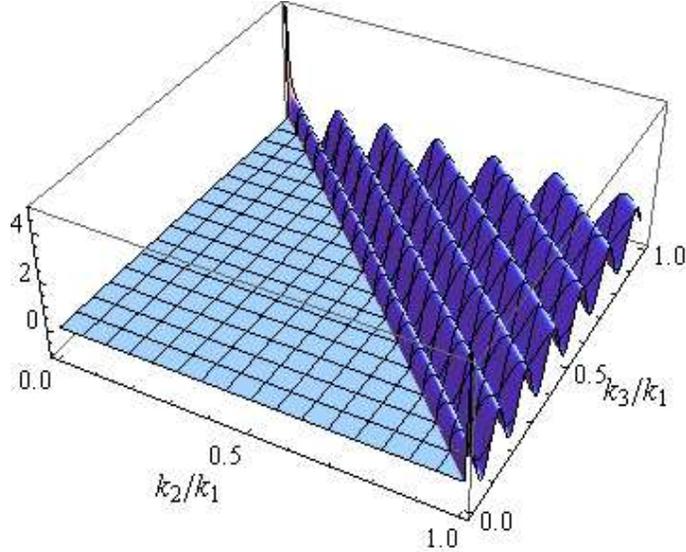, width=9cm}
\end{center}
\caption{The running and shape of the resonance bispectrum (\ref{S_res_full}) with $C=100$.}
\label{Fig:res_running_shape}
\end{figure}

\begin{figure}[t]
\begin{center}
%\epsfxsize=10cm
%\epsfbox{Fep.eps}
\epsfig{file=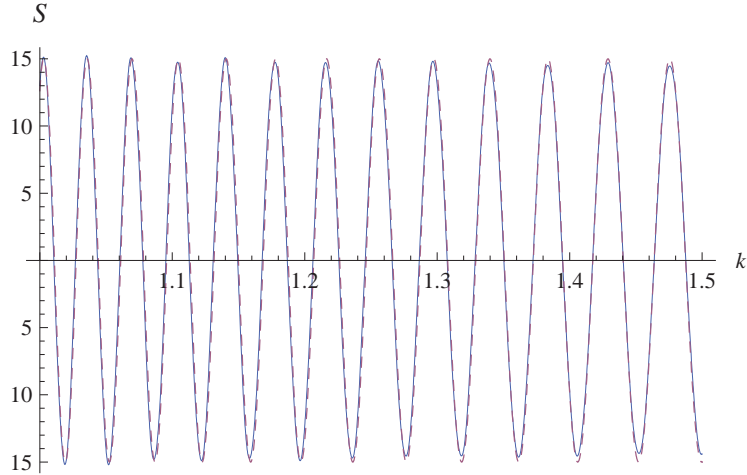, width=10cm}
\end{center}
\medskip
\caption{Numerical result (solid line) of the bispectrum running for the example (\ref{Vres_ex}) ($m=3\times 10^{-6}\mpl, c=5\times 10^{-7}, \Lambda=0.0007 \mpl, \phi\approx 15\mpl$), compared with the ansatz (\ref{ansatz_res}) (dashed line).}
\label{Fig:res_running}
\end{figure}

As mentioned, we have derived this ansatz from the last term in (\ref{res_3pt_terms}). Other terms will become important in the squeezed limit. The full integration (\ref{res_3pt_terms}) has been worked out in \cite{Flauger:2010ja}, and the leading order results are
\bea
S^{\rm res} = f^{\rm res}_{NL} \left[ \sin \left( C\ln (K/k_*) \right) + \frac{1}{C} \sum_{i\ne j} \frac{k_i}{k_j} \cos\left( C\ln (K/k_*) \right) + \CO\left(\frac{1}{C^2} \right) \right] ~,
\label{S_res_full}
\eea
where $\CO(1/C^2)$ terms are neglected because we need $1/C=H/\omega \ll 1$ for large resonance.
The numerical coefficient in (\ref{fNL_res}) turns out to be $\sqrt{\pi}/(8\sqrt{2})$.
As we can see, the extra terms satisfy the symmetry we mentioned and indeed give large corrections in the very squeezed limit, e.g.~$k_3 < k_1 H/\omega$. These terms also ensure a consistency condition that we will study in Sec.~\ref{Sec:consistency}.
An example is plotted in Fig.~\ref{Fig:res_running_shape}. The spike at the very squeezed limit is due to the second term in (\ref{S_res_full}).
Overall, we see that the leading {\em shape} of this bispectrum is quite trivial, being almost a function of $K$ only, until it gets to the very squeezed limit. The most distinctive feature of this type of non-Gaussianities is the {\em running} behavior captured in (\ref{ansatz_res}). Unfortunately, this ansatz is not factorizable if the $K$-range is too large.

More arbitrary scale-dependence can be introduced if the features are applied over a finite range, or with varying periodicity and amplitude.

As a useful comparison, the resonant running here and sinusoidal running that we studied in the last subsection are clearly distinguishable from each other observationally. The resonant running oscillates with periods that are always much smaller than the local scale, $\Delta K \ll K$; the frequency has a specific scale-dependence, $\Delta K/K={\rm const.}$; and the frequency in the power spectrum ($\sim \sin(C \ln k)$ in $k$-space) is exactly the same as that in the bispectrum ($\sim \sin(C \ln K)$ in $K$-space).
In contrast, the bispectrum of the sinusoidal running oscillates with a fixed period that approximately equals to the scale at the location of the sharp feature, $\Delta K \sim k_*$; the frequency is scale-independent; and the power spectrum ($\sim \sin(2k/k_*)$ in $k$-space) has twice an oscillatory frequency of the bispectrum ($\sim \sin(K/k_*)$ in $K$-space).

As an illustration, we look at an example,
\begin{equation}
V(\phi) =
\frac{1}{2}m^2 \phi^2 \left[1 +
c~\sin\left(\frac{\phi}{\Lambda}\right)\right] ~.
\label{Vres_ex}
\end{equation}
In this example, the inflaton is rolling over the small but periodic ripple laid on the potential. This induces an oscillatory component in the slow-roll parameters with an amplitude $\dot \eta_A \approx \sqrt{6}cm\phi/\Lambda^2$ and a frequency
$\omega \approx \dot\phi/\Lambda \approx 2m/(\sqrt{6}\Lambda)$.
So we have
\bea
f_{NL}^{\rm res} \sim \frac{c\mpl^3}{\Lambda^{5/2} \phi^{1/2}} ~.
\eea
and
\bea
C\approx 2/(\phi\Lambda) ~.
\eea

A numerical example is shown in Fig.~\ref{Fig:res_running}. As we can see, the ansatz (\ref{ansatz_res}) gives a very accurate fit to the actual running behavior. The mode function and power spectrum are the superposition of the usual unperturbed solution and a small oscillatory component \cite{Chen:2008wn,Flauger:2009ab,Hannestad:2009yx}. We can choose parameters so that the size of the ripples on the power spectrum is small, but bispectrum is made large. This is because the non-Gaussianities rise more quickly if we increase the frequency, while the mode function has difficulty responding efficiently when the external source oscillates too fast. This mechanism may be realized in terms of brane inflation \cite{Bean:2008na} where the periodic feature comes from duality cascade in warped throat \cite{Hailu:2006uj}, or the monodromy inflation \cite{Silverstein:2008sg,McAllister:2008hb} where the periodic feature comes from instanton effects \cite{Flauger:2009ab,Hannestad:2009yx}.

\subsection{Folded shape: a non-standard vacuum}
\label{Sec:folded}

In this subsection, we study the effect of non-standard vacuum on the primordial non-Gaussianities. We consider a different wave-function from the Bunch-Davies vacuum when modes are well within the horizon. To start, let us first discuss several motivations for this case.

\begin{itemize}

\item
A non-Bunch-Davies vacuum can actually occur much more simply than it might sound like. Any deviation from the attractor solution of the inflaton generically generates a component of non-Bunch-Davies vacuum. This is because a general mode function is a superposition of two components, $c_1(\bk) u(\bk,t) + c_2(\bk) u^*(\bk,t)$, and in attractor solution we choose one of the component asymptotic to the Bunch-Davies vacuum. A disturbance will generically introduce a mixture with the other component. In this sense, we have already encountered such a case when we studied the effect of a sharp feature in Sec.~\ref{Sec:sin}. Indeed, after the inflaton crosses the sharp feature, the oscillatory behavior in the power spectrum is precisely due to the superposition of the second non-Bunch-Davies component for some finite $k$-range. For an {\em infinitely sharp} change, such a disturbance with a small amplitude extends to all $k$ that have not exited the horizon at the time of sharp feature. An analytical illustration can be found in Sec.~5.3 of Ref.~\cite{Bean:2008na}. The location of the sharp feature can become superhorizon at the present time, but its influence has extended to much smaller scales and becomes observable.
The resonance case in Sec.~\ref{Sec:res} is another type of example. An analytical illustration can be found in Sec.~3.3 of Ref.~\cite{Flauger:2009ab}.
For non-Gaussianities studied in Sec.~\ref{Sec:sin} \& \ref{Sec:res}, we only concentrated on the effects caused by the features in slow-roll parameters. In analytical analyses, we approximated the mode function by the Bunch-Davies component and ignored the disturbance. The study of this subsection can be regarded as the complementary analyses on the effect of a different mode component.

\item
In inflationary background, modes can be quantized in terms of time-dependent creation and annihilation operators, $a_\bk(t)$ and $a^\dagger_{-\bk}(t)$. The Bunch-Davies vacuum is defined as the vacuum annihilated by $a_\bk(t)$ as $t\to -\infty$. If a different adiabatic vacuum is defined which is annihilated by $a_\bk(t_0)$ at a finite $t_0$, we introduce a non-Bunch-Davies component. For example, see \cite{Martin:2000xs,Danielsson:2002kx}. The origin and magnitude of such a component have been debated and studied by many papers, often under the name of the "trans-Planckian effect". See Ref.~\cite{Brandenberger:2005be,Greene:2005aj} for summary and references.

\item
There are inflation models where the scale of new physics can be very low. In particular, in warped space it is proportional to the exponentially small warp factor. In some DBI inflation models \cite{Chen:2005ad,Bean:2007eh}, the speed limit of the inflaton and the scale of new physics are both related to the warp factor in such a way that the local warped new scale can drop near or even below the Hubble energy scale in certain epoch of inflation. Clearly the simple scalar field Bunch-Davies vacuum is no longer sufficient. Such models further open up the possibilities of vacuum choices.

\end{itemize}

After these discussions, let us now focus on a specific simple problem \cite{Chen:2006nt}. We modify the wave-function of the Bunch-Davies vacuum by a small second component and examine its consequence for the three-point function calculated in Sec.~\ref{Sec:Eq}. We consider the general single field inflation with a small sound speed $c_s$ or a large $\lambda/\Sigma$ \cite{Chen:2006nt,Meerburg:2009fi}.

\begin{figure}[t]
\begin{center}
%\epsfxsize=10cm
%\epsfbox{Fep.eps}
\epsfig{file=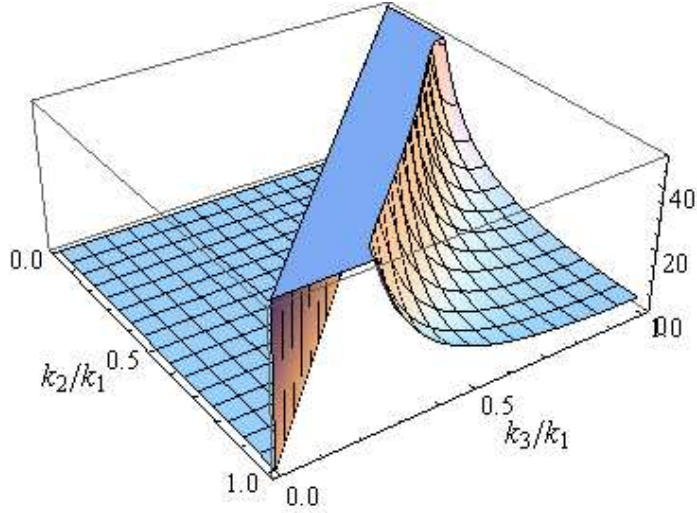, width=10cm}
\end{center}
\caption{Shape of $\tilde S_\lambda$ (truncated).}
\label{Fig:shape_nonBD1}
\end{figure}

\begin{figure}[ht]
\begin{center}
%\epsfxsize=10cm
%\epsfbox{Fep.eps}
\epsfig{file=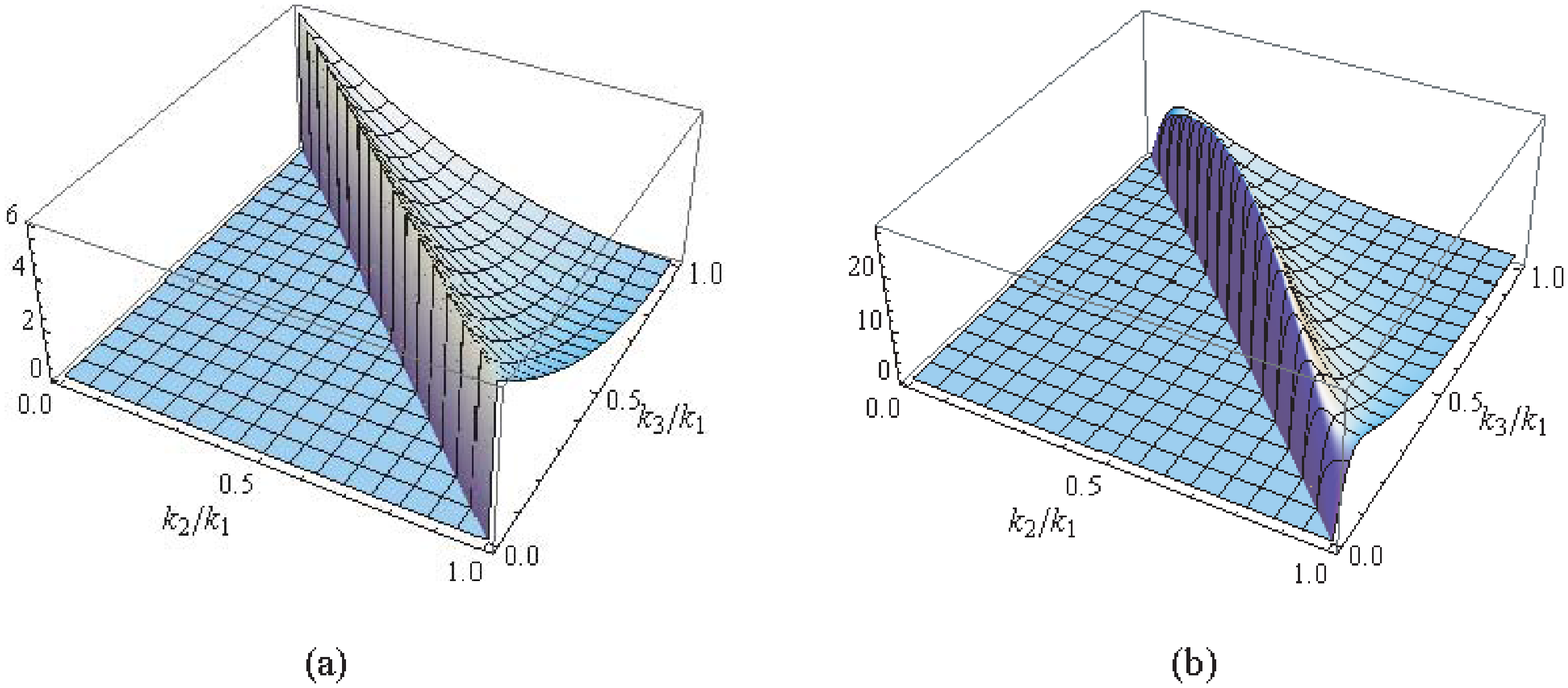, width=15cm}
\end{center}
\caption{Two ansatz for the folded shape. (a) Eq.~(\ref{ansatz_fold}), (b) Eq.~(\ref{ansatz_fold_2}) with $k_c/k_1=0.1$.}
\label{Fig:shape_ansatz_fold}
\end{figure}

So the mode function is
\bea
u_k(\tau) = \frac{i H}{\sqrt{4\epsilon c_s k^3}}
\left[ (1+i k c_s\tau)e^{-i k c_s\tau} + C_- (1-i k c_s\tau)e^{i k c_s\tau} \right] ~,
\label{mode_nonBD}
\eea
where $|C_-| \ll 1$ and can be $k$-dependent. In the first example above, the extra component starts at a specific time in the past. In the second class of examples, it may start either at a specific time or specific energy scale. To see a common feature without addressing these model-dependent issues, we look at the simple limit where the $\tau$ in (\ref{mode_nonBD}) can go all the way to $-\infty$. The computation of the correlation function is essentially the same as in Sec.~\ref{Sec:Eq}. The leading order correction to the bispectra is obtained by replacing one of the three $u_k(\tau)$ in the integrand by its $C_-$ component. So it simply replaces one of the $k_i$'s in the shapes with $-k_i$. For example, the correction to $S_\lambda$ is
\begin{align}
\tilde S_\lambda &= \left( \frac{1}{c_s^2} - 1 - \frac{2\lambda}{\Sigma} \right) \frac{3k_1k_2k_3}{2}
\nonumber \\
& \times \left( \frac{ {\rm Re}(C_-(k_3)) }{(k_1+k_2-k_3)^3} + \frac{ {\rm Re}(C_-(k_2)) }{(k_1-k_2+k_3)^3} + \frac{ {\rm Re}(C_-(k_1)) }{(-k_1+k_2+k_3)^3} \right) ~.
\label{nonBD_lambda}
\end{align}

The shape of $\tilde S_\lambda$ is shown in Fig.~\ref{Fig:shape_nonBD1}.
The most important feature of this shape is the enhancement at the folded triangle limit, e.g.~$k_1+k_2-k_3=0$. The detailed form of enhancement is model dependent. For example, it is different for another shape $S_c$. The divergence in this folded limit occurs due to our simple limit of taking $\tau$ to $-\infty$. Imposing some kind of cutoff at the lower limit of $\tau$ will eliminate this divergence, although as mentioned the detailed modification will be highly model dependent. For example, a simple constant cutoff $\tau_c$ will introduce a factor of
$1+(\half x_c^2-1) \cos x_c - x_c \sin x_c $ for each of the three terms in (\ref{nonBD_lambda}), where $x_c\equiv (k_1+k_2-k_3) c_s \tau_c$ or its cyclic. Very close to the folded limit, $\tilde K c_s |\tau_c| \ll 1$ ($\tilde K=k_1+k_2-k_3$ or its cyclic), this regulates away the divergence; away from the folded limit, $\tilde K c_s |\tau_c| \gg 1$, these extra factors are unity on average but with oscillations. These oscillation can be either physical, or regarded to be zero if $x_c$ is within a regulation scale which exists since the non-Bunch-Davies component is present for a finite time in the past.

The case for slow-roll inflation is qualitatively similar, and more examples of the bispectra shapes and the observational prospects are discussed in \cite{Holman:2007na,Meerburg:2009ys}.
In this case, the proportional parameter for the bispectra amplitude is no longer enhanced by $1/c_s^2$ or $\lambda/\Sigma$, but $<\CO(1)$.

In order to facilitate the data analyses, a simple ansatz has been proposed in Ref.~\cite{Meerburg:2009ys},
\bea
S_{\rm ansatz,1}^{\rm fold} = 6 \left(\frac{k_1^2}{k_2k_3} + {\rm 2~perm.}\right)
- 6 \left( \frac{k_1}{k_2} + {\rm 5~perm.} \right) +18 ~,
\label{ansatz_fold}
\eea
which represents certain important features of this kind of bispectra.
It has a smooth rising behavior in the folded limit.
This ansatz is plotted in Fig.~\ref{Fig:shape_ansatz_fold}(a).
Since the real shape has a model dependent cutoff, it remains open questions how sensitive this is to data analyses and how well the ansatz (\ref{ansatz_fold}) represents it.
We can also write down an ansatz which is more directly motivated from the example (\ref{nonBD_lambda}) and the comments after that equation,
\bea
S_{\rm ansatz,2}^{\rm fold} = k_1k_2k_3 \frac{k_1+k_2-k_3}{(k_c+k_1+k_2-k_3)^4} + {\rm 2~perm.} ~,
\label{ansatz_fold_2}
\eea
where the cutoff scale $k_c=1/(c_s \tau_c)$ is a parameter.
For $k_1+k_2-k_3\gg k_c$ and cyclic, we have neglected the oscillatory part and only taken the average. In this ansatz we can change the powers in the numerator and denominator to model model-dependent variations. The relation $(k_c+k_1+k_2-k_3)^{-n}=(\Gamma(n))^{-1} \int_0^\infty dt~ t^{n-1} e^{-(k_c+k_1+k_2-k_3)t}$ may be used to factorize the ansatz.
This ansatz is plotted in Fig.~\ref{Fig:shape_ansatz_fold}(b).

Another type of non-Bunch-Davies vacuum, namely an $n$-particle state built on the normal Bunch-Davies vacuum, was studied in \cite{Martin:1999fa,Gangui:2002qc} and the non-Gaussianities were found to be unobservable.

\section{Quasi-single field inflation}
\label{Sec:quasi}
\setcounter{equation}{0}

Having considered single field inflation, we now relax the condition on the number of fields. At least during inflation, we only need to consider quantum fluctuations of light fields, since if the mass of fields are very heavy, (here the relevant scale is $m\gg H$), they contribute only classically and determine the classical inflaton trajectory. Multiple light fields can arise naturally if we consider the inflation models as the consequence of a UV completed framework. However, as discussed in Sec.~\ref{Sec:Model_building}, due to the back-reaction from the inflationary background, the mass of light fields are naturally of order $H$. The potential with such a shape is too steep for slow-roll inflation.

Therefore, as a natural step beyond the single field, let us consider slow-roll models with one inflationary direction, and one or more other directions that have mass neither much heavier nor much lighter than $H$. We will call the quanta in the inflationary direction as the inflaton and its mode the curvature mode, and the others isocurvaton and isocurvature modes. We call these models the {\em quasi-single field inflation models} \cite{Chen:2009we,Chen:2009zp}.

Note that the thematic order in this review is not chronological. The non-Gaussianities in this type of models were not computed until very recently for a couple of reasons. If the mass of particles is of order $\CO(H)$ or larger, the amplitude of these fields decay exponentially in time after horizon-exit. So they would not seem to be important for super-horizon perturbations even if they couple to the curvature mode. As we will see, however, their amplitudes at or near the horizon-exit are enough to make them interesting. What really suppresses their contribution is the fast oscillation behavior present for $m\gg H$. Methodologically, isocurvature-to-curvature transition for non-Gaussianities was studied restrictively in the regime of super-horizon classical evolution in multi-field space \cite{Salopek:1990jq,Bartolo:2001cw,Bernardeau:2002jy,Bernardeau:2002jf,Rigopoulos:2005xx,Rigopoulos:2005ae,Seery:2005gb,Vernizzi:2006ve}, which we shall explain in more details in the next section. However, for quasi-single field inflation models, a full quantum computation in the in-in formalism is necessary to properly include the contributions from both the horizon exit and the superhorizon evolution.

\subsection{Intermediate shapes: massive isocurvatons}
\label{Sec:int}

There are potentially different ways massive isocurvatons can be coupled to the inflaton. We currently do not have a general approach in terms of model building. So what we shall do is to first study this problem through a simple example, and then discuss the features of the results that can be regarded as generic signatures of this class of models \cite{Chen:2009we,Chen:2009zp}.

\begin{figure}[t]
\begin{center}
\epsfig{file=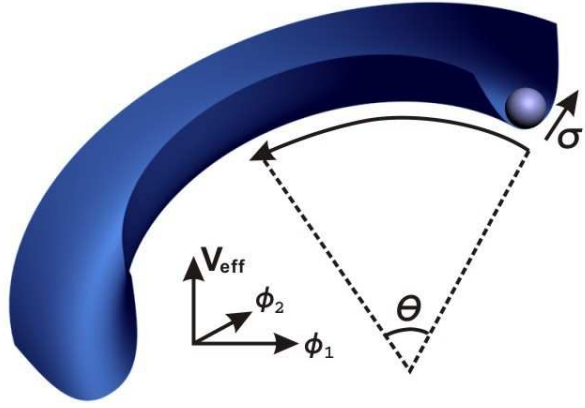, width=8cm}
\end{center}
\caption{Quasi-single field
  inflation with turning trajectory. The field $\theta$ and $\sigma$ are in the polar coordinates. The $\theta$ is the inflationary
  direction with a slow-roll potential. The $\sigma$ is the isocurvature direction, which typically has mass of order $H$.}
\label{Fig:turn}
\end{figure}

We consider the case where the inflaton is turning constantly by going around (a fraction of) a circle with radius $R$ in the angular $\theta$ direction. See Fig.~\ref{Fig:turn}.
All the parameters, such as $R$ and couplings, are assumed to be constant during the turning. We call this assumption the constant turn case.
In the $\theta$ direction the potential is the usual slow-roll potential $V_{\rm sr}(\theta)$. The field in the radial direction is denoted as $\sigma$ and has mass of order $H$, and lifted by the potential $V(\sigma)$. For such a turning trajectory, it is convenient to write the action in terms of fields in the polar coordinates, $\theta$ and $\sigma$, instead of in the Cartesian coordinates,
\bea
S_m = \int d^4x \sqrt{-g} \left[
-\half ( R+\sigma)^2 g^{\mu\nu} \partial_\mu \theta \partial_\nu \theta
- \half g^{\mu \nu} \partial_\mu \sigma \partial_\nu \sigma
- V_{\rm sr}(\theta) - V(\sigma) \right] ~.
\label{action:quasi-single}
\eea
The potential $V(\sigma)$ balances off the centrifugal force necessary for the turning and traps the field at the bottom of the effective potential,
$V_{\rm eff}(\sigma) = -\half \dot\theta_0^2 (R+\sigma)^2 + V(\sigma)$.
We define the minimum of this effective potential to be $\sigma=0$. We expand the effective potential as
\bea
V_{\rm eff} = {\rm const.} + \half \left(V''-\dot\theta_0^2 \right) \sigma^2 + \frac{1}{6} V'''\sigma^3 + \cdots ~,
\eea
where $\dot\theta_0$ is the turning angular velocity and the primes on $V$ denote derivatives with respective to $\sigma$.

To study the perturbation theory, we perturb the fields in the spatially flat gauge,
\bea
\theta(\bx,t) = \theta_0(t) + \delta\theta(\bx,t) ~, ~~~~~
\sigma(\bx,t) = \delta\sigma(\bx,t) ~,
\label{fieldpert_quasisingle}
\eea
and obtain the following Hamiltonian,
\bea
\CH_0 = a^3 \left[ \half R^2 \dot {\delta\theta_I}^2 +
  \frac{R^2}{2a^2}
  (\partial_i \delta\theta_I)^2
+ \half \dot{\delta\sigma_I}^2 + \frac{1}{2a^2} (\partial_i
\delta\sigma_I)^2 + \half m^2 \delta \sigma_I^2
\right] ~,
\label{CH0_quasi}
\eea
\bea
\CH^I_2 &=& -c_2 a^3 \delta\sigma_I \dot{\delta\theta_I} ~,
\label{CH2_quasi}
\\
\CH^I_3 &=&  c_3 a^3 \delta\sigma_I^3 ~,
\label{CH3_quasi}
\eea
where
\bea
c_2 = 2 R \dot\theta_0 ~, \quad c_3= \frac{1}{6} V''' ~, \quad m^2= V'' + 7\dot \theta_0^2
\eea
are all constants.
Terms suppressed by $\CO(\epsilon)$ have been ignored in this gauge. The curvature perturbation $\zeta$ is most transparent in another gauge, the uniform inflaton gauge, where
\bea
\theta(\bx,t) = \theta_0(t) ~,
\quad
\sigma(\bx,t) = \sigma_0(t) + \delta\sigma(\bx,t) ~,
\eea
and the spatial metric is
\bea
h_{ij}(\bx,t) = a^2(t) e^{2\zeta(\bx,t)} \delta_{ij} ~.
\eea
In this gauge, $\zeta$ appears in the metric as the space-dependent rescale factor and the fluctuations in the inflaton is shifted away. The relation between $\zeta$ and $\delta\theta$ is the gauge transformation. At the leading order this is
\bea
\zeta \approx - \frac{H}{\dot \theta_0} \delta\theta ~.
\label{zeta_deltatheta}
\eea
We will calculate the correlation functions in terms of $\delta\theta$ and then use this relation to convert them to those of $\zeta$.
The full perturbation theory that one obtains in the uniform inflaton gauge justifies the above omission of several $\CO(\epsilon)$ terms in the spatially flat gauge \cite{Chen:2009zp}.

There are several important points for this Hamiltonian.

First, the kinematic Hamiltonian (\ref{CH0_quasi}) describes two free fields in the inflationary background. One is massless and has the familiar mode function,
\bea
u_\bk = \frac{H}{R\sqrt{2k^3}} ( 1+i k\tau)e^{-i k\tau} ~.
\label{mode_u_quasi}
\eea
Another is massive and the mode function is
\bea
v_\bk = -i e^{i(\nu+\half)\frac{\pi}{2}} \frac{\sqrt{\pi}}{2} H
  (-\tau)^{3/2} H^{(1)}_\nu (-k\tau) ~,
\label{mode_v1_quasi}
\eea
where
\bea
\nu = \sqrt{9/4-m^2/H^2} ~.
\eea
For $0\le m/H\le 3/2$, the amplitude of the mode $v_\bk$ decays as $(-\tau)^{-\nu+3/2}$ after horizon-exit $k\tau\to 0$. The lighter the isocurvaton is, the slower it decays. At the $\nu\to 3/2$ (i.e.~$m/H\to 0$) limit, the amplitude is frozen.
For $m/H>3/2$, $\nu$ becomes imaginary, the mode $v_\bk$ not only contains a decay factor $(-\tau)^{3/2}$ but also an oscillation factor $\tau^\nu$. This oscillation is marginal for $m\sim H$, but if $m \gg H$, it causes cancellation in the integrals of the correlation function and is equivalent to factors of Boltzmann-like suppression $\sim e^{-m/H}$. We will consider the case $0\le \nu \le 3/2$.

Second, there is a sharp contrast between the $V'''_{\rm sr}$ for the slow-roll inflaton field and the $V'''$ for the massive field $\sigma$ in the non-inflationary direction. The former has to be very small, $\sim \CO(\epsilon^2) P_\zeta^{1/2} H$, in order to maintain the smallness of the slow-roll parameters. (Here we use $\epsilon$ to denote collectively all slow-roll parameters, $\epsilon\equiv -\dot H/H^2$, $\eta \equiv \dot\epsilon/\epsilon H$, and $\xi\equiv \dot\eta/\eta H$.) Consequently it contributes $\CO(\epsilon^2)$ to the $f_{NL}$ of bispectrum in slow-roll inflation, generally smaller than the $\CO(\epsilon)$ contributions from the other terms in the same model. However, for quasi-single field inflation, the direction orthogonal to slow-roll does not have to satisfy the slow-roll conditions, and $V'''$ is almost unconstrained. For example, in the inflationary background, it can be of order $H$; and similarly, $V''''$ can be of order one, etc. This isocurvaton self-interaction (\ref{CH3_quasi}) becomes the source of large non-Gaussianities.

\begin{figure}[t]
\center
\epsfig{file=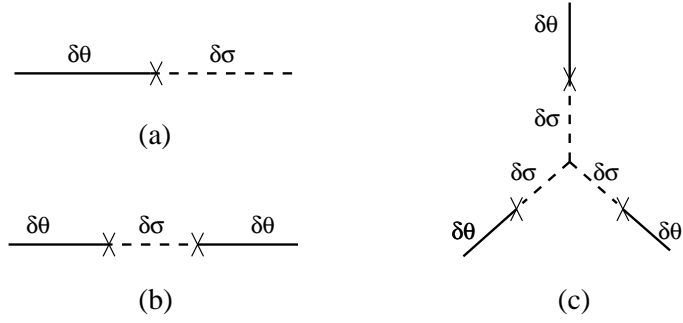, width=0.55\textwidth}
\caption{Feynman diagrams for the transfer vertex
  (a), corrections to the power spectrum from isocurvature modes (b), and the
  leading bispectrum (c).}
\label{Fig:fdiag_quasi}
\end{figure}

Third, the coupling between the isocurvaton and inflaton appears as a form of a two-point vertex operator in (\ref{CH2_quasi}). We treat this term as part of the interaction Hamiltonian, and it is represented by the transfer vertex in Fig.~\ref{Fig:fdiag_quasi} (a). The strength of the coupling is determined by the turning angular velocity $\dot\theta_0$ in this model. This coupling is responsible for the transformation of the isocurvature perturbations, in particular their large non-Gaussianities, to the curvature perturbation.

We calculate correlation functions corresponding to the Feynman diagrams Fig.~\ref{Fig:fdiag_quasi} in terms of the in-in formalism, which we reviewed in Sec.~\ref{Sec:in-in}. As an illuminating example to illustrate the different advantages of the three forms of the in-in formalism, we recall from Sec.~\ref{Sec:in-in} that the three-point function can be written in the following forms. The original definition (\ref{inin_summary}) and (\ref{inin_factorizedeven}), which we refer to as the factorized form, leads to
\bea
\langle \delta\theta^3 \rangle
&=&
-12 c_2^3 c_3 u_{p_1}^*(0) u_{p_2}(0) u_{p_3}(0)
\cr &\times&
{\rm Re} \Bigg[ \int_{-\infty}^0 d\ttau_1~ a^3(\ttau_1)
  v_{p_1}^*(\ttau_1) u'_{p_1}(\ttau_1)
\int_{-\infty}^{\ttau_1} d\ttau_2~ a^4(\ttau_2) v_{p_1}(\ttau_2)
v_{p_2}(\ttau_2) v_{p_3}(\ttau_2)
\cr &\times&
\int_{-\infty}^0 d\tau_1~ a^3(\tau_1) v_{p_2}^*(\tau_1)
u_{p_2}^{\prime *} (\tau_1)
\int_{-\infty}^{\tau_1} d\tau_2~ a^3(\tau_2) v_{p_3}^*(\tau_2)
u_{p_3}^{\prime *} (\tau_2) \Bigg]
\cr &\times&
(2\pi)^3 \delta^3(\sum_i \bp_i) + {\rm 9~other~similar~terms}
\cr
&+& {\rm 5~permutations~of~} \bp_i ~.
\label{FacForm_term}
\eea
The perturbation theory here starts from the fourth order.
The reorganized commutator form (\ref{inin_commutator}) leads to
\bea
\langle \delta\theta^3 \rangle
&=& 12 c_2^3 c_3 u_{p_1}(0) u_{p_2}(0) u_{p_3}(0)
\cr
&\times&
{\rm Re} \Bigg[ \int_{-\infty}^0 d\tau_1 \int_{-\infty}^{\tau_1}
  d\tau_2 \int_{-\infty}^{\tau_2} d\tau_3 \int_{-\infty}^{\tau_3}
  d\tau_4~
\prod_{i=1}^4 \left( a^3(\tau_i) \right)
\cr
&\times&
a(\tau_2)
\left(u_{p_1}^{\prime}(\tau_1) - c.c. \right)
\left( v_{p_1}(\tau_1) v_{p_1}^*(\tau_2) - c.c. \right)
\cr
&\times&
\left( v_{p_3}(\tau_2) v_{p_3}^*(\tau_4) u_{p_3}^{\prime *}(\tau_4) -
c.c. \right)
v_{p_2}(\tau_2) v_{p_2}^*(\tau_3) u_{p_2}^{\prime *}(\tau_3)
\Bigg]
\cr
&\times&
(2\pi)^3 \delta^3(\sum_i \bp_i)
+  {\rm 2~other~similar~terms}
\cr
&+& {\rm 5~permutations~of~} \bp_i ~.
\label{ComForm_term}
\eea

In the IR ($\tau\to 0$), each of the ten terms in the factorized form diverge as $\tau^{3-6\nu}$ for $3/2>\nu>1/2$ ($0<m<\sqrt{2}H$); while in the commutator form, various subtractions off the complex conjugates and the requirement that the final result has to be real make such divergence explicitly disappear.

In the UV ($\tau \to -\infty$),
each factor of the multiple integral that integrates from $-\infty$ to $0$ has a definite convergent direction if we choose one of the two contour tilts, $\tau_i \to -\infty(1\pm i\epsilon)$, accordingly. Or more efficiently, by a Wick rotation $\tau_i \to \pm i z_i$. This would have been the case for the commutator form if we can break up the integrand into individual terms. However in order to achieve the explicit IR convergence, as we saw above, these terms have to be grouped; but then they have contradicting convergence directions.

\begin{figure}[t]
\begin{center}
\epsfig{file=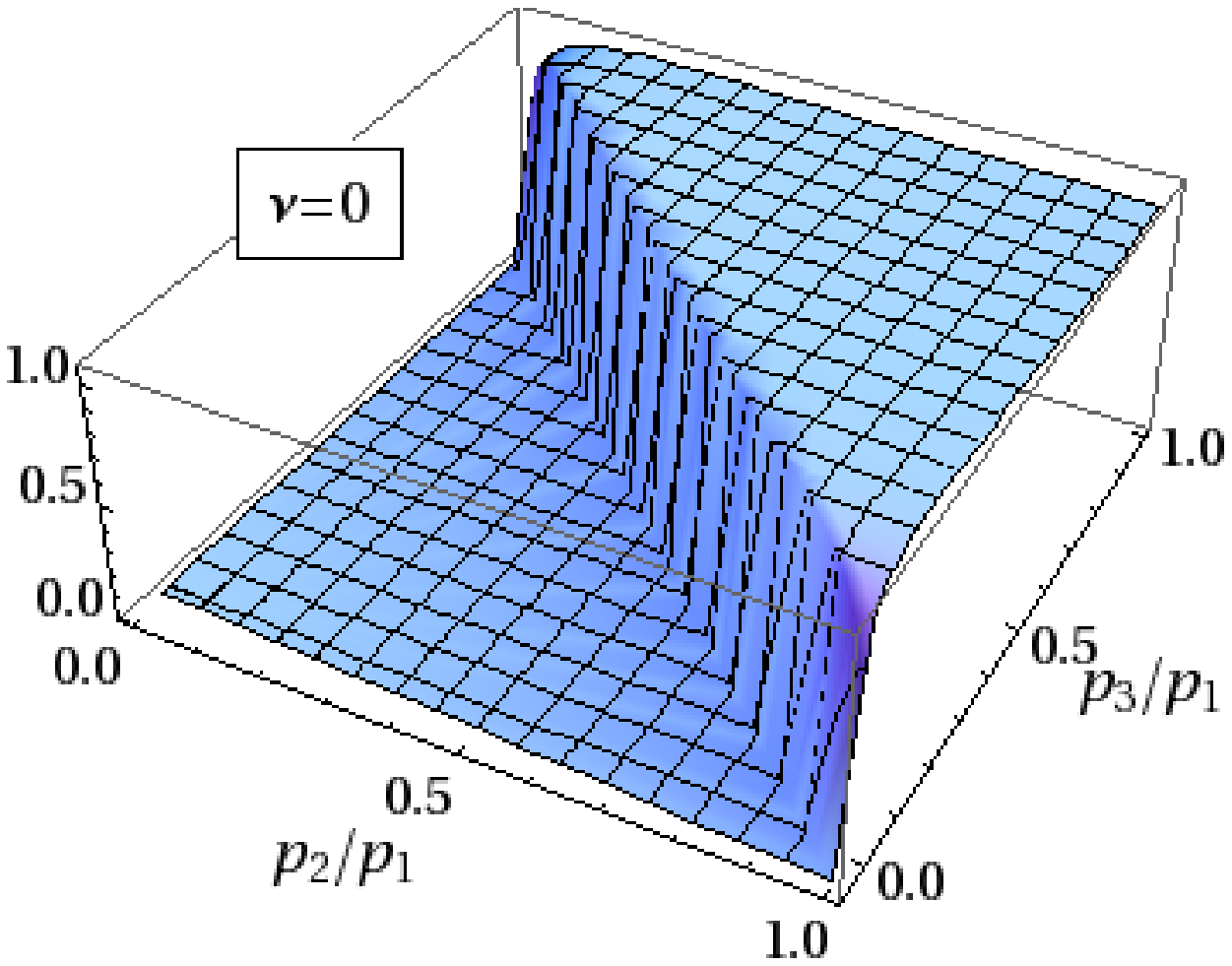, width=0.4\textwidth}
\epsfig{file=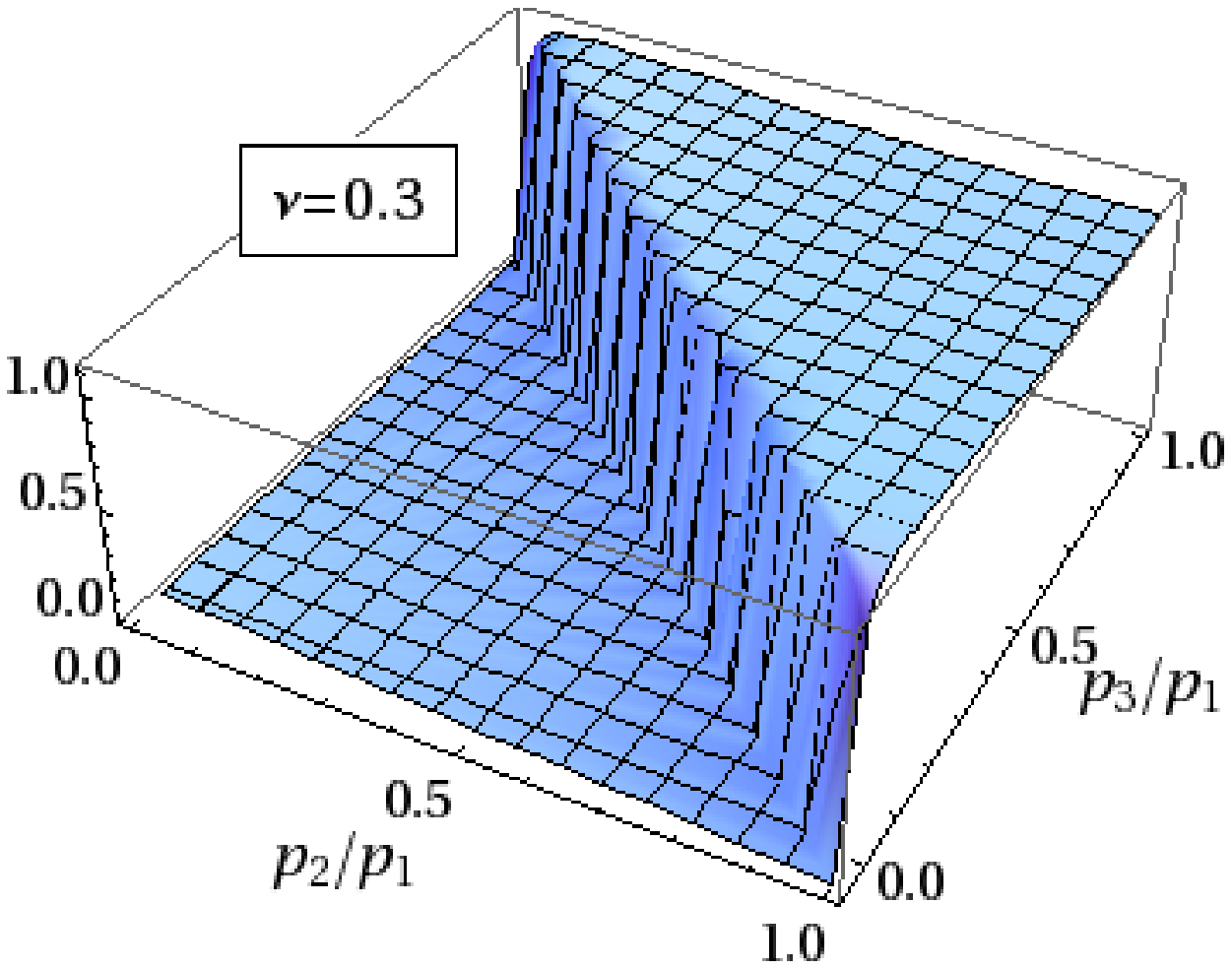, width=0.4\textwidth}
\epsfig{file=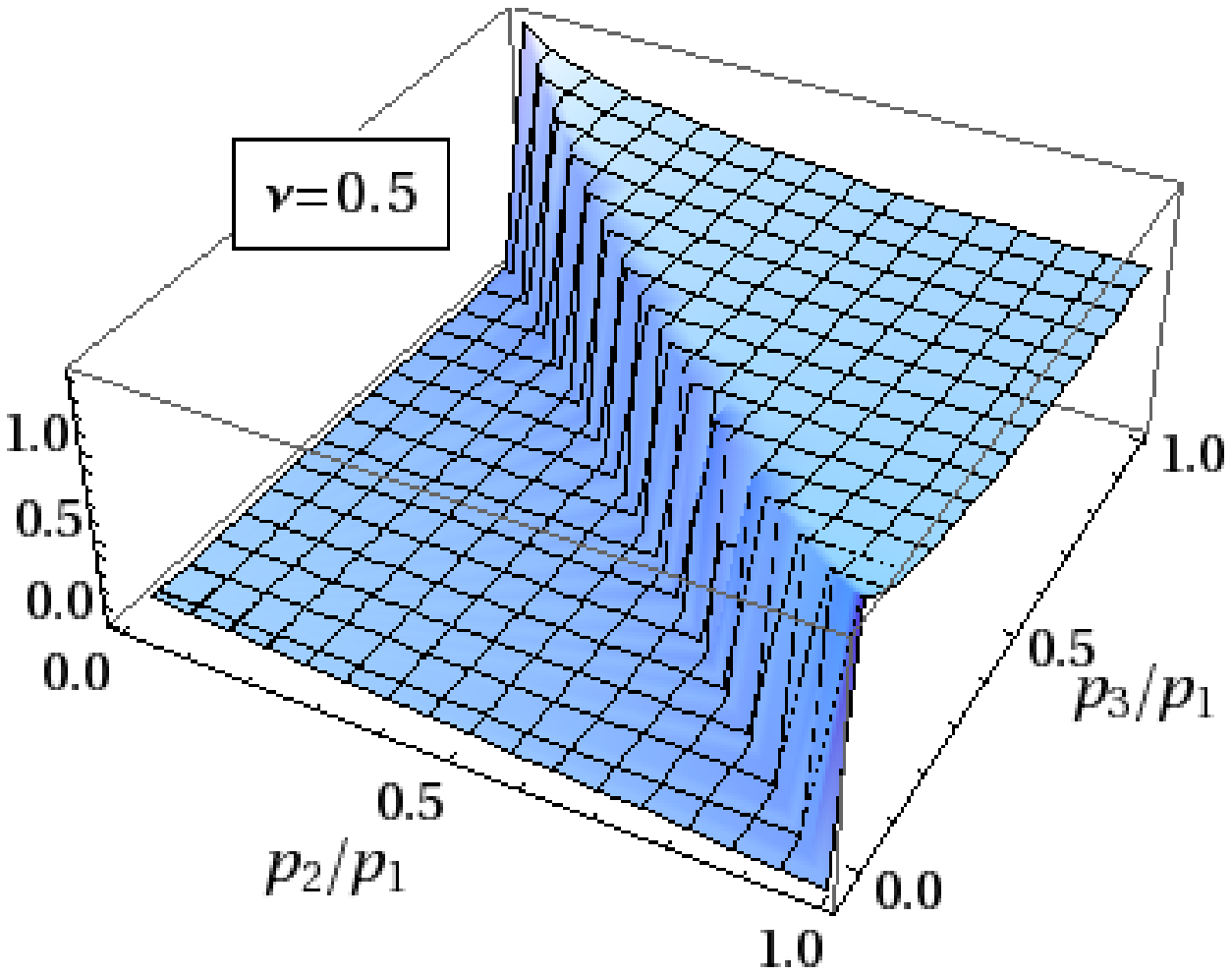, width=0.4\textwidth}
\epsfig{file=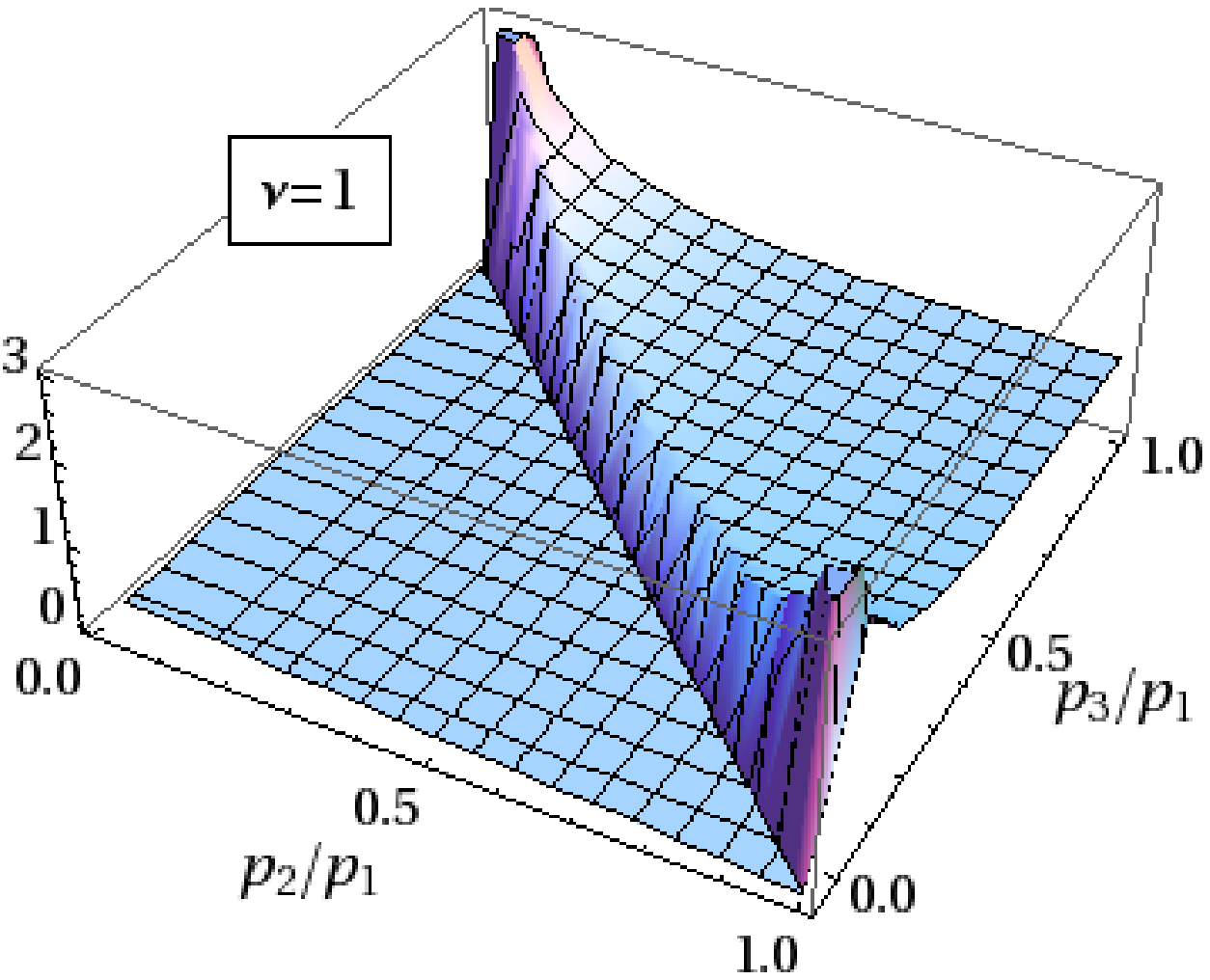, width=0.4\textwidth}
\end{center}
\caption{
Numerical results for the shapes of bispectra with intermediate forms. We plot
$S$ with $\nu=0,~ 0.3,~ 0.5,~ 1$. The plot is
normalized such that $S=1$ for $p_1=p_2=p_3$.}
\label{Fig:shape_int}
\end{figure}

To take advantage of both forms, we introduce a cutoff $\tau_c$, and write the IR part ($\tau_c<\tau\le 0$) of the integrals in terms of the commutator form, and the UV part ($\tau<\tau_c$) in terms of the factorized form, in the following mixed form (\ref{inin_mixedform}),
\begin{align}
\sum_i \int_{\tau_c}^0 d\tau_1 \cdots \int_{\tau_c}^{\tau_{i-1}} d\tau_i
~{\rm \{commutator~form\}} ~
\int_{-\infty}^{\tau_c} d\tau_{i+1} \cdots
\int_{-\infty}^{\tau_{n-1}} d\tau_n
~{\rm \{factorized~form\} } ~.
\label{inin_mixedform_tau}
\end{align}
This shows explicitly both convergence behavior of the correlation function. Combining with Wick-rotations of the integration contours in the UV, this form provides an efficient way to evaluate the correlation functions numerically.
The shapes of bispectra are presented in Fig.~\ref{Fig:shape_int} for $\nu=0, 0.3, 0.5, 1$.

\begin{figure}[t]
\begin{center}
%\epsfxsize=10cm
%\epsfbox{Fep.eps}
\epsfig{file=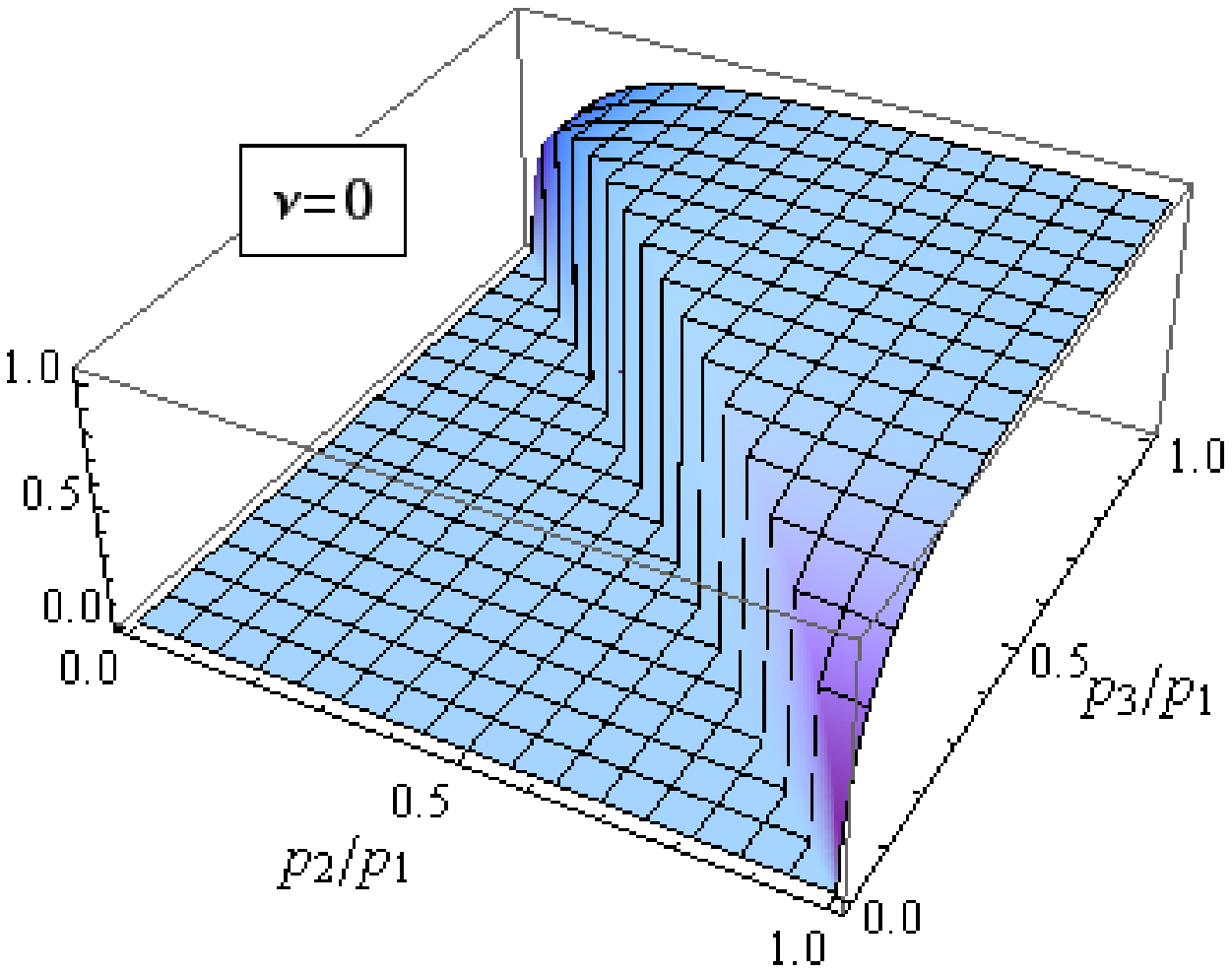, width=0.4\textwidth}
\epsfig{file=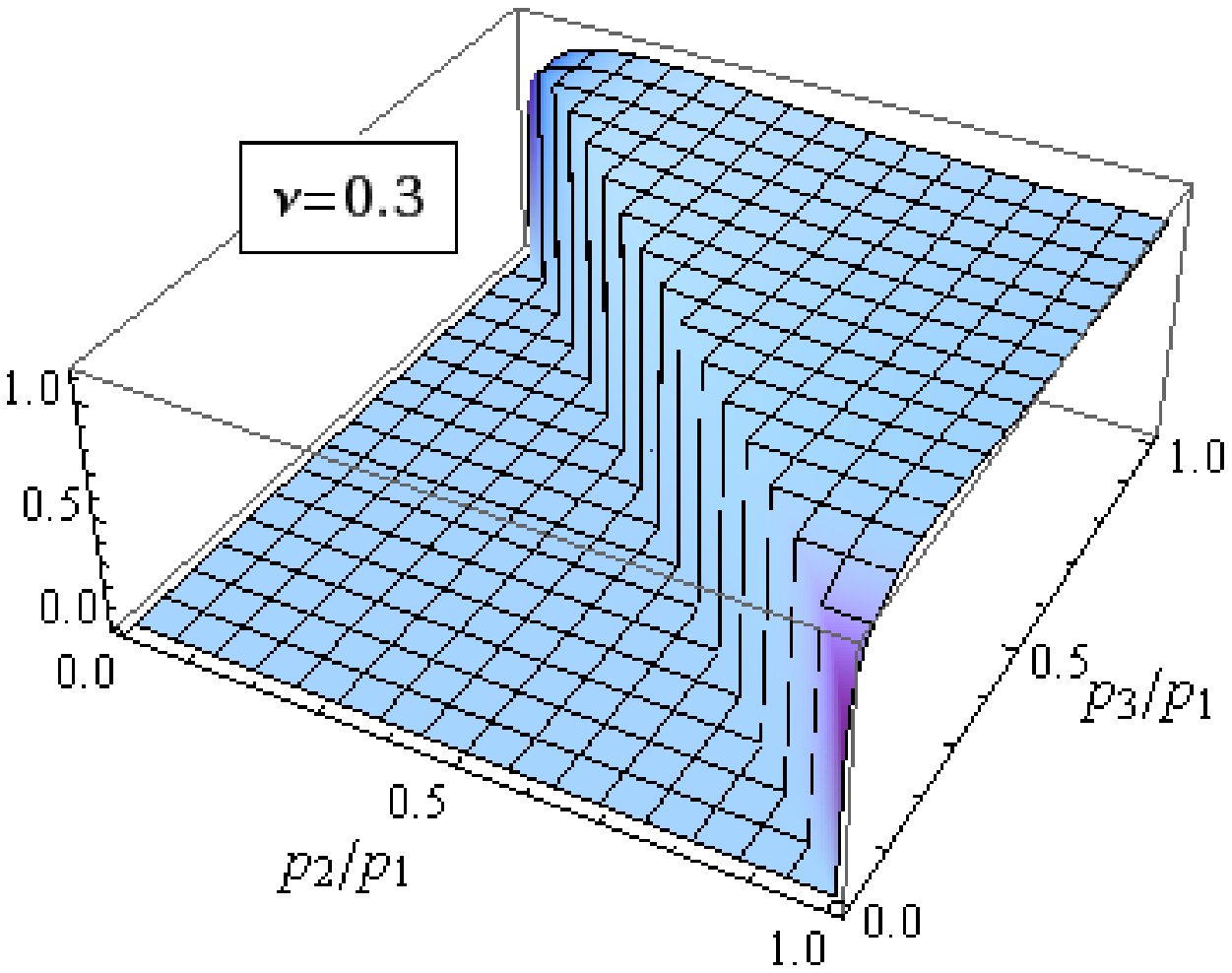, width=0.4\textwidth}
\epsfig{file=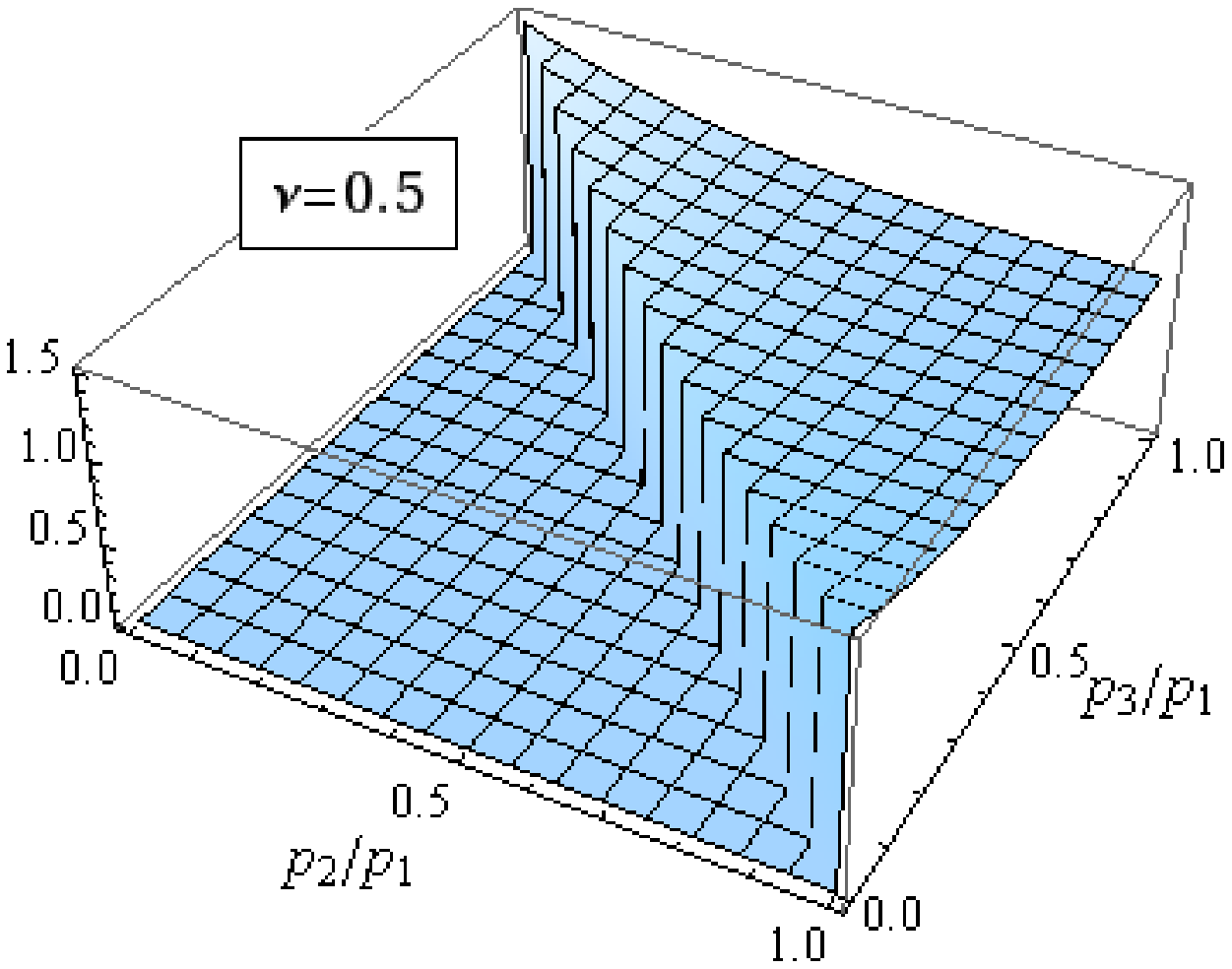, width=0.4\textwidth}
\epsfig{file=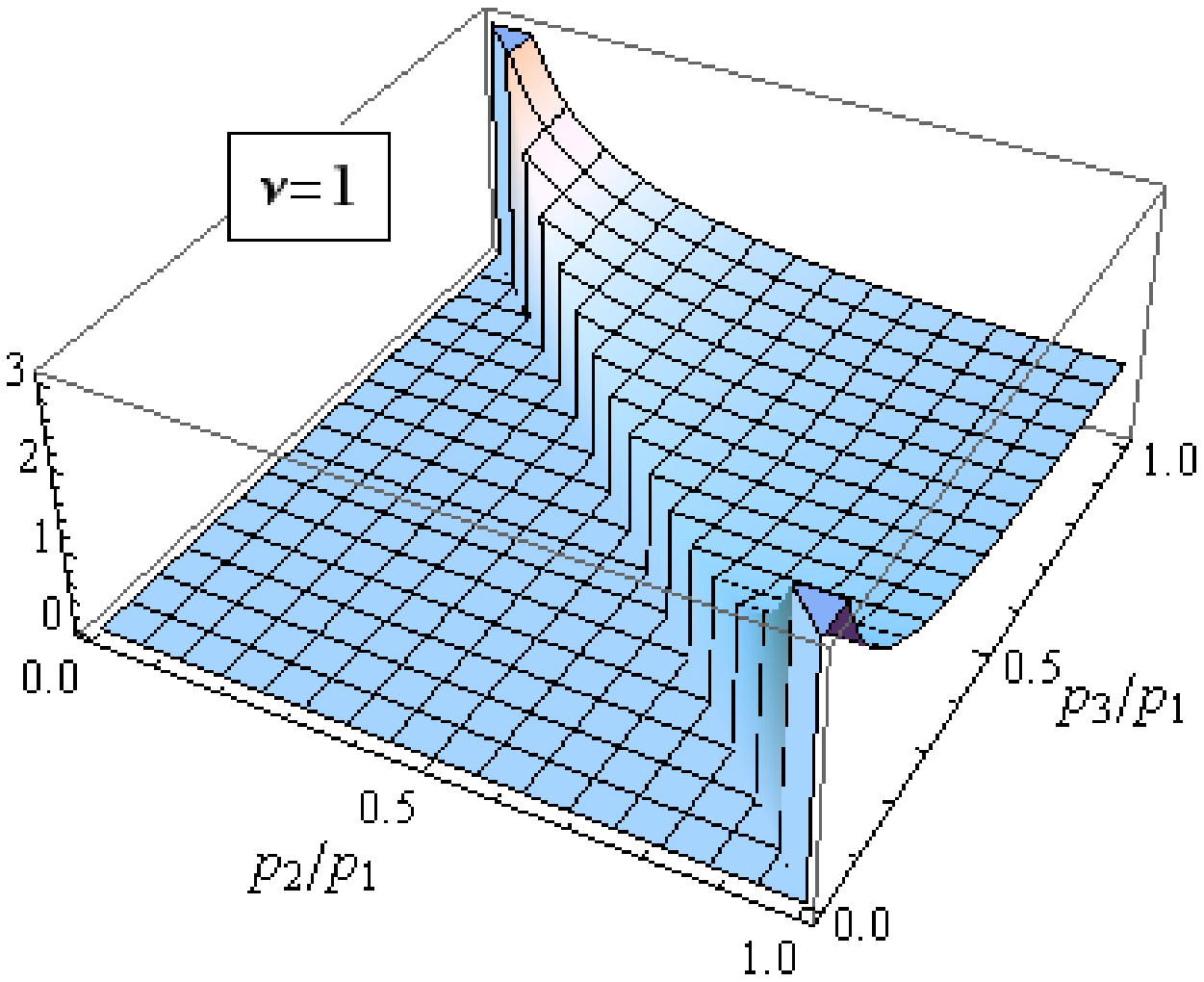, width=0.4\textwidth}
\end{center}
\caption{Shape ansatz (\ref{ansatz_int}) for the intermediate forms.}
\label{Fig:ansatz_int}
\end{figure}

To better understand the shapes analytically, we can work out the squeezed limit ($p_3\ll p_1=p_2$) of the three-point function,
\bea
\langle \delta\theta^3 \rangle \to
\frac{c_2^3 c_3}{HR^6}
\frac{1}{p_1^{\frac{7}{2}-\nu} p_2 p_3^{\frac{3}{2}+\nu}}
s(\nu) ~(2\pi)^3 \delta^3 (\sum_i \bp_i) ~,
\label{slimit_quasi}
\eea
where $s(\nu)$ is a $\nu$-dependent numerical number.

Recall that the squeezed limit of $S$ for the equilateral shape goes as $p_3/p_1$, while for the local shape $(p_3/p_1)^{-1}$.
Both the numerical results in Fig.~\ref{Fig:shape_int} and the analytical results in Eq.~(\ref{slimit_quasi}) show that here we have a one-parameter family of shapes, $\sim (p_3/p_1)^{1/2-\nu}$, lie between the two. We call them the {\em intermediate shapes}.

The physical origin of such shapes can be understood as follows, and should be a generic signature for the quasi-single field inflation models. As we have seen, the large equilateral non-Gaussianity arises because the interacting modes cross the horizon around the same time. The shape of bispectrum peaks at the equilateral limit where the modes all have comparable wavelengths. As we will see in Sec.~\ref{Sec:multifield}, the large local non-Gaussianity arises due to the classical non-linear evolution of superhorizon modes in the multifield space; so the interactions are causally disconnected and behave local in position space. This is non-local from the momentum space point of view. So
the shape of bispectrum peaks at the squeezed limit.
Now for quasi-single field inflation, the large non-Gaussianities come from the massive isocurvaton. Depending on the mass, these modes either decay right away after they exit the horizon (for $m>\sqrt{2}H$), or survive for a long time at the super-horizon scales (for $m<\sqrt{2}H$). In the former case, the generation and transfer of non-Gaussianities maximize for modes that are exiting the horizon around the same time, resulting in quasi-equilateral shapes; in the latter case, the generation and transfer of non-Gaussianities happen in a superhorizon fashion, resulting in quasi-local shapes.
In this regard, let us look more closely at the special limit $m/H \to 0$ ($\nu\to 3/2$).

In this massless limit, an infrared cutoff to the integrals are necessary. Otherwise the transfer will last for ever for the constant turn case. The cutoff corresponds to the ending of the turning. Let us discuss the following two cases. First, we still keep $V'''$ large. Our analyses still apply in this case. Interestingly, the shape of the bispectrum goes to that of the local form in this limit. As we will explain in Sec.~\ref{Sec:local}, this is a generic signature of a massless isocurvaton. The infrared e-fold cutoff will introduce some running in the $f^{\rm int}_{NL}$ because different modes experience different turning e-folds. Second, we would like to make the isocurvature directions flat so this becomes a two-field slow-roll inflation models. Such models were intensively studied and it is known that the isocurvature modes can be transferred to the curvature mode by turning. However, since $V'''\sim \CO(\epsilon^{3/2}) H^2/\mpl$ is required to maintain the small slow-roll parameters, the contribution we computed here generates too small non-Gaussianity. We expect contributions from other terms are small as well. So it is much more difficult to generate large non-Gaussianities in such models, essentially because imposing the slow-roll conditions in all directions are too restrictive.

To connect with data analyses, guided by the numerical results and analytical squeezed limit, we can use the following ansatz to describe the full family of shapes,
\bea
S_{\rm ansatz}^{\rm int} = \frac{3^{\frac{9}{2}-3\nu}}{10}
\frac{f_{NL}^{\rm int}(p_1^2+p_2^2+p_3^2)(p_1p_2p_3)^{\frac{1}{2}-\nu}}{(p_1+p_2+p_3)^{\frac{7}{2}-3\nu}} ~.
\label{ansatz_int}
\eea
These shapes are shown in Fig.~\ref{Fig:ansatz_int}. Comparing with Fig.~\ref{Fig:shape_int}, we can see that they match quite well except near $\nu=0$, around which it is better represented by another ansatz in Ref.~\cite{Chen:2009zp}. The size of this bispectrum is
\bea
f_{NL}^{\rm int} = \alpha(\nu) P_\zeta^{-1/2} \left( \frac{\dot\theta_0}{H} \right)^3 \left( -\frac{V'''}{H} \right) ~,
\eea
where $P_\zeta \approx 6.1 \times 10^{-9}$ and $\dot\theta_0$ is the turning angular velocity. The $\alpha(\nu)$ is a positive numerical number which, depending on $\nu$, can give an additional enhancement factor of order $N_f$ ($N_f$ is the turning e-folds).
Since $(\dot\theta_0/H)^2$ and $V'''/H$ are the expansion parameters in the perturbation theory, they have to be small to trust our calculation. Nonetheless this is not the model-building requirement.

The fluctuations of more massive ($m > \CO(H)$) fields may become important if they play a role later in the reheating \cite{Mulryne:2009ci,Gong:2009dh}. Such cases typically require some tunings for special conditions, so that the highly suppressed fluctuation amplitude can become important.

\section{Multifield inflation}
\label{Sec:multifield}
\setcounter{equation}{0}

As we have seen in Sec.~\ref{Sec:int}, if we take the isocurvaton mass to zero in quasi-single field inflation while keep the nonlinear self-couplings of the isocurvaton $V'''$ large, the shape of the large bispectrum in the squeezed limit approaches the local form. The local form is in fact the earliest and most well-studied example of non-Gaussianities \cite{Salopek:1990jq,Gangui:1993tt,Verde:1999ij,Komatsu:2001rj}, although it was first found to be small as we have seen in Sec.~\ref{Sec:Nogo}. As we will explain in this section, a large local form is a signature of massless isocurvatons that have large non-linear evolution in multifield space.
We have arrived this shape from the in-in formalism by taking the massless limit. But if we stay in this limit, there is an easier formalism, the $\delta N$ formalism \cite{Starobinsky:1986fxa,Sasaki:1995aw,Lyth:2005fi}, in which the underlying physics of the local shape becomes transparent.

\subsection{Local shape: massless isocurvatons}
\label{Sec:local}

We recall that, in single field inflation, if we use the uniform inflaton gauge where there are no fluctuations in the inflaton field, the scalar perturbation $\zeta$ enters in the scale factor as $a^2 e^{2\zeta}$. For superhorizon modes, $\zeta$ is frozen. If we look at the different comoving superhorizon patches, they are causally disconnected from each other. So they evolve independently and locally in space. In such a gauge, the only difference is a space-dependent scale factor. This is also called the separate universe picture. The primordial curvature perturbations manifest themselves as the different number of expansion e-fold, $\delta N$, at different positions.

We would like to generalize this picture to the multifield case in the following $\delta N$ formalism. We will resort to a simple version of $\delta N$ formalism stated below, which is of course a consequence of the in-in formalism, but formulated from a simple perspective which clearly illustrates the points in this section. Otherwise, as we will explain, in the most general sense the $\delta N$ formalism should be defined as the in-in formalism written in terms of specified gauges.

\begin{itemize}

\item
We consider a set of scalars $\phi_i$ during inflation. Inflaton is one of them but can be different linear combinations of $\phi_i$'s as a function of time, and the other orthogonal fields are called the isocurvatons. All the modes that we are eventually interested in should all have become {\em superhorizon} when the initial slice (specified below) is chosen. We look at different horizon-size patches and label them with the coarse-grained comoving coordinate $\bx$. In the in-in formalism, the superhorizon modes behave as the c-number time-dependent background for each comoving patch. So we evolve these patches independently and classically.

\item
We pick an initial spatially flat slice, on which there is no scalar fluctuations in the metric and all the fluctuations are in the scalar fields $\phi_{0i} + \delta \phi_i (\bx)$. We {\em assume} that we know the statistics of such fluctuations.

\item
We pick the final uniform density slices. Relative to the unperturbed and perturbed initial spatially flat slices, we have, respectively, the unperturbed and perturbed final uniform density slices. For single field inflation, these two final surfaces are the same. For multifield models, they are generally different.
Such final slices have the properties that the universe has the same energy densities and field configurations everywhere on them.
They can be chosen during either the inflation or the reheating.
After that, every separated universe will have the same evolution. The only difference is the scale factor. This is the analogy of the uniform inflaton gauge in single field inflation. We study the cases where such slices exist.

\item
We evolve the unperturbed $\phi_{0i}$ in the initial slice classically to the unperturbed final slice, and denote the number of e-folds as $N_0(\phi_{0i})$. This is the unperturbed e-fold number.
We evolve the perturbed $\phi_{0i}+ \delta\phi_i(\bx)$ in the initial slice classically to the perturbed final slice, and denote the number of e-folds as $N(\phi_{0i}+\delta\phi_i(\bx))$. The difference between them
\bea
\delta N = N(\phi_{0i}+\delta\phi_i(\bx)) - N_0(\phi_{0i}) ~,
\label{deltaN_def}
\eea
is the curvature perturbation $\zeta$. Here $N_0$ is a constant that can be shifted to make $\langle \delta N \rangle$ zero.

\item
We expand
\bea
\delta N = N_i \delta\phi_i + \half N_{ij} \delta\phi_i \delta\phi_j + \cdots ~,
\eea
where the subscripts on $N$ denote the partial derivatives with respect to $\phi_i$. For example, $N_{ij} = (\partial N/\partial \phi_i)(\partial N/\partial \phi_j)$. Repeated indices are summed over.
The correlation functions of $\zeta$ can then be computed as the classical averages of the products, such as,
\begin{align}
\langle \zeta(\bx_1) \zeta(\bx_2) \rangle &= N_i N_j \langle \delta\phi_i(\bx_1) \delta\phi_j(\bx_2) \rangle ~,
\\
\langle \zeta(\bx_1) \zeta(\bx_2) \zeta(\bx_3) \rangle &= \half N_{ij} N_k N_l \langle \delta\phi_i(\bx_1) \delta\phi_j(\bx_1) \delta\phi_k(\bx_2) \delta\phi_l(\bx_3) \rangle + {\rm 2 ~ perm.} ~.
\label{deltaN_3pt}
\end{align}

\item
We have assumed that the statistics of the $\delta\phi_i(\bx)$ are known on the initial slice. But this is not always easy to get. So we will consider the simple cases
where this statistics can be approximated as Gaussian. Otherwise, calculating such initial statistics requires using the full quantum mechanical in-in formalism.

Most generally, one identifies $\delta N$ with the scalar curvature $\zeta$ in the uniform inflaton gauge; and the relation between $\delta N$ and $\delta\phi(\bx)$ in the $\delta N$ formalism is the gauge transformation between the uniform inflaton gauge and the spatially flat gauge. Calculating the correlation functions for $\zeta$ becomes calculating those for $\delta\phi(\bx)$ using the in-in formalism. An example is the one we have seen in Sec.~\ref{Sec:int}.

\item
So far we have not used the condition that the isocurvatons are massless ($m\ll H$). If they are massive, after horizon exit the modes decay. So in the superhorizon classical regime, where the $\delta N$ formalism is supposed to be useful, we are back in the single field inflation. Sub and near horizon perturbations should be computed by the full in-in formalism. Therefore having massless isocurvatons opens up classical multifield space in which we can have sizable $\delta N$ defined in (\ref{deltaN_def}).

\end{itemize}

\begin{figure}[t]
\begin{center}
%\epsfxsize=10cm
%\epsfbox{Fep.eps}
\epsfig{file=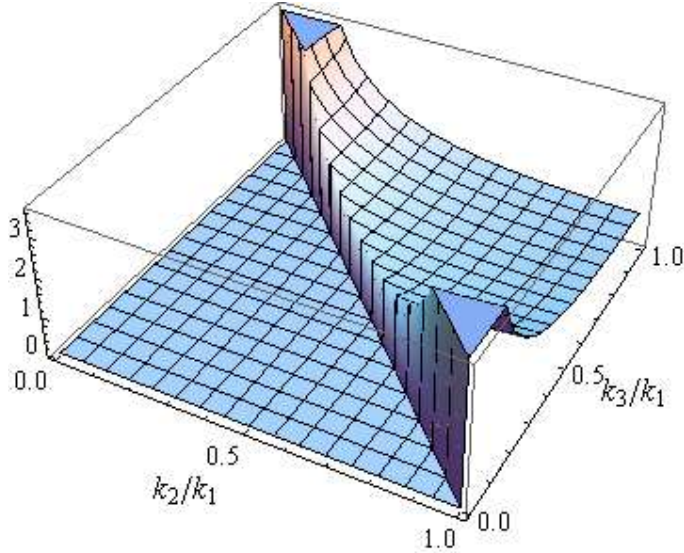, width=9cm}
\end{center}
\caption{Shape of the local form (\ref{ansatz_loc}).}
\label{Fig:shape_loc}
\end{figure}

Now let us consider the Gaussian fluctuations $\delta\phi_i$. From Sec.~\ref{Sec:inflation}, we know that for massless scalars,
\bea
\langle \delta\phi_i(\bk_1) \delta\phi_j(\bk_2) \rangle
= \frac{H_*^2}{2k_1^3} (2\pi)^3 \delta^3(\bk_1+\bk_2) \delta_{ij} ~,
\eea
where $H_*$ is the Hubble parameter when the corresponding mode exits the horizon.
If the scalars are not exactly massless, $H$ will have a running dependence on $k_1$ caused by the decay of the amplitude.
Using
\bea
\delta N(\bk) = N_i \delta\phi_i(\bk) + \half N_{ij} \int \frac{d^3\bk'}{(2\pi)^3} \delta\phi_i(\bk-\bk') \delta\phi_j(\bk') ~,
\eea
we get the power spectrum
\bea
\langle \zeta(\bk_1) \zeta(\bk_2) \rangle = (2\pi)^5 \frac{P_\zeta}{2k_1^3} \delta^3(\bk_1+\bk_2) ~,
\eea
where
\bea
P_\zeta = \left( \frac{H_*}{2\pi} \right)^2 N_i^2 ~,
\eea
and the bispectrum
\bea
\langle \zeta(\bk_1) \zeta(\bk_2) \zeta(\bk_3) \rangle
= N_{ij}N_i N_j \frac{H_*^4}{4}
\left( \frac{1}{k_1^3 k_2^3} + {\rm 2~perm.} \right)
(2\pi)^3 \delta^3(\sum \bk_i) ~.
\eea
According to the definition (\ref{3pt_def}), the shape function is
\bea
S^{\rm loc} = \frac{3}{10} f^{\rm loc}_{NL}
\left( \frac{k_1^2}{k_2k_3} + {\rm 2~perm.} \right) ~,
\label{ansatz_loc}
\eea
where
\bea
f_{NL}^{\rm loc} = \frac{5}{6} \frac{N_{ij} N_i N_j}{(N_l^2)^2} ~.
\label{fNL_deltaN}
\eea
As usual, we have ignored a mild running from the power spectrum $P_\zeta$.
The shape (\ref{ansatz_loc}) is called the local shape, which we plot in Fig.~\ref{Fig:shape_loc}. This form is already factorizable.

The physics of this shape can be understood from the derivation above. As explicitly demonstrated in (\ref{deltaN_def})-(\ref{deltaN_3pt}), this non-Gaussianity is generated locally in position space for superhorizon modes. After Fourier transform, it becomes non-local in momentum space. That is the reason that the shape peaks at the squeezed limit.

If the perturbation $\delta\phi(\bx,t)$ on the initial spatially flat slice cannot be approximated as Gaussian, the shapes of final bispectra can be more complicated.

\medskip

$\bullet$ {\bf An example: the curvaton model.}
We use the curvaton model \cite{Linde:1996gt,Lyth:2001nq,Moroi:2001ct,Lyth:2002my,Bartolo:2003jx,Huang:2008ze,Li:2008jn,Enqvist:2008gk,Huang:2008bg} as an example to illustrate the generation of large local non-Gaussianity. We also use it to demonstrate the $\delta N$ formalism.

In this model, we assume that during inflation there is another light field $\sigma$ with the potential
\bea
V(\sigma) = \half m_\sigma^2 \sigma^2 ~,
\label{curvaton_potential}
\eea
and $m_\sigma \ll H$. This field is called the curvaton field for reasons that will become clear shortly. The energy density of the curvaton field is negligible initially. During inflation, it fluctuates and obtains the primordial amplitude $\sigma_* = \sigma_0 + \delta\sigma(\bx)$, where $*$ denotes its value at the horizon-exit and after that the amplitude is approximately frozen.
These perturbations are Gaussian for the potential (\ref{curvaton_potential}) with the canonical kinetic term, but can be more complicated otherwise. Here we study the simple Gaussian case.
After inflation, these $\sigma$-modes remain frozen until the Hubble parameter drops below $m_\sigma$. Then the $\sigma$-field starts to oscillate around the bottom of the potential and behavior as matter. The Universe is still radiation-dominated. The fraction of the matter energy density quickly grows, because the matter dilutes as $a^{-3}$ while radiation $a^{-4}$. The $\sigma$-field decays to radiation when it reaches its lifetime.

Another assumption of the curvaton model is that the primordial fluctuations in the inflaton field is much smaller than what is needed to achieve $\zeta\sim \CO(10^{-5})$, although their total energy density may still be the dominant one. So the primordial curvature perturbation is contributed by the fluctuations in the $\sigma$ field, hence the name curvaton field.

At the initial spatially flat slice $t_0$, we denote the radiation and curvaton density as $\rho_{r0}$, $\rho_{m0}$, respectively, and the scale factor as $a_0$. Both components initially redshift as radiation. This lasts until the Hubble parameter reaches $m_\sigma$ at $t_1$. We denote the scale factor at $t_1$ as $a_1$. The Friedman equation at $t_1$ is
\bea
3 \mpl^2 m_\sigma^2 = \left(\frac{a_0}{a_1}\right)^4 (\rho_{r0} + \rho_{m0}) \approx \left(\frac{a_0}{a_1}\right)^4 \rho_{r0} ~.
\label{curvaton_evol_1}
\eea
After this the curvaton starts to oscillate and behave as matter.
Denote the decay rate of the curvaton as $\Gamma$. We use the sudden decay approximation and assume that they decay instantaneously at the epoch $H=\Gamma$, because a process that has the time scale $T$ falls into the Hubble expansion epoch with $H=1/T$. We denote this instant as $t_2$. The Friedman equation at $t_2$ is
\bea
3 \mpl^2 \Gamma^2 =
\left( \frac{a_0}{a_2} \right)^4 \rho_{r0} + \left( \frac{a_0}{a_1} \right)^4 \left( \frac{a_1}{a_2} \right)^3 \rho_{m0} ~.
\label{curvaton_evol_2}
\eea
Because at $t_2$ the universe has the same Hubble parameter (hence the same energy density) everywhere, this is the final uniform density slice. After that, both components become radiation and the evolution everywhere is the same. So we want to work out the expansion efolds $N$ from $t_0$ to $t_2$ as a function of the initial field value $\sigma$.
From (\ref{curvaton_evol_2}), we get
\bea
e^{-4N} + e^{-3N} \alpha = {\rm const.} ~,
\label{curvaton_evol_N}
\eea
where $\alpha \equiv (a_0/a_1)(\rho_{m0}/\rho_{r0})$. From (\ref{curvaton_evol_1}) we can solve for $a_0/a_1$ which is independent of $\sigma$ at the leading order, and from (\ref{curvaton_potential}) we know that $\rho_{m0}$ is proportional to $\sigma^2$. Therefore $\alpha$ is proportional to $\sigma^2$. Also note that $r_\rho \equiv (a_2/a_0)\alpha = e^N \alpha = \rho_m/\rho_r|_{t_2}$ is the ratio of the energy density between the curvaton and the rest of the radiation at $t_2$. Using these simple facts, we can differentiate (\ref{curvaton_evol_N}) with respect to $\sigma$ once and twice, and get
\bea
f^{\rm loc}_{NL} = \frac{5}{6} \frac{N_{\sigma\sigma}}{N_\sigma^2} = \frac{5}{3 r_\rho} - \frac{5(4+9 r_\rho)}{12(4+ 3r_\rho)} ~.
\eea
In terms of the definition
$r \equiv 3\rho_{m}/(4 \rho_r + 3\rho_m) |_{t_2} = 3r_\rho/(4+ 3r_\rho)$ often used in the literature,
\bea
f^{\rm loc}_{NL} = \frac{5}{4r} - \frac{5}{3} - \frac{5r}{6} ~.
\eea
So Large local non-Gaussianity arises if $r \ll 1$.
Note that although (\ref{curvaton_evol_N}) only depends on $\sigma$, this is a multifield model because the curvaton takes effects during the reheating.
In some simple models in which the curvaton leads to non-adiabatic perturbations between dark matter and photons, $r$ is tightly constrained by observations \cite{Komatsu:2008hk}.

\medskip

The large local form has been studied most extensively in the past. Variety of possibilities exist. They all share the common feature that non-Gaussianities are generated by some massless isocurvaton fields which acquire the superhorizon evolution during the inflation. For example,
in multifield slow-roll inflation a turning trajectory \cite{Gordon:2000hv} can transfer non-Gaussianities from other directions to the inflaton direction \cite{Salopek:1990jq,Bartolo:2001cw,Bernardeau:2002jy,Bernardeau:2002jf,Rigopoulos:2005xx,Rigopoulos:2005ae,Seery:2005gb,Vernizzi:2006ve}. But it is found to be very difficult to make non-Gaussianities large essentially because the very restrictive slow-roll conditions in all directions.
In modulated reheating \cite{Dvali:2003em,Kofman:2003nx} or preheating \cite{Enqvist:2004ey,Jokinen:2005by,Barnaby:2006km,Chambers:2007se,Bond:2009xx} scenarios, the role of isocurvatons are played by the massless fields which control the couplings during the reheating or preheating. Thus they create a large local non-Gaussianity in a similar fashion as the curvaton model.

Local form is also found in different contexts, such as models with special types of massive gauge fields that acquire super-horizon evolution \cite{Bartolo:2009pa,ValenzuelaToledo:2009af,Dimastrogiovanni:2010sm}, some non-local theories of inflation \cite{Barnaby:2007yb,Barnaby:2008fk}, and certain cyclic universe scenario as alternatives to inflation \cite{Buchbinder:2007at,Koyama:2007if,Khoury:2008wj,Lehners:2009ja,Lehners:2009qu}.

The current CMB constraint on the local bispectrum is $-10 < f_{NL}^{\rm local} < 74$ \cite{Komatsu:2010fb}.
Current constraint from large scale structure gives $-29 < f_{NL}^{\rm local} < 70$ \cite{Slosar:2008hx}.
Variety of methods have been invented to measure the local and other different forms of non-Gaussianities \cite{Yadav:2007yy,Smith:2009jr,Senatore:2009gt,Dalal:2007cu,Desjacques:2009jb,Verde:2009hy,Roncarelli:2009pp,Rudjord:2009mh,Rudjord:2009au,Smidt:2009ir,Cabella:2009rp,Bucher:2009nm,LoVerde:2007ri,Sefusatti:2009xu}.

\section{Summary and discussions}
\label{Sec:sum_and_dis}
\setcounter{equation}{0}

\subsection{Summary}
\label{Sec:summary}

In this subsection, we summarize the main results of Sec.~\ref{Sec:Single} - \ref{Sec:multifield}.
Non-Gaussianities, conceptually being the expectation values of perturbations in a time-dependent background, are defined by the first-principle in-in formalism. Physically, having large primordial non-Gaussianities means that there are large non-linear interactions of some sort determined by certain dynamics during inflation.
Measuring them tells us the nature of the dynamics.

\begin{itemize}

\item {\em Equilateral shape and higher derivative kinetic terms}.

In single field inflation, the long wavelength modes that already exited the horizon are frozen. They cannot have large interactions with short wavelength modes that are still within the horizon.
When modes are well within the horizon, they oscillate and the contributions to non-Gaussianities average out.
Therefore the only chance to have large interaction is when all modes have similar wavelengths and exit the horizon at about the same time.
Theories with higher derivative kinetic terms provide such interaction terms.
This is why the resulting bispectrum shape peaks at the equilateral limit in the momentum space.
It drops to zero at the squeezed limit $k_3\ll k_1=k_2$ as $k_3/k_1$.
It happens that, when these higher derivative terms become important enough so that the inflationary mechanism is no longer slow-roll, these large non-Gaussianities become observable.
The forms of the bispectra are given in Eq.~(\ref{S_lam}) and (\ref{S_c}), and the shapes are plotted in Fig.~\ref{Fig:shape_eq1} and \ref{Fig:shape_eq2}. The factorizable ansatz that is used to represent them in data analyses is given in Eq.~(\ref{ansatz_eq}) and plotted in Fig.~\ref{Fig:shape_ansatz_eq}.

\item {\em Sinusoidal running and sharp feature}.

A sharp feature, in a potential or internal field space, introduces a sharp change in the slow-roll parameters, or the generalized slow-variation parameters. This can boost the magnitudes of time-derivatives of some parameters by several orders of magnitude while still keep the power spectrum viable. These time-derivatives act as couplings in the interaction terms. So they enhance the non-Gaussianities among the modes which are near the horizon-exit. How deep they affect the modes inside the horizon depends on how sharp the changes are.

The changes in these parameters can be roughly approximated as delta-functions in time.
Correlation functions involve integrations of products of the slow-variation parameters and the mode functions. The latter contain oscillatory behavior $\sim e^{-i K\tau}$, where the comoving momentum $K$ is $k_1+k_2+k_3$ for bispectra and $\tau$ is the conformal time. The delta-function specifies a scale $\tau_*$. This is why after integration the bispectrum contains a sinusoidal factor $\sim \sin (K/k_*)$, where $k_*=-1/\tau_*$ is the momentum of the mode that is near the horizon-exit at the time of the feature.
So the most important property of this type of non-Gaussianity is this characteristic running.
A numerical result of the running behavior is plotted in Fig.~\ref{Fig:sharp_running}. An ansatz is given in Eq.~(\ref{ansatz_sharp}) and (\ref{sin_envelop}).

\item {\em Resonant running and periodic features}.

The periodic features do not have to be sharp. They introduce a small background oscillatory component in the slow-variation parameters. On the other hand, the mode functions are also oscillatory before they exit the horizon. Their frequencies are high when they lie deep inside the horizon and become lower as their wavelengths get stretched by the inflation. They are frozen after the wavelengths become comparable with the horizon size $H^{-1}$. This means that their frequencies continuously scan through the range from $\mpl$ (or some other large fundamental scale) to $H$. Therefore as long as the background oscillatory frequency $\omega$ satisfies $H<\omega<\mpl$, at some point during the evolution the small oscillatory component in the slow-variation parameters will resonant with the oscillatory behavior of the mode functions, and cause a large constructive contribution to the integration.

The periodicity of the features leads to a periodic-scale-invariance in density perturbations. Namely, they are scale invariant if we rescale all momenta by a discrete e-fold $2\pi nH/\omega$, where $n$ is an integer. This is why the most important feature of this non-Gaussianity is a running behavior $\sim \sin ( C \ln K + {\rm phase} )$, where $C=\omega/H$. This leads to the ansatz (\ref{ansatz_res}). The full expression is given in (\ref{S_res_full}) and plotted in Fig.~\ref{Fig:res_running_shape}. A numerical result is plotted in Fig.~\ref{Fig:res_running}.

\item {\em Folded shape and non-Bunch-Davies vacuum}.

The usual mode function of the Bunch-Davies vacuum has the positive energy mode $\sim e^{-ik\tau}$. Now we consider a non-Bunch-Davies vacuum by adding a small component of negative energy mode $\sim e^{ik\tau}$. The three-point function involves an integration of the product of three mode functions with momentum $k_1$, $k_2$ and $k_3$. So the leading correction to the Bunch-Davies results is
to replace one of the $k_i$'s with $-k_i$. The usual $K=k_1+k_2+k_3$ in $e^{-iK\tau}$ becomes $\tilde K = k_1+k_2-k_3$ and its cyclic. This effect is most important if factors of $\tilde K$ appear in the denominators after the $\tau$-integration. Hence the most important feature of this type of modification is to enhance the non-Gaussianity in the folded triangle limit. An example of these bispectra is given in Eq.~(\ref{nonBD_lambda}) and plotted in Fig.~\ref{Fig:shape_nonBD1}. Ansatz that partially capture this feature are given in Eq.~(\ref{ansatz_fold}) and (\ref{ansatz_fold_2}), and plotted in Fig.~\ref{Fig:shape_ansatz_fold}.

\item {\em Intermediate shapes and massive isocurvatons}.

All mechanisms discussed so for single field inflation apply to multifield inflation.
We now consider new effects caused by introducing more fields to inflation models. These extra fields are called isocurvatons.

Since light fields typically acquire a mass of order $H$, the Hubble parameter, we first consider the quasi-single field inflation models where there is one massless inflaton while the isocurvatons have mass of order $H$ instead of massless.

Unlike multifield slow-roll inflation, where each flat direction only has small non-linear terms in order to satisfy the slow-roll conditions, massive directions are not inflationary direction and are free to have large non-linear self-interactions. These non-linear interactions can be transferred to the curvature mode through couplings and source the large non-Gaussianity.

The massive isocurvaton eventually decays after horizon exit simply because they are diluted by the expansion. How fast it decays depends on its mass. If the mass is heavier, $m>\sqrt{2}H$, it decays faster. So the interactions can only happen when all modes are all closer to the horizon exit. This is closer to the case of the equilateral shape that we encountered above, and results in bispectra with quasi-equilateral shapes. If the mass is lighter, $m<\sqrt{2}H$, it decays slower. More non-Gaussianity is generated in the super-horizon scales. This is closer to the case of the local shape that we will come to below, and results in bispectra with quasi-local shapes.
Overall, at the squeezed limit $k_3\ll k_1=k_2$, the bispectrum shapes behave as $(k_3/k_1)^{1/2-\nu}$, where $\nu$ goes from $0$ to $3/2$ (corresponding to $m$ from $3H/2$ to $0$) in the example we studied.
In particular, if we take the massless limit while keeping the cubic self-interactions of isocurvaton large, we get a large bispectrum that has the same squeezed limit shape as the local one. Therefore, we have a one-parameter family of shapes that lie between the local and equilateral shape.

The numerical results of these shapes are presented in Fig.~\ref{Fig:shape_int}. A simple ansatz is given in Eq.~(\ref{ansatz_int}) that represents this family of shapes quite well, and is plotted in Fig.~\ref{Fig:ansatz_int}.

\item {\em Local shape and massless isocurvatons}.

The fluctuation amplitudes of massless scalars do not decay after the horizon exit, and therefore this opens up a multifield space for the superhorizon evolution. For superhorizon modes, we can use the separate universe picture and study the classical behavior of different patches of universe. These patches are separated by horizons and evolve independently of each other. So the evolution is local in space.

Non-Gaussianities are generated when this multifield evolution is nonlinear, and any nonlinearity arising in the separate universe picture should also be local in space. A locality in position space translates to a non-locality in momentum space. This is why the resulted local shape bispectrum peaks at the squeezed limit.
The behavior is $(k_3/k_1)^{-1}$ for $k_3\ll k_1=k_2$.
This bispectrum is given in Eq.~(\ref{ansatz_loc}) and the shape is plotted in Fig.~\ref{Fig:shape_loc}.

\end{itemize}

In all cases, the power spectra are either approximately scale-invariant so indistinguishable from the simplest slow-roll models, or modified with features that can be made small enough to satisfy the current observational constraints.

Large bispectra generically implies large trispectra, i.e.~the four-point correlation functions. But trispectra contain more information and can be large even if bispectra are small. Experimentally, trispectra are more difficult to detect, but contain much more shape configurations. Each category above should have interesting extensions to trispectra. See Ref.~\cite{Chen:2009bc,Arroja:2009pd,Huang:2006eha,Arroja:2008ga,Gao:2009gd,Gao:2009at,Mizuno:2009mv} for the equilateral case and \cite{Seery:2006vu,Seery:2006js,Seery:2008ax,Adshead:2009cb,Bartolo:2009kg,ValenzuelaToledo:2009nq} for the local case.

It is certain that this list will grow in future works, providing more refined and diverse connections between theories and experiments.

\subsection{A consistency condition}
\label{Sec:consistency}

As we have seen, in single field inflation, the mode that has exited the horizon is frozen. This is characterized by a constant $\zeta$ over a horizon size patch. The physical meaning of the constant $\zeta$ is a small rescaling of the scale factor. This is the only effect that the superhorizon mode has on modes with much shorter wavelength. This fact is used by Maldacena to derive a consistency condition \cite{Maldacena:2002vr} for the three-point correlation functions in the squeezed limit for single field inflation.

\medskip

{\bf $\bullet$ Consistency condition.}
In the squeezed limit $k_3\ll k_1=k_2$, $k_3$ is the superhorizon mode that exited the horizon and acts as a zero-mode modulation to the two remaining modes. The correlation $\langle \zeta_{k_1} \zeta_{k_2} \zeta_{k_3} \rangle$ is an average of the following quantity
\bea
\langle \zeta_{k_1} \zeta_{k_2} \rangle_{\zeta_{k_3}} \zeta_{k_3}
\label{2pt_modulation}
\eea
over different $\zeta_{k_3}$. We will shift the average $\langle \zeta_{k_3} \rangle$ to zero by definition. Here the two-point average $\langle \zeta_{k_1} \zeta_{k_2} \rangle_{\zeta_{k_3}}$ is evaluated with different local scalings determined by the shift $\zeta_{k_3}$. If the two-point function is exactly scale-invariant, $\langle \zeta_{k_1} \zeta_{k_2} \rangle_{\zeta_{k_3}}$ is just a constant. So the 3pt vanishes because $\langle \zeta_{k_3} \rangle=0$. The non-zero contribution comes from the breaking of the scale-invariance. To see this, we expand the two-point average in terms of a long wavelength mode $\zeta_{k_4}$ near the scale $\langle \zeta_{k_4} \rangle =0$,
\bea
\langle \zeta_{k_1} \zeta_{k_2} \rangle_{\zeta_{k_4}} =
\langle \zeta_{k_1} \zeta_{k_2} \rangle_{0}
+ \frac{d \langle \zeta_{k_1} \zeta_{k_2} \rangle}{d\zeta_{k_4}} \Big|_{0} \zeta_{k_4}
+ \half \frac{d^2\langle \zeta_{k_1} \zeta_{k_2} \rangle}{d\zeta_{k_4}^2} \Big|_0 \zeta_{k_4}^2
+ \cdots ~.
\label{2pt_expand}
\eea
Multiply this with $\zeta_{k_3}$ and average over $\zeta_{k_3}$. The first term contributes zero since this is the scale-invariant component. The second term gives
\bea
\frac{d \langle \zeta_{k_1} \zeta_{k_2} \rangle}{d\zeta_{k_4}} \Big|_{0} \langle \zeta_{k_3} \zeta_{k_4} \rangle ~.
\eea
To get non-zero average, $\bk_3+\bk_4=0$ is needed.
Using the relation $d\zeta_{k_4} = - d\ln k$, we get
\bea
\langle \zeta_{k_1} \zeta_{k_2} \zeta_{k_3} \rangle
\to -
\frac{ d\langle \zeta_{k_1} \zeta_{k_2} \rangle }{d \ln k} \Big|_{0}
\langle \zeta_{k_3} \zeta_{k_3} \rangle
~.
\label{leading_order_consistency}
\eea
The higher order terms in (\ref{2pt_expand}) give
\bea
\half \frac{d^2\langle \zeta_{k_1} \zeta_{k_2} \rangle}{(d\ln k)^2} \Big|_{0}
\langle \zeta_{k_3} \zeta_{k_4}^2 \rangle
-\frac{1}{6} \frac{d^3 \langle \zeta_{k_1}\zeta_{k_2} \rangle}{(d \ln k)^3} \Big|_{0} \langle \zeta_{k_3} \zeta_{k_4}^3 \rangle
+ \cdots ~,
\label{higher_order_consistency}
\eea
where $\bk_3 + 2\bk_4=0$, $\bk_3 + 3\bk_4=0$, and so on have to be satisfied respectively for each term to get non-zero average.
If we only consider the tree-level three-point function, these higher order terms
can be truncated since they involve more factors of $P_\zeta$ and should be related to the loop diagram contributions to $\langle \zeta_{k_1} \zeta_{k_2} \zeta_{k_3} \rangle$. The tree diagram is $\CO(P_\zeta^2)$.

To connect the averages we used here with the correlation functions that we defined in previous sections, we need to restore the phase factors.
Here the two-point average $\langle \zeta_{k_1} \zeta_{k_2} \rangle_{\rm here}$ is performed with the special point $k_1=k_2$ in the phase space. To connect this with the previous definition of $\langle \zeta_{k_1} \zeta_{k_2} \rangle$, we need to include the phase space in the neighborhood. Namely, $\langle \zeta_{k_1} \zeta_{k_2} \rangle = \langle \zeta_{k_1} \zeta_{k_2} \rangle_{\rm here} (2\pi)^3 \delta^3(\bk_1+\bk_2)$, so from the definition (\ref{2pt_def}) we have $\langle \zeta_{k_1} \zeta_{k_2} \rangle_{\rm here} = (2\pi)^2 P_\zeta(k_1)/2k_1^3$. Similarly, $\langle \zeta^3 \rangle = \langle \zeta^3 \rangle_{\rm here} (2\pi)^3 \delta^3 (\sum \bk_i)$.
With the usual definition of the spectrum index $n_s-1 \equiv d\ln P_\zeta/d\ln k$, from (\ref{leading_order_consistency}) we get the following consistency condition \cite{Maldacena:2002vr}
\bea
\langle \zeta_{k_1} \zeta_{k_2} \zeta_{k_3} \rangle
\to
- (n_s-1) \frac{1}{4k_1^3 k_3^3} P_\zeta(k_1) P_\zeta(k_3)
(2\pi)^7 \delta^3 (\sum_i \bk_i) ~.
\label{consistency}
\eea

Although originally derived for slow-roll inflation, the only assumption is the single field. So this applies to any single field inflation models and has important physical implications that we discuss shortly \cite{Creminelli:2004yq}. Note that the derivation of this relation (\ref{consistency}) does not rely on the smallness of the slow-variation parameters either.
For the general single field inflation models that we studied in Sec.~\ref{Sec:Eq}, at tree level this has been checked with explicit results to three different orders \cite{Cheung:2007sv,Chen:2006nt} including the slow-roll limit \cite{Maldacena:2002vr}. For resonance models, this is checked to the leading order \cite{Flauger:2010ja}.

There are three types of interesting corrections to the condition (\ref{consistency}).

Firstly, as mentioned, the right hand side of (\ref{consistency}) should receive corrections from loop contributions. These loop contributions are associated with higher derivatives of the two-point function. The terms (\ref{higher_order_consistency}), together with (\ref{leading_order_consistency}), provide the corresponding consistency conditions at different orders of $P_\zeta$. Note that for such orders, the correlation functions such as $\langle \zeta_{k_1} \zeta_{k_2} \rangle$ on the right hand side should also include loop corrections.

Secondly, when we assume that the only effect of the frozen superhorizon mode on the much shorter scale is a constant background rescaling, we assume that there is no interaction when these modes are all within the horizon.\footnote{I would like to thank Yi Wang for helpful discussions on this point.} However, large subhorizon interaction is possible in some cases, such as in Sec.~\ref{Sec:res} and \ref{Sec:folded}. Such interactions disappear below a new length scale at subhorizon, only then the above assumption becomes valid. For example in the resonance model, for $H/\omega < k_3/k_1 \ll 1$, the modes $\bk_1$, $\bk_2$ and $\bk_3$ are guaranteed to resonant with the oscillatory background at some point when all of them are within the horizon. Only if $k_3/k_1 \ll H/\omega$, such resonance will not happen. This is why the consistency condition is satisfied only in the {\em very} squeezed limit ($ k_3 \ll k_1/C$ with $C=\omega/H \gg 1$).
For the squeezed region $k_1/C < k_3 \ll k_1$, the left hand side of the condition is larger than the right hand side by a factor of $Ck_3/2k_1$.
For the non-Bunch-Davies vacuum case, a similar scale is determined by the UV cutoff $\tau_c$, from which the non-Bunch-Davies vacuum starts to take effect.

Thirdly, even after the long wavelength mode exits the horizon, as long as $k_3/k_1$ is not infinitely small, there is still dependence of the two-point function on the derivative of this long wavelength mode, in addition to the overall constant shift. This introduces a different type of finite $k_3/k_1$ corrections. They start from the second order in $k_3/k_1$, because the first order corresponds to the first spatial derivative of the long wavelength mode and should vanish due to isotropy \cite{Creminelli:2004yq}. These corrections will be amplified by the associated amplitude $f_{NL}^{\rm non-loc}$, and give an additive correction $f_{NL}^{\rm non-loc}(k_3/k_1)^2$ to the $n_s-1$ on the right hand side of the condition. For a large $f_{NL}^{\rm non-loc}$, therefore, the condition needs to be satisfied in a {\em very} squeezed limit. The equilateral bispectra (\ref{S_lam}) and (\ref{S_c}) are this type of examples.

The consistency condition (\ref{consistency}) can be straightforwardly generalized to higher order correlation functions \cite{Seery:2006vu,Huang:2006eha}.
We emphasize that this condition only applies to single field inflation models. For inflation models involving more than one field, as we have seen, non-Gaussianities can be transferred from the isocurvature directions which do not respect this relation.

\medskip

{\bf $\bullet$ Physical implication.}
Besides providing consistency checks for analytical computations, the condition also has interesting physical implications. In the following, we discuss the scale invariant cases \cite{Creminelli:2004yq}, as well as the feature cases and loop corrections, ending with some cautionary remarks.

This consistency relation implies that the tree-level bispectrum in the squeezed limit is determined by the power spectrum and spectral index. We distinguish the following two cases. For the scale-invariant case, $n_s-1$ is of order $\CO(\epsilon)$ and the right hand side of (\ref{consistency}) takes the local form. Indeed, as we have seen, for the single field inflation models where the non-Gaussianities are large, they take the equilateral forms which vanish in the infinitely squeezed limit. For the non-scale-invariant case, especially the highly oscillatory case such as the resonance model, the power spectrum can be highly oscillatory and $n_s-1$ becomes large. This can still be consistent with observations since the large $n_s-1$ is also highly oscillatory and therefore may escape a detection so far. But such a running non-Gaussianity is orthogonal to the scale-invariant forms.

For the loop diagrams,
in the scale-invariant case, these terms are suppressed by higher orders of slow-variation parameters from, e.g.~$d^2\langle \zeta_{k_1}\zeta_{k_2} \rangle/(d\ln k)^2$, and higher orders of $\zeta$ from, e.g.~$\langle \zeta_{k_3}^3 \rangle$;
in the non-scale-invariant example, the extra terms are still highly oscillatory.

In summary, a detection of an approximately scale-invariant local non-Gaussianity in the infinitely squeezed triangle limit with $f^{\rm loc}_{NL}>\CO(\epsilon)$ can rule out all single field inflation models.

In experiments, however, the triangle cannot be perfectly squeezed. So it is an important question how squeezed it should be to achieve the above goal. For example, in the third type of corrections we discussed previously in this subsection, we need $f^{\rm non-loc}_{NL} (k_3/k_1)^2$ to be smaller than $n_s-1$ for the consistency condition to hold, so that the contaminations from whatever non-local $f_{NL}$ is small. Assuming the primordial local form is practically detectable only if $f^{\rm loc}_{NL} > \CO(1)$, we at least need $f^{\rm non-loc}_{NL} (k_3/k_1)^2 < \CO(1)$. For the class of the general single field models we studied in Sec.~\ref{Sec:Single}, if the other forms of non-Gaussianities, such as the equilateral one, can be constrained below $f_{NL}^{\rm non-loc} \sim \CO(10)$, a squeezed configuration with $k_3/k_1<0.1$ will be enough for our purpose. However, a completely model-independent statement is much trickier, because there may be bispectra with very large amplitude but orthogonal to any known bispectra that have been constrained experimentally.
Besides that, in the second type of corrections, large finite-$k_3/k_1$ corrections can also arise due to subhorizon interactions.
Therefore, as a cautionary remark, if we would like to rule out all single field inflation models in a rigorous model-independent fashion with a detection of scale-invariant local non-Gaussianity, we have to keep in mind the caveat that there may be single field models which only respect the consistency condition in a very squeezed region beyond the experimental reach.

\subsection{Superpositions}
\label{Sec:superposition}

Different shapes and runnings of non-Gaussianities can be superimposed in inflation models. For example,

\begin{itemize}

\item {\em Mixing shapes}.

It is possible that different non-Gaussianity generation mechanisms are from different components in a model, or at different stages during inflation. So two or more different shapes can get mixed, and the final shape can be rather different. For example, in Fig.~\ref{Fig:mixed_eq_loc}, we plot a mixed shape between the local and equilateral shape. Notice that this is different from the intermediate shapes, since obviously the squeezed limit is always dominated by the local form. Examples of such models are discussed in Ref.~\cite{RenauxPetel:2009sj}.

\begin{figure}
\begin{center}
%\epsfxsize=10cm
%\epsfbox{Fep.eps}
\epsfig{file=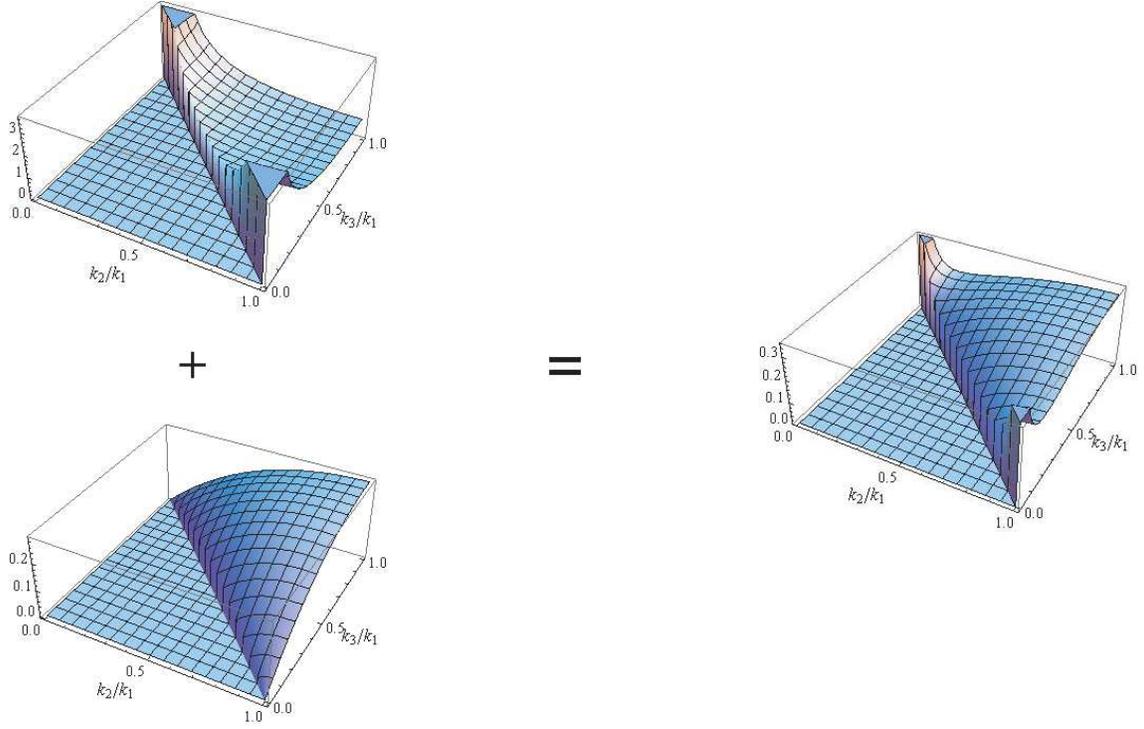, width=15cm}
\end{center}
\medskip
\caption{A mixing of the equilateral (Fig.~\ref{Fig:shape_eq2}) and local shape (Fig.~\ref{Fig:shape_loc}).}
\label{Fig:mixed_eq_loc}
\end{figure}

\item {\em Mixing shape and running}.

The shapes can also be mixed with runnings. Same as the power spectrum, the non-Gaussianities generically have some mild scale dependence. But a more dramatic case is the superposition with a strong running, such as the sinusoidal or resonant running. For example, an inflaton passing through features frequently and turning constantly at the same time on a potential landscape can generate a bispectrum which is a superposition of the resonant running and intermediate shape, as we illustrate in Fig.~\ref{Fig:mixed_int_res}. Clearly, these two signals are orthogonal to each other very well, and have to be picked up separately through different methods in data analyses.

\begin{figure}
\begin{center}
%\epsfxsize=10cm
%\epsfbox{Fep.eps}
\epsfig{file=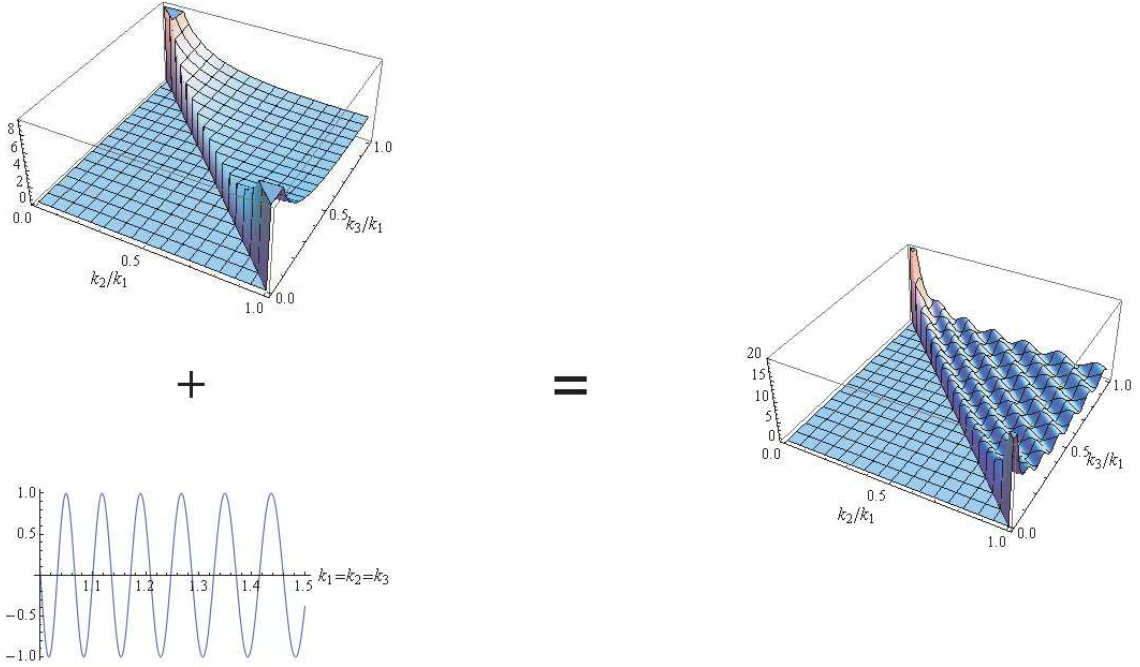, width=15cm}
\end{center}
\medskip
\caption{A mixing of an intermediate shape [$\nu=7/6$ in (\ref{ansatz_int})] and a resonant running (\ref{ansatz_res}).}
\label{Fig:mixed_int_res}
\end{figure}

\item {\em Orthogonalization}.

If a non-Gaussianity is the linear superposition of several base components, one can generally perform a change of bases to make the new bases orthogonalized.
For example, as we have seen in Sec.~\ref{Sec:Eq}, the leading large bispectrum has two components, $S_\lambda$ and $S_c$. The two shapes are very similar, and represented by the equilateral ansatz in data analyses. However since they do have small difference, one can subtract their similarities and get a new orthogonalized base component \cite{Senatore:2009gt}. The orthogonalization is defined by the shape correlator such as (\ref{correlator}). Using this definition, the new bases can be chosen as
\bea
S_1 \approx S_\lambda + 0.22 S_c
\label{newbase_S1}
\eea
and $S_2 = S_c$.
[Note that the $S_\lambda$ and $S_c$ used here do not include the prefactors $(1/c_s^2 - 1- 2\lambda/\Sigma)$ and $(1/c_s^2-1)$ in (\ref{S_lam}) and (\ref{S_c}).]
Their shapes are shown in Fig.~\ref{Fig:ortho_eq}. Notice that $S_1$ is half positive and half negative.
Because $S_1$ is not of the simplest factorizable type,
the following simple ansatz has been proposed to represent $S_1$ in data analyses \cite{Senatore:2009gt},
\bea
S_{\rm ansatz}^{\rm orth} = -18 \left(\frac{k_1^2}{k_2k_3} + {\rm 2~perm.}\right)
+18 \left( \frac{k_1}{k_2} + {\rm 5~perm.} \right) - 48 ~.
\label{ansatz_orth}
\eea
We plot the shape of this ansatz in Fig.~\ref{Fig:shape_ansatz_orth}.
The current CMB constraint on this orthogonal ansatz is
$-410 < f_{NL}^{\rm orth} < 6$ \cite{Komatsu:2010fb}.

\begin{figure}
\begin{center}
%\epsfxsize=10cm
%\epsfbox{Fep.eps}
\epsfig{file=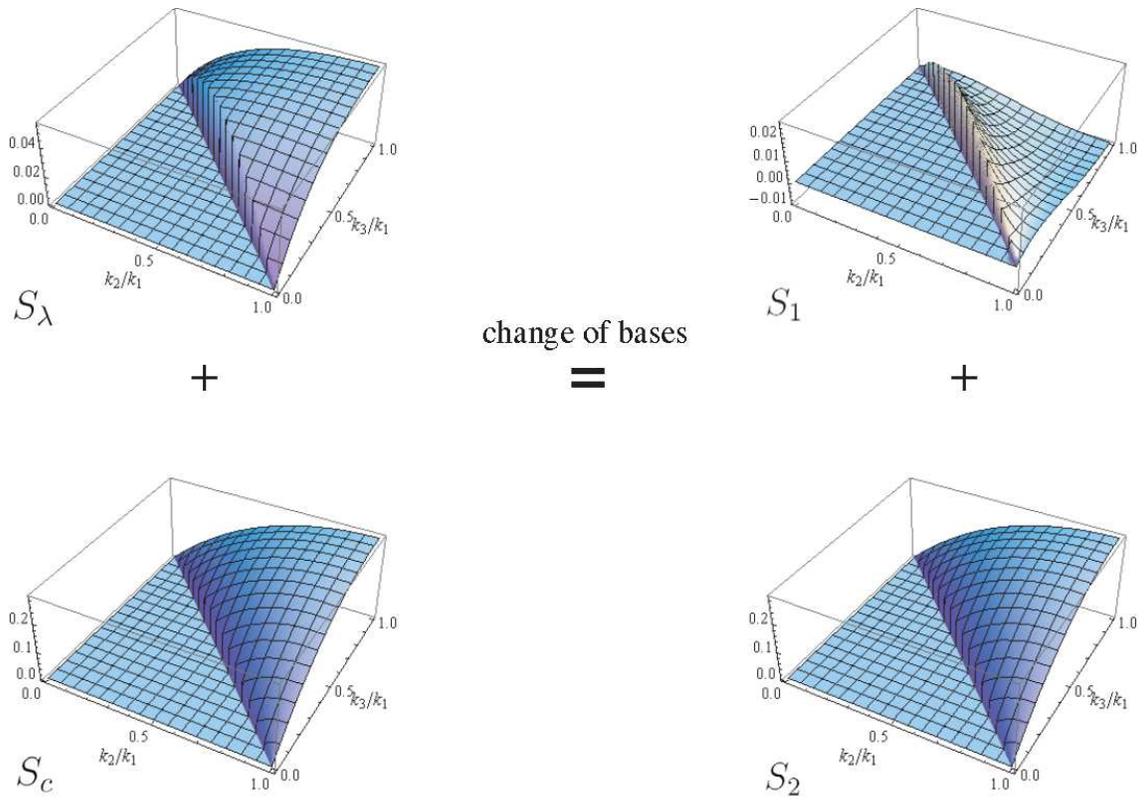, width=15cm}
\end{center}
\medskip
\caption{Orthogonalization of two shapes in Sec.~\ref{Sec:Eq} (Fig.~\ref{Fig:shape_eq1} and Fig.~\ref{Fig:shape_eq2}) through a change of bases, $c_\lambda S_\lambda + c_c S_c = c_1 S_1 + c_2 S_2$.}
\label{Fig:ortho_eq}
\end{figure}

\begin{figure}
\begin{center}
%\epsfxsize=10cm
%\epsfbox{Fep.eps}
\epsfig{file=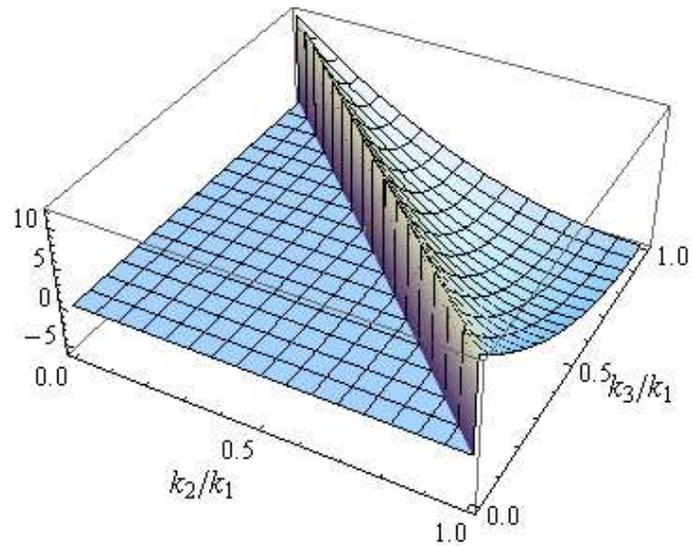, width=9cm}
\end{center}
\medskip
\caption{An ansatz $-S_{\rm ansatz}^{\rm orth}$ in (\ref{ansatz_orth}) for the orthogonal shape $S_1$ in (\ref{newbase_S1}). Note we added a minus sign in this plot.}
\label{Fig:shape_ansatz_orth}
\end{figure}

\begin{figure}
\begin{center}
%\epsfxsize=10cm
%\epsfbox{Fep.eps}
\epsfig{file=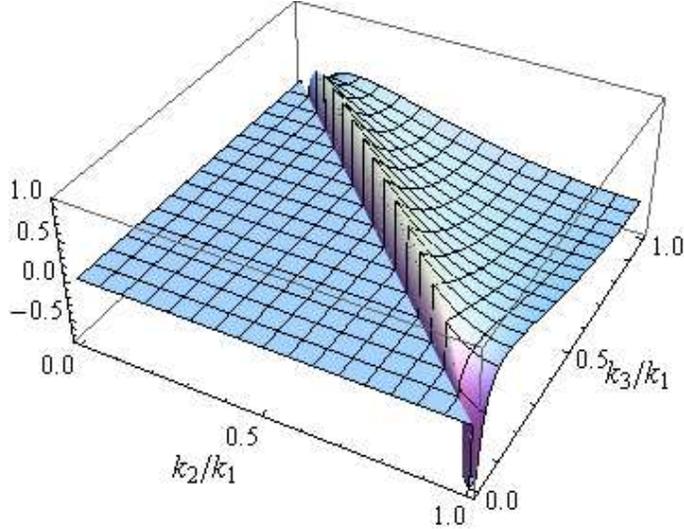, width=9cm}
\end{center}
\medskip
\caption{Another factorizable orthogonal ansatz (\ref{ansatz_orth2}).}
\label{Fig:shape_ansatz_orth2}
\end{figure}

For known examples of general single field inflation, such as the DBI and k-inflation, we generically get equilateral shapes. This is also clear from their physical origin that we have emphasized. The orthogonal shape relies on a delicate cancellation between the two generic shapes. In principle one can do this since the required parameter space is allowed in our effective field theory of general single field inflation in Sec.~\ref{Sec:Eq}, and this may provide guidance to future model building.
For example, one may fine-tune the parameters in the k-inflation models \cite{ArmendarizPicon:1999rj,Li:2008qc}. Therefore,
unlike the previous cases, the direct motivation here is more oriented to data analyses. The advantage of this operation is that it makes full use of data, which impose constraints on both components. In addition, as a bonus, the ansatz for the equilateral (\ref{ansatz_eq}), folded (\ref{ansatz_fold}) and orthogonal (\ref{ansatz_orth}) shapes are not linearly independent. As we can see, they all happen to be the equilateral ansatz shifted by a constant shape ansatz ($S={\rm const.}$) \cite{Fergusson:2008ra}. Constraining two orthogonal bases provide efficient constraints on all three of them.

Let us do a more data-analysis oriented exercise. We would like to construct an ansatz that is orthogonal to both local and equilateral ansatz, since both were well constrained by data. (Note that $S_{\rm ansatz}^{\rm orth}$ in (\ref{ansatz_orth}) is not quite orthogonal to the local ansatz, with a correlation $\sim -0.48$). To do this we start with a trial shape $S_{\rm trial}$, and demand the new orthogonal shape
\bea
S_{\rm ansatz}^{\rm orth,2} = S_{\rm trial} + c_1 S^{\rm loc} + c_2 S^{\rm eq}_{\rm ansatz}
\eea
be orthogonal to both the local and equilateral ansatz,
\bea
S_{\rm ansatz}^{\rm orth,2} \cdot S^{\rm loc} = 0 ~,
\quad
S_{\rm ansatz}^{\rm orth,2} \cdot S^{\rm eq}_{\rm ansatz} =0 ~.
\label{orth_cond}
\eea
The simplest factorizable trial shapes can be either the constant shape or the local-like shape $k_1/k_2+{\rm 5~perm.}$, and both give the same result. Let us use the constant shape $S_{\rm trial}=1$ as an example. Solving the conditions (\ref{orth_cond}) gives $c_1=-0.0953$ and $c_2=-0.204$. So the new orthogonal ansatz is
\bea
S_{\rm ansatz}^{\rm orth,2} = 1.19 \left(\frac{k_1^2}{k_2k_3} + {\rm 2~perm.}\right)
-1.22 \left( \frac{k_1}{k_2} + {\rm 5~perm.} \right) + 3.44 ~.
\label{ansatz_orth2}
\eea
The numerical details may change slightly depending on the detailed definition and computation of the inner product (\ref{inner_product}).
The shape is plotted in Fig.~\ref{Fig:shape_ansatz_orth2}. It is somewhat exotic but the ansatz is simple. By construction, this ansatz is much more orthogonal to the local form (with correlation $\sim 0$) than the $S_{\rm ansatz}^{\rm orth}$ currently used in Ref.~\cite{Senatore:2009gt,Komatsu:2010fb}. It also happens to have reasonably large correlation ($\sim 0.86$) with the orthogonal shape in single field inflation (the $S_1$ shown in Fig.~\ref{Fig:ortho_eq}), similar to that ($\sim -0.91$) between $S_1$ and  $S_{\rm ansatz}^{\rm orth}$.
Obviously, other choices of trial shapes can result in more exotic orthogonal shapes.

One can perform a similar orthogonalization for the two shapes in (\ref{slow_roll_S}), now they are both local to start with.
More generally, if a non-Gaussianity has more base components, we can orthonormalize all of them one by one, in the sense of the Gram-Schmidt process.

\end{itemize}

\subsection{Conclusion}

The field of primordial non-Gaussianity is growing rapidly in recent years, with simultaneous progress from the experimental results, data analyses methods, non-linear cosmology theories, physical model buildings, computational techniques, and theoretical formalisms. The progress that we have seen so far is no doubt just a beginning.

In this review, we have studied the primordial non-Gaussianities coming from the inflation models, especially various mechanisms that can produce observable large non-Gaussianities with viable power spectra. We emphasized the fingerprints that different underlying physics leave on non-Gaussian profiles, which break the degeneracy of model building. We described the physical pictures and presented their effective Lagrangians to the extent that they can be recognized when encountered in the inflation model building in a more fundamental theory. We also derived the resulting bispectra and represented them in terms of simple ansatz to the extent that they can be useful to data analyses. With the current rapid progress, we anticipate much more future developments along these lines through refinements and discoveries in both theories and experiments.

The standard model of cosmology -- the Big Bang theory with $\Lambda$CDM -- is now established better than ever, with the precision data coming from the cosmic microwave background and large scale structures. New data will continue to flow from many ongoing and forthcoming experiments. Although Nature does not seem to be obligated to provide us any more information beyond the standard model, exciting possibilities exist that would help us to understand the origin of the Big Bang. These include the more detailed deviations from the scale-invariance of the power spectrum, the primordial gravitational waves that we may detect from the CMB polarization, the isocurvature perturbations between matter and radiation, and the primordial non-Gaussianities.
Without these types of data, the number of theoretical models with degenerate observational consequences proliferate with time and it will be hard to understand the microscopic nature of the inflation beyond our current knowledge, as well as to distinguish inflation from other possible alternatives.
As we have reviewed, primordial non-Gaussianity -- the collider in the very early universe -- is one of the few hopes. It is becoming a target of many modern experiments. We do not know which cards Nature is hiding from us, but we are hoping and preparing for the best.

\medskip
\section*{Acknowledgments}

I would like to thank Rachel Bean, Richard Easther, Girma Hailu, Bin Hu, Min-xin Huang, Shamit Kachru, Eugene Lim, Hiranya Peiris, Sash Sarangi, Gary Shiu, Henry Tye, Yi Wang and Jiajun Xu for valuable collaborations and sharing their insights on the works reviewed here. I would also like to thank James Fergusson, Michele Liguori, David Lyth, David Seery, Paul Shellard and David Wands for very helpful discussions. I am supported by the Stephen Hawking advanced fellowship.

\end{spacing}

\end{document}